\documentclass{aa}  

\usepackage{graphicx}
\usepackage{txfonts}
\usepackage{lipsum}
\usepackage{lscape}             
\usepackage{placeins}           

\usepackage{xcolor}    
\usepackage[dvipsnames]{xcolor}
\usepackage[colorlinks=true,
            linkcolor=blue,
            citecolor=blue,
            urlcolor=blue]{hyperref}

\newcommand{\gaia}{\it Gaia}
\newcommand{\kepler}{\it Kepler}

\newcommand{\sphaverage}{$\langle S_\mathrm{\!ph} \rangle$}
\newcommand{\prot}{$P_\mathrm{rot}$}
\newcommand{\porb}{$P_\mathrm{orb}$}
\newcommand{\deltalogprot}{$\delta \log{P_\mathrm{rot}}$}
\newcommand{\CaiiIRT}{\ion{Ca}{ii} IRT}
\newcommand{\logRIRT}{$\log{R^{\prime}_{\mathrm{IRT}}}$}

\newcommand{\teff}{$T_\mathrm{eff}$}

\begin{document}
    \titlerunning{Enhanced chromospheric magnetic activity after the intermediate-period gap}
    \authorrunning{Godoy-Rivera et al.}

   \title{Hints of enhanced magnetic activity after the intermediate rotation period gap as traced by the chromospheric \ion{Ca}{ii} infrared triplet}



   \author{
        Diego Godoy-Rivera\inst{1,2}\fnmsep\thanks{Corresponding author: \email{godoyrivera.astro@gmail.com}}
        \and 
        Savita Mathur\inst{1,2}
        \and
        Tyler Richey-Yowell\inst{3}
        \and
        Ângela R. G. Santos\inst{4,5,6}
        \and
        Rafael A. García\inst{6}
        \and
        Desmond H. Grossmann\inst{1,2}
        \and
        Zachary R. Claytor\inst{7}
        \and
        Paul G. Beck\inst{1,2}     
        }

   \institute{
            Instituto de Astrofísica de Canarias (IAC), 38205 La Laguna, Tenerife, Spain
            \and Universidad de La Laguna (ULL), Departamento de Astrofísica, 38206 La Laguna, Tenerife, Spain
            \and Lowell Observatory, 1400 W. Mars Hill Road, Flagstaff, AZ 86001, USA
            \and Instituto de Astrofísica e Ciências do Espaço, Universidade do Porto, CAUP, Rua das Estrelas, PT4150-762 Porto, Portugal
            \and Departamento de F\'isica e Astronomia, Faculdade de Ciências, Universidade do Porto, Rua do Campo Alegre 687, PT4169-007 Porto, Portugal
            \and
            Université Paris-Saclay, Université Paris Cité, CEA, CNRS, AIM, 91191 Gif-sur-Yvette, France
            \and
            Space Telescope Science Institute, 3700 San Martin Drive, Baltimore, MD 21218, USA
            }


 
  \abstract{   
  For low-mass stars ($M$ < 1.4 M$_{\odot}$), the connection between stellar rotation and magnetic activity governs stellar spin-down, shapes the environments of their exoplanets, and provides an age-diagnostic via magneto-gyro-chronology. Recently, unexpected phenomena known as the intermediate rotation period gap and the rotational stalling have been discovered. These are likely due to internal angular momentum redistribution, and mark departures from a smooth spin-down evolution. These rotational features have been shown to cause enhanced magnetic activity on the photosphere, as measured by the photometric index from light curves ({\sphaverage}), in both cluster and field stars. However, their influence on other magnetic activity proxies, and particularly in field stars, remains poorly understood. In this work, we study the impact of the intermediate-period gap on chromospheric magnetic activity as traced by the \ion{Ca}{ii} infrared triplet (IRT) index. We target the stars observed by the {\kepler} mission, as this is the largest and most reliable sample of field stars with measured rotation periods sensitive to the gap. We calculate the {\CaiiIRT} index for the {\kepler} stars using the spectroscopic information from the {\gaia} mission data release three (DR3). We study the rotation-activity relation as a function of location on the Hertzsprung-Russell (HR) diagram and spectral type, finding that K dwarfs are more active than G dwarfs, which in turn are more active than F dwarfs. For main-sequence stars, we find that chromospheric magnetic activity is also enhanced after the intermediate-period gap, mirroring its effect on the photospheric {\sphaverage} index. Our work reveals that the intermediate-period gap marks a genuine transition in stellar magnetic behavior, not only at the photosphere but also at the chromosphere. This highlights the need to account for its signatures across activity proxies, as well as its impact on exoplanet habitability and the age-rotation-activity relation.
  }


   \maketitle
   \nolinenumbers
\section{Introduction}
\label{sec:introduction}

Stellar magnetic activity is a fundamental ingredient in how low-mass main-sequence (MS) stars, namely those with masses $\lesssim 1.4 M_{\odot}$ and convective envelopes, evolve and interact with their surroundings (e.g., \citealt{gudel07,stassun14,airapetian20}). Activity is tightly connected to rotation through dynamo processes operating in their convective envelopes and interiors (e.g., \citealt{brun17,charbonneau20}). Magnetized stellar winds remove angular momentum and brake the stellar surface \citep{parker58,schatzman62,weber67,kawaler88}, so rotational spin-down links surface magnetic activity to the internal transport of angular momentum and, more generally, to the long-term evolution of stellar rotation (e.g., \citealt{denissenkov10,gallet13,gallet15,vansaders13}). 

Along their evolution, stars inhabit different activity and rotation regimes. They are born as rapid rotators in the saturated regime, where magnetic activity is roughly independent of rotation rate (e.g., \citealt{vilhu87,stauffer94,jeffries11}). As they age and spin down, stars transition to the unsaturated regime where rotation and activity are closely connected (e.g., \citealt{pizzolato03,wright11,wright18}), virtually forgetting their initial conditions and populating the converged rotational sequence. In this latter regime, both rotation and magnetic activity decay in a broadly systematic way ($\propto$ age$^{-1/2}$; \citealt{skumanich72}), enabling the use of rotation and activity as age diagnostics through gyro-chronology (e.g., \citealt{barnes03,barnes07,epstein14,garcia14a,angus15,angus19,godoyrivera21a,bouma23,stassun24,vanlane25}) and magneto-chronology (e.g., \citealt{mamajek08,gondoin18,lorenzooliveira18,mathur23,ye24,carvalhosilva25}). 

Beyond stellar physics, magnetic activity sets the conditions in which exoplanets form and evolve, shaping star–planet interactions (e.g., \citealt{strugarek18,ahuir21,allan26,pezzotti26}) and influencing the potential habitability of planetary systems (e.g., \citealt{vidotto13,richeyyowell19,richeyyowell23,johnstone21,see25}).

A detailed understanding of the rotation–activity connection has therefore become central to models of stellar and planetary evolution (e.g., \citealt{matt15,tripathi21}). The Skumanich-like spin-down picture for low-mass stars described above ($\propto$ age$^{-1/2}$) has been challenged over the past decade by new photometric and spectroscopic observations (e.g., \citealt{angus15,vansaders16,garraffo18,hall21,saunders24}). Early hints in open clusters suggested that rotational sequences did not always follow the expected smooth age progression (e.g., \citealt{meibom11,agueros18}). This behavior was later studied in more detail by \citet{curtis19}, who showed that the rotational sequences of the open clusters Praesepe and NGC~6811 largely overlap for K dwarfs despite their different ages (0.7 vs. 1.0 Gyr, respectively), revealing a spin-down stalling phase in which stars remain rotating more rapidly than predicted by standard gyro-chronology relations (see also \citealt{curtis20,gruner20,dungee22}; Bernizzoni et al. in preparation). Further departures from  Skumanich-like spin-down have been identified in field stars, in the form of a bimodal distribution of rotation periods separated by a sparsely populated gap, hereafter referred to as the intermediate-period gap (e.g., \citealt{mcquillan14,reinhold20,gordon21,lu22}).

The physical origin of the intermediate-period gap and the spin-down stalling are still under discussion, with several studies proposing the internal redistribution of angular momentum as a mechanism to explain them (e.g., \citealt{spada20,spada26,lu24,han26}). In this scenario, angular momentum is transferred from the rapidly rotating core to the slowly rotating envelope, temporarily compensating the envelope magnetic braking, and resulting in a stalled surface rotation. From an observational point of view, these phenomena should leave a measurable imprint on stellar magnetic activity. If the surface rotation does not follow the standard spin-down, dynamos are expected to respond, and activity diagnostics should reflect this modified rotational history. Indeed, several investigations of magnetic activity around these features have been published recently (e.g., \citealt{corsaro21,see21,chahal23,degott25}).

Hints of changes in magnetic activity associated with the above were reported by \citet{santos24}, based on the coronal X-ray luminosities of \citet{wright11}, in terms of a change in the slope of the rotation-activity relation of field stars around the intermediate-period gap, as well as by \citet{richeyyowell22} in terms of the chromospheric UV flux of cluster K dwarfs showing a prolonged saturation near the age of their spin-down stalling phase. Further evidence has been provided by studying the photospheric spot-filling fractions of cluster stars (\citealt{cao23}), and the photometric variability of both field and cluster stars (\citealt{mathur25} and \citealt{santos25}, respectively; see also Borg et al. in preparation). While these works have encompassed different atmospheric layers and stellar settings, enhanced magnetic activity in the chromosphere as probed by field stars has only been detected in \ion{Ca}{ii}~H\&K \citep{ye25}. Expanding such detection to other chromospheric activity diagnostics would provide supporting evidence that intermediate-period gap marks a universal transition in magnetic behavior across the stellar atmosphere and a range of ages.

Chromospheric activity indicators offer a sensitive probe of magnetic fields in the outer atmospheres of cool stars (\citealt{hall08,linsky17}; Godoy-Rivera et al. in preparation). Among them, the \ion{Ca}{ii} infrared triplet (IRT) has emerged as a powerful tracer of chromospheric activity in low-mass stars (e.g., \citealt{shine72,linsky79,martin17,zills24}). This index presents the advantage of being accessible in large modern spectroscopic surveys, being less affected by interstellar absorption than traditional optical lines, and having an intrinsically higher signal-to-noise ratio at longer wavelengths than the widely used \ion{Ca}{ii}~H\&K index in cool stars (e.g., \citealt{fritzewski21,lanzafame23,strassmeier25}). With the goal of building a comprehensive picture between rotational evolution and chromospheric diagnostics, we investigate whether the intermediate-period gap imprints a measurable signature on magnetic activity as traced by the \ion{Ca}{ii} IRT in low-mass field stars.

This paper is organized as follows. In Sect.~\ref{sec:data_targetrotation} we describe the target sample and its photometric rotation and photospheric magnetic activity proxies. In Sect.~\ref{sec:data_calcium} we present the chromospheric \ion{Ca}{ii} IRT magnetic activity index. In Sect.~\ref{sec:characterization} we assess the sample overlap between these photospheric and chromospheric activity proxies, and examine their degree of agreement. In Sect.~\ref{sec:discussion} we analyze the relation between rotation and the \ion{Ca}{ii} IRT index, and its dependence on spectral type, Rossby number, and stellar multiplicity. We conclude in Sect.~\ref{sec:conclusions}.
\section{Target list and photometric rotation-activity data}
\label{sec:data_targetrotation}

\begin{table}
\caption{Target list, stellar properties, and activity indices.}
\label{tab:table_catalog}
\centering                         
\scriptsize 
\begin{tabular}{ll}
\hline
\hline
Column & Description\\
\hline 
KIC & KIC ID\\
{\gaia} DR3 & {\gaia} DR3  source ID\\
$M_{G_0}$ & Absolute de-reddened $G$-band magnitude\\
$(BP-RP)_0$ & De-reddened $(BP-RP)$ color\\
Summary Flag CMD & Identifies Joint-MS and Joint-Evolved targets\\
Spectral Type Joint MS & Identifies spectral type of Joint-MS targets\\
Flag Binary Union & Flags the targets identified as binary candidates\\
{\prot} & Rotation period\\
$\sigma_{P_\mathrm{rot}}$ & Rotation period error\\
{\sphaverage} & Photometric activity proxy\\
$\sigma_{\langle S_\mathrm{\!ph} \rangle}$ & Photometric activity proxy error\\
$\delta \log{P_\mathrm{rot}}$ & Difference of logarithmic period (star minus gap)\\
($\mathrm{Ro}/\mathrm{Ro}_{\odot}$) & Rossby number (solar-normalized)\\
$\log{R'_{\mathrm{IRT}}}$ & Calcium activity index\\
$\sigma_{\log{R'_{\mathrm{IRT}}}}$ & Calcium activity index error\\
{\porb} & Orbital period\\
$\sigma_{P_\mathrm{orb}}$ & Orbital period error\\
$\mathrm{e}$ & Orbital eccentricity\\
$\sigma_{\mathrm{e}}$ & Orbital eccentricity error\\
\hline
\\
\end{tabular}
\tablefoot{The full table is available in Sect.~``Data availability.''}
\end{table}

We focused on the stars observed by {\kepler} \citep{borucki10}, a mission that provided long-baseline ($\sim 4$ years), high-precision photometric light curves. The {\kepler} targets were recently characterized using the third data release (DR3) of the {\gaia} mission \citep{gaia16,gaia21,gaia23a}, allowing a detailed analysis of their color-magnitude diagram (CMD; Sect.~\ref{subsec:data_targetrotation_targetlist}). Combined with the {\kepler} observations, this sample offers the most comprehensive view of photometric  activity and rotation for field stars in a regime sensitive to the intermediate-period gap (Sect.~\ref{subsec:data_targetrotation_ProtSph} and ~\ref{subsec:data_targetrotation_gap}). We report our target list in Table~\ref{tab:table_catalog}, together with the main properties used throughout this work.
\subsection{Target sample}
\label{subsec:data_targetrotation_targetlist}

\begin{figure}[ht]
    \centering
    \includegraphics[width=0.91\hsize]{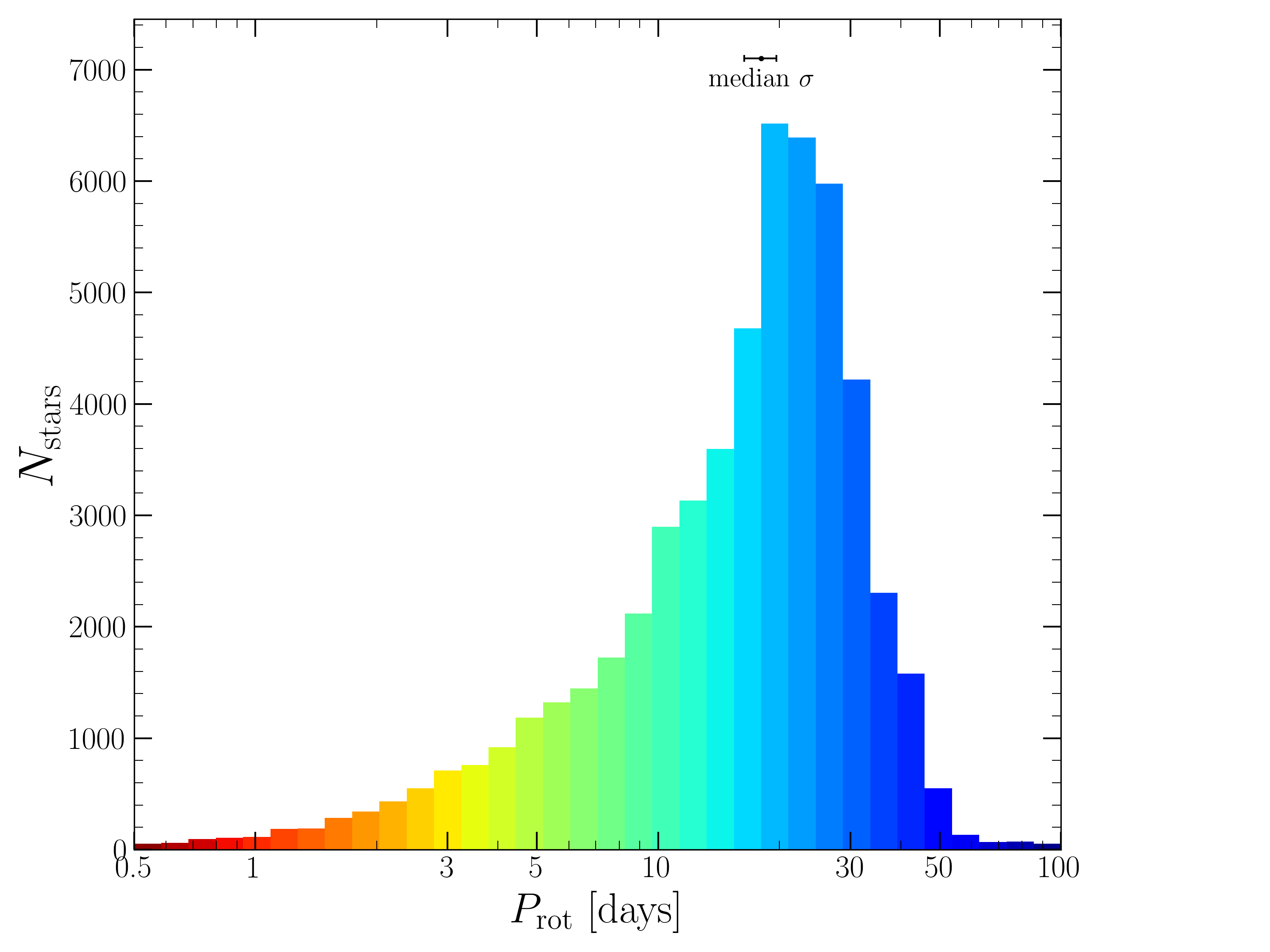}\\
    \includegraphics[width=0.91\hsize]{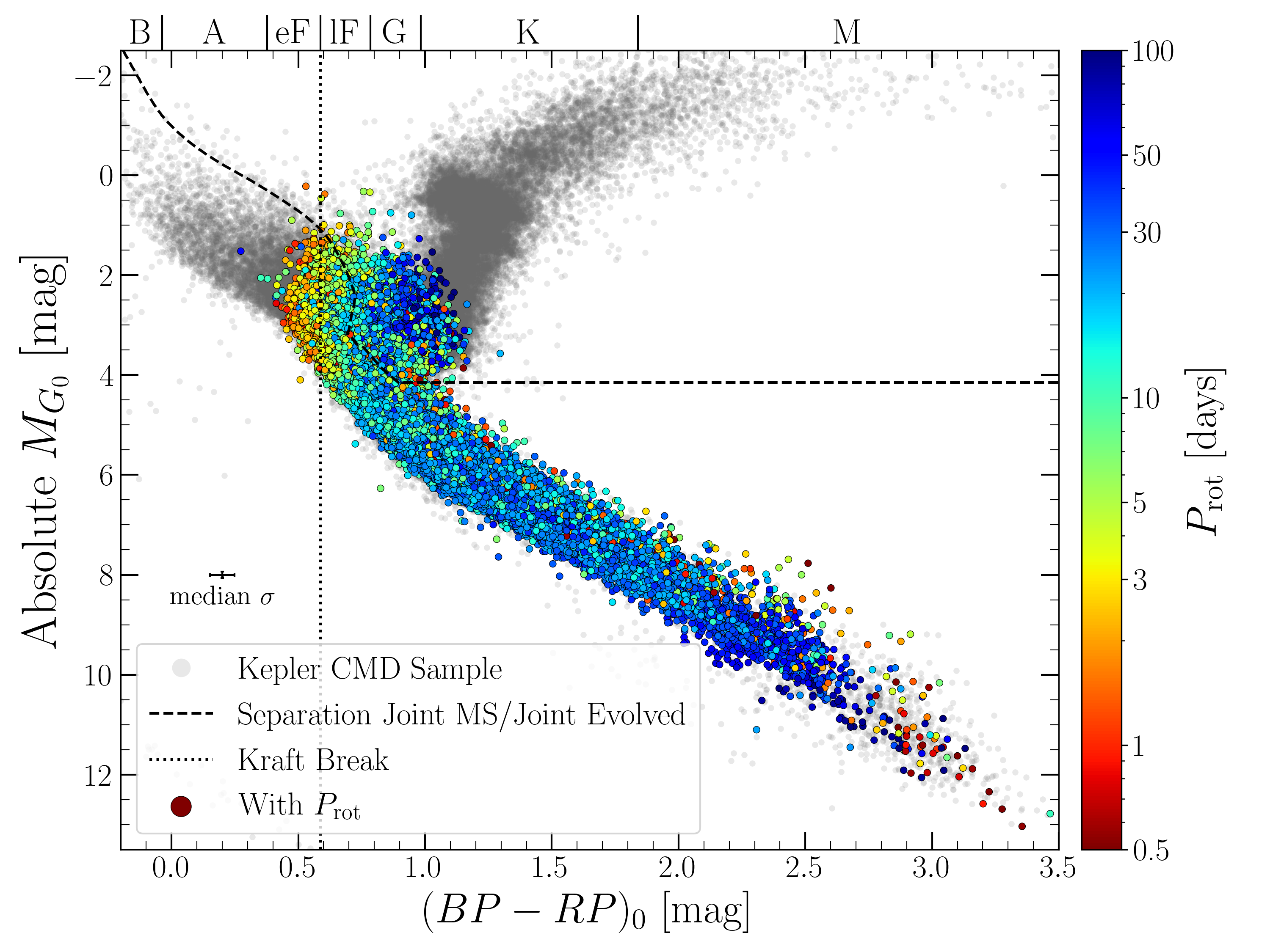}\\ 
    \includegraphics[width=0.91\hsize]{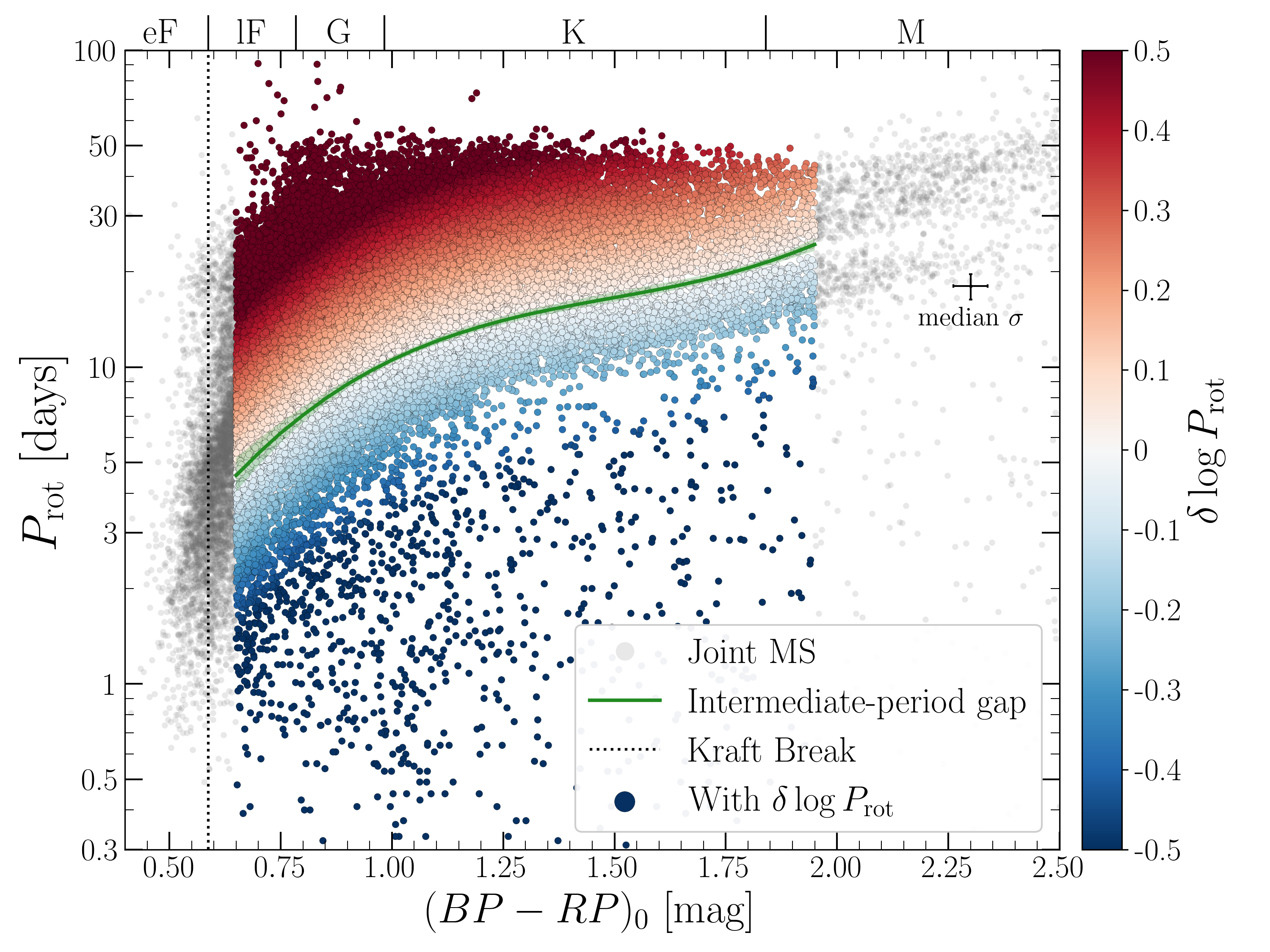} 
    \caption{Characterization of the rotating {\kepler} stars. Top: distribution of rotation periods. Middle: CMD with the full {\kepler} sample shown in grey, and the subset with measured rotation periods shown colored by their {\prot} values. The dashed line shows the separation between Joint-MS and Joint-Evolved targets, and the vertical dotted line illustrates the Kraft break. Bottom: {\prot} vs. color diagram for the Joint-MS targets. The intermediate-period gap is shown as the green line (see Table~\ref{tab:gap_vs_color}). The stars in the color range of the period-gap are color-coded by their {\deltalogprot} values (see Equation~\ref{eqn:delta_log_Prot}). Typical uncertainties are illustrated by the median $\sigma$ symbol.}
    \label{fig:Figure_data_targetrotation}
\end{figure}

We defined the target sample as the 196,762 stars observed by {\kepler} characterized in \citet{godoyrivera25} (see also \citealt{mathur17}). These stars have {\kepler} input catalog (KIC; \citealt{brown11}) IDs with {\gaia} DR3 counterparts in the \texttt{one-to-one} \texttt{Gaia-Kepler.fun}\footnote{\url{https://gaia-kepler.fun/}} crossmatch. They predominantly span apparent magnitudes between $12 < G < 16$ mag, with a median distance of $\approx 1.1$ kpc. The middle panel of Fig.~\ref{fig:Figure_data_targetrotation} shows the absolute and de-reddened CMD of the sample, which includes solar-like stars, upper MS stars, and evolved phases such as the red giant branch (RGB) and red clump (RC).

To streamline our analysis, we divided the {\kepler} CMD sample into two broad categories, namely ``Joint MS'' and ``Joint Evolved''. Based on the analysis of \citet{godoyrivera25}, the Joint-MS category comprises the stars located along the MS (specifically their \texttt{Dwarf}, \texttt{Photometric~Binary}, \texttt{Overlap Dwarf/Subgiant}, and \texttt{Uncertain~MS} regions) and amounts to 131,258 targets (66.7\% of the sample), while the Joint-Evolved category comprises post-MS stars (specifically their \texttt{Subgiant} and \texttt{Giant~Branch} regions) and amounts to 47,999 targets (24.4\% of the sample). This separation provided a straightforward CMD classification\footnote{We note that the Joint MS and Joint Evolved categories amount to $\approx$ 91.1\% of the total {\kepler} sample, with the remaining $\approx$ 8.9\% being composed of stars that failed the quality cuts for a reliable CMD placement and categorization (see Table~1 in \citealt{godoyrivera25}).}, which we illustrate as the dashed line in the middle panel of Fig.~\ref{fig:Figure_data_targetrotation}. We report the resulting classification as the ``Summary Flag CMD'' column in Table~\ref{tab:table_catalog}.

We note that throughout this work, whenever possible, we chose to characterize our sample in the de-reddened {\gaia} DR3 color and absolute magnitude space. While several literature catalogs report stellar parameters (such as effective temperature and luminosity) for the {\kepler} targets (e.g., \citealt{zhang25}), {\gaia} DR3 is the best current source to ensure a homogeneous, precise, and unbiased characterization for the full sample \citep{godoyrivera25}.
\subsection{Rotation periods and photometric magnetic activity}
\label{subsec:data_targetrotation_ProtSph}

Rotation periods and magnetic activity proxies have been measured for {\kepler} solar-like targets thanks to active regions moving into and out of view as stars rotate (e.g., \citealt{mcquillan14,sagynbayeva25}). In this work, we adopted the values of rotation period ({\prot}) and average photometric activity proxy ({\sphaverage}) from \citet{santos19} and \citet{santos21}. These were derived from the long-cadence {\kepler} light curves\footnote{Available at MAST via: \url{https://archive.stsci.edu/doi/resolve/resolve.html?doi=10.17909/t9-mrpw-gc07}.}, calibrated following \citet{garcia11}, \citet{garcia14b}, and \citet{pires15}. From these, {\prot} was obtained from the rotationally modulated variability, and {\sphaverage} was calculated as the mean of the standard deviations of the light curve measured over segments of length $5\times${\prot} \citep{mathur14a}. While other literature catalogs report complementary (or even larger) samples of rotation periods and/or activity proxies for the {\kepler} stars (e.g., \citealt{long23,reinhold23,claytor25,degott25,kamai25}), we prioritized the values by \citet{santos19} and \citet{santos21}. This choice was motivated by their {\prot} and {\sphaverage} measurements being derived in a homogeneous and self-consistent manner, thus ensuring a uniform treatment across the sample. Importantly, the above definition of {\sphaverage} ensures its physical interpretation as a proxy for photospheric magnetic activity \citep{mathur14a,mathur14b,salabert16,salabert17,kashyap26}.

We crossmatched our target list (Sect.~\ref{subsec:data_targetrotation_targetlist}) with the \citet{santos19} and \citet{santos21} catalog, and found 54,881 targets in common (27.9\% of the sample). The resulting {\prot} distribution is shown in the top panel of Fig.~\ref{fig:Figure_data_targetrotation}. The distribution peaks at $\sim 20$ days and spans a broad range of values (from below 1 up to 100 days), consistent with a field population (e.g., \citealt{vansaders19}). In the middle panel of Fig.~\ref{fig:Figure_data_targetrotation}, the CMD of this sample is color-coded by {\prot}, which shows the color-dependence of the rotation periods of low-mass stars. This is also illustrated in the bottom panel of Fig.~\ref{fig:Figure_data_targetrotation}. Early-type solar-like stars tend to be faster rotators than later-type stars. This is particularly true for the {\kepler} sample, which is mostly comprised of relatively old stars that have already transitioned into the unsaturated regime and the converged rotational sequence (e.g., \citealt{santos24}). Above the Kraft break \citep{kraft67}, corresponding to F5V stars (\citealt{beyer24}; $(BP-RP)_0 \lesssim$ 0.587 mag or {\teff} $\gtrsim$ 6,550 K from \citealt{pecaut13}), illustrated as the vertical dotted lines in the middle and bottom panels of Fig.~\ref{fig:Figure_data_targetrotation}, stars tend to remain rapid rotators ({\prot}$\lesssim$ 5 days) due to an inefficient magnetic braking related to their thin convective layers (e.g., \citealt{matt15}). Below the Kraft break, $(BP-RP)_0 \gtrsim$ 0.587 mag, we see the effects of spin-down due to angular momentum losses in stars with convective envelopes. On the Joint-Evolved zone of the CMD, we see the impact of structural effects (i.e., radius expansion) as stars leave the MS and evolve along the RGB.
\subsection{Intermediate-period gap}
\label{subsec:data_targetrotation_gap}

In part of the color range probed by the bottom panel of Fig.~\ref{fig:Figure_data_targetrotation}, the {\prot} distribution of the {\kepler} MS stars is bimodal. This underdensity of stars at intermediate rotation periods is known as the intermediate-period gap (e.g., \citealt{mcquillan13,mcquillan14,davenport17,davenport18,santos19,santos21}). Near this dearth region, stars tend to have lower {\sphaverage} values than stars with longer periods. This behavior is counter-intuitive, as in the unsaturated regime, faster rotators are generally expected to show higher activity (see Sect.~\ref{sec:introduction}). Consequently, in the {\sphaverage}-{\prot} diagram, the gap corresponds to a local minimum in activity (see Fig.~B.6 in \citealt{santos23}). This has been used to determine the location of the gap as a function of {\teff} (e.g., \citealt{reinhold20,santos24,santos25}).

Throughout this work, we followed the above definition of the intermediate-period gap, namely the {\prot} value at which there is a local minimum in {\sphaverage}. To maximize homogeneity (see Sect.~\ref{subsec:data_targetrotation_targetlist}), we determined the location of the gap as a function of de-reddened {\gaia} DR3 color instead of effective temperature. We followed the procedure described in Appendix A of \citet{santos25}, adapted to $(BP-RP)_0$ colors instead of {\teff}. The resulting parametrization is reported in Table~\ref{tab:gap_vs_color} and shown by the green solid line (and shaded region around it) in the bottom panel of Fig.~\ref{fig:Figure_data_targetrotation}. The gap parametrization was done for the range $0.65\leq(BP-RP)_0\leq 1.95$. Beyond the bluest boundary, the {\prot} distribution is not found to be bimodal, and an activity minimum is difficult to define. Beyond the reddest boundary, the sample sizes become increasingly smaller, which hampers the detection of the local minimum in {\sphaverage}. These are also reflected in the increasing uncertainty of the gap location towards the bluer and redder extremes. To account for the uncertainties on the rotation periods, de-reddened colors, and local minima identification, we performed bootstrapping by varying the rotation period values, the $(BP-RP)_0$ values, the size of the $(BP-RP)_0$ bins, and the size of the $\log P_\text{prot}$ bins. We followed the steps detailed in \citet{santos25}, with the size of the $(BP-RP)_0$ bins being randomly selected between 0.01, 0.02, 0.03, 0.04, and 0.05 mag.

\begin{table}
\caption{Intermediate-period gap ($\pm 1\sigma$) as a function of de-reddened {\gaia} DR3 $(BP-RP)_0$ color.}
\label{tab:gap_vs_color}
\centering                         
\scriptsize 
    \begin{tabular}{cccc}
    	\hline
    	\hline
        $(BP-RP)_0$ & $P_\text{\!rot,gap-16$^{\text{th}}$}$ & $P_\text{\!rot,gap-50$^{\text{th}}$}$ & $P_\text{\!rot,gap-84$^{\text{th}}$}$\\
        \hline
    0.65 & 0.608 & 0.658 & 0.713 \\
	0.70 & 0.689 & 0.728 & 0.772 \\
    0.75 & 0.762 & 0.792 & 0.826 \\
    0.80 & 0.826 & 0.848 & 0.874 \\
    0.85 & 0.883 & 0.900 & 0.919 \\
       \ldots & \ldots & \ldots & \ldots \\
       \hline
    \end{tabular}
\tablefoot{The full table is available in Sect.~``Data availability.''}
\end{table}

For later analysis (see Sect.~\ref{sec:discussion}), it is important to know how distant a given star is, in terms of rotation period, from the intermediate-period gap. For this purpose, we followed \citet{santos25} and calculated the parameter
\begin{equation}
    \delta \log{P_\mathrm{rot}} = \log{P_\mathrm{rot,star}} - \log{P_\mathrm{rot,gap}},
\label{eqn:delta_log_Prot}
\end{equation}
which is the difference between a star's logarithmic {\prot} minus the logarithmic period of the intermediate-period gap evaluated at the star's {\gaia} color. In the bottom panel of Fig.~\ref{fig:Figure_data_targetrotation}, the Joint-MS stars within the color range of the intermediate-period gap are color-coded by their {\deltalogprot} values.
\section{Calcium infrared index}
\label{sec:data_calcium}

The {\CaiiIRT} ($\lambda = 850.03$, $854.44$, and $866.45$ nm) is a well established diagnostic of chromospheric stellar activity (e.g., \citealt{shine72,linsky79,dempsey93,soderblom93,andretta05,busa07,huang24}), with strong correlations with other activity indices such as \ion{Ca}{ii}~H\&K \citep{martin17}. In {\gaia} DR3, a {\CaiiIRT} activity index was reported for $2\times10^6$ stars \citep{lanzafame23}, based on medium resolution spectra ($\lambda/\Delta \lambda \sim 11,500$) collected by the radial velocity spectrometer (RVS) between the wavelengths 845 to 872 nm for targets brighter than apparent $G=13$ mag (for reference, the {\kepler} targets have a median apparent $G=14.6$ mag; \citealt{godoyrivera25}). For each target, the observed spectrum was normalized and compared with a template spectrum representing an inactive photosphere. The template spectrum was subtracted from the observed one to isolate the chromospheric contribution, and the activity index was calculated by averaging over the core of the three lines. This is reported as $\alpha$, found as the \texttt{activityindex\_espcs} parameter in the \texttt{gaiadr3.astrophysical\_parameters} table.

\begin{figure}[ht]
    \centering
    \includegraphics[width=0.91\hsize]{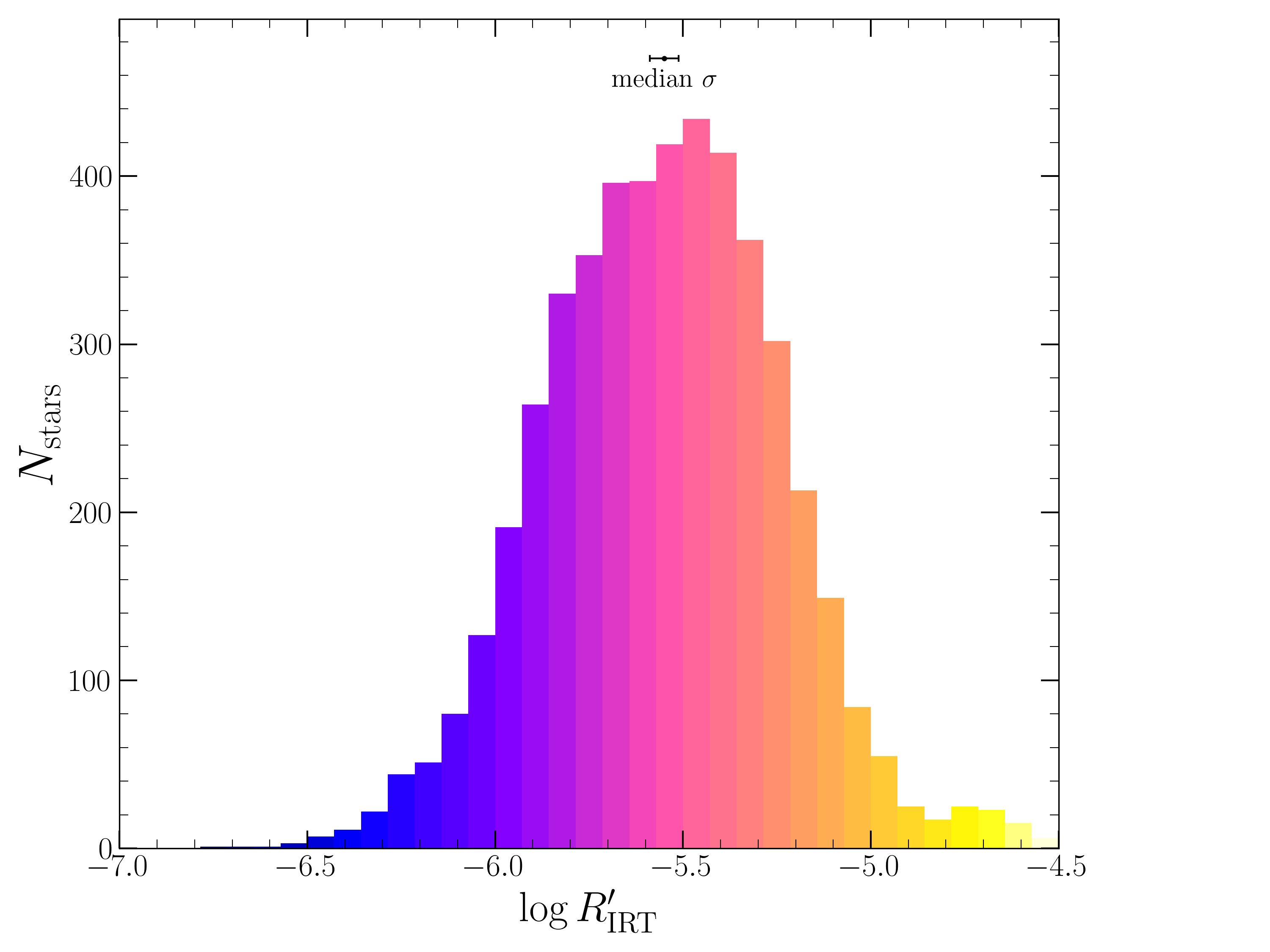}\\
    \includegraphics[width=0.91\hsize]{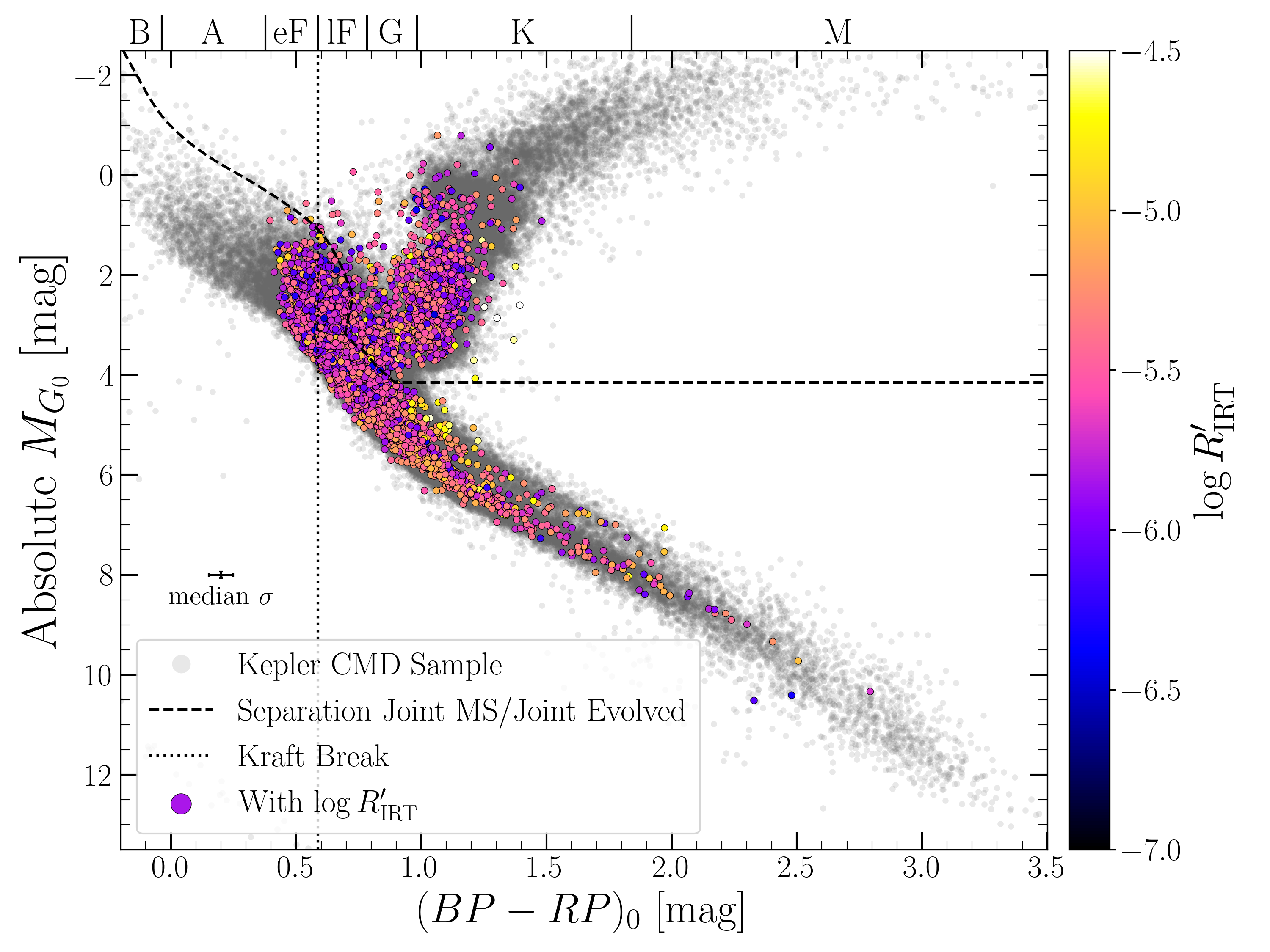}\\
    \includegraphics[width=0.91\hsize]{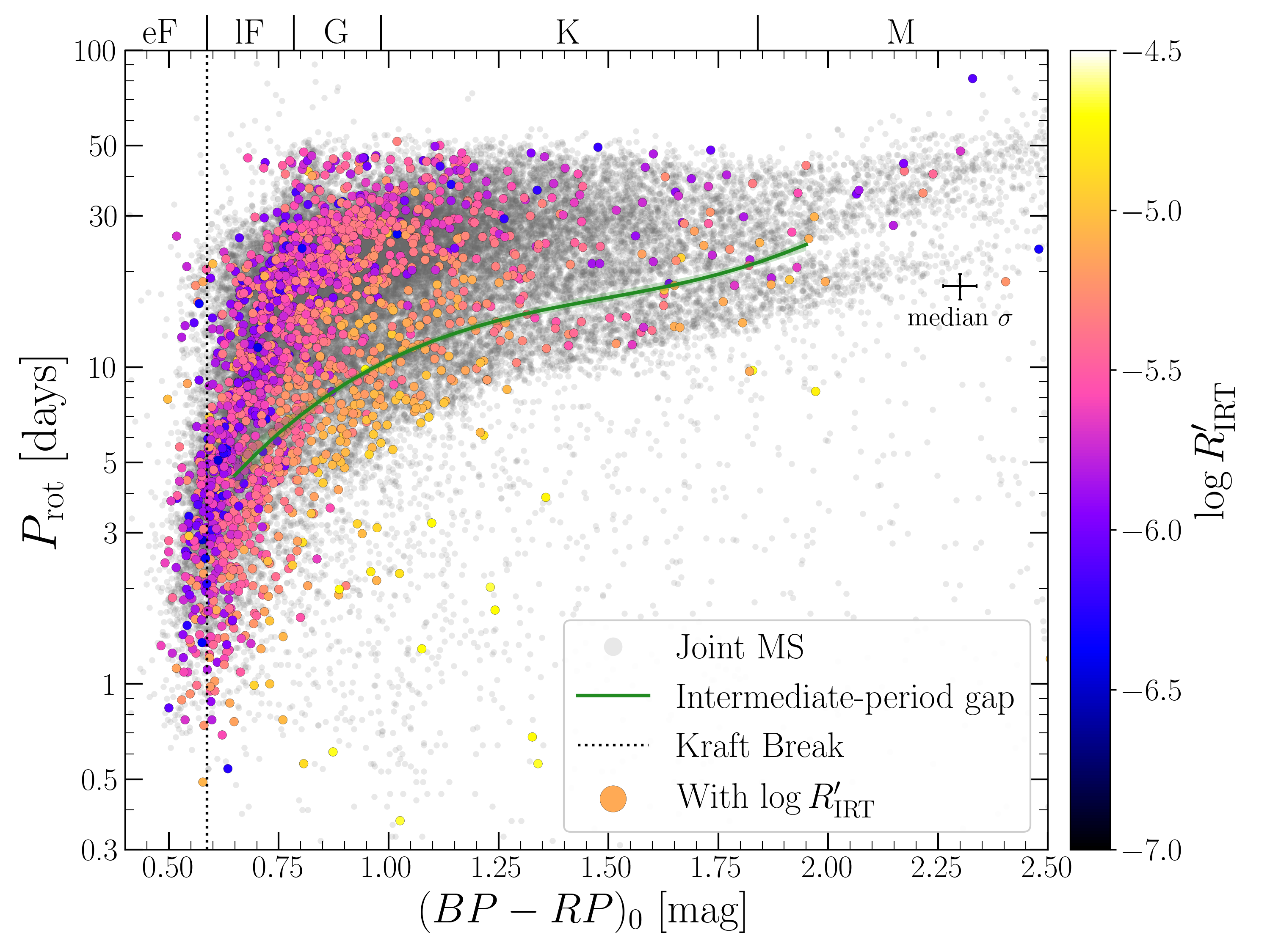}
    \caption{Analogous to Fig.~\ref{fig:Figure_data_targetrotation}, for the {\CaiiIRT} activity index. In all panels the color-coding indicates {\logRIRT} values.}
    \label{fig:Figure_data_calcium}
\end{figure}

We followed \citet{lanzafame23} and converted the $\alpha$ values to the {\logRIRT} index as
\begin{equation}
\log{R^{\prime}_{\text{IRT}}} = \left(C_0 + C_1 \theta + C_2 \theta^2 + C_3 \theta^3 \right) + \log{\alpha},
\label{eqn:calcium_central}
\end{equation}
and propagated uncertainties as
\begin{equation}
\sigma_{\log{R^{\prime}_{\text{IRT}}}} =\sqrt{ \bigg[ \left(C_1 + 2 C_2 \theta + 3 C_3 \theta^2 \right) \sigma_{\theta} \bigg]^2 + \left[\frac{\sigma_{\alpha}}{\alpha \ln(10)} \right]^2 },
\label{eqn:calcium_sigma}
\end{equation}
where $C_{0,1,2,3}$ are the metallicity-dependent coefficients reported in Table 1 of \citet{lanzafame23}, $\sigma_{\alpha}$ is the activity index uncertainty (\texttt{activityindex\_espcs\_uncertainty}), $\theta = \log T_{\text{eff}}$, and $\sigma_{\theta} = \frac{\sigma_{T_{\text{eff}}}}{\ln(10) T_{\text{eff}}}$. For this computation, we took the corresponding set of input temperatures, either \texttt{gspspec} or \texttt{gspphot}, used in the calculation of the template spectrum (reported as \texttt{activityindex\_espcs\_input}). Given that the  $C_{0,1,2,3}$ coefficients are reported in discrete metallicity bins, we interpolated between them and evaluated the coefficients at the corresponding (\texttt{gspspec} or \texttt{gspphot}) metallicity of each star. To ensure reliability on the chromospheric activity index, we followed \citet{breton25} and \citet{freund25}, and applied a 3$\sigma$ quality cut, namely $\alpha/\sigma_{\alpha} > 3$. 

\begin{figure}[ht]
    \centering
    \includegraphics[width=0.7\hsize]{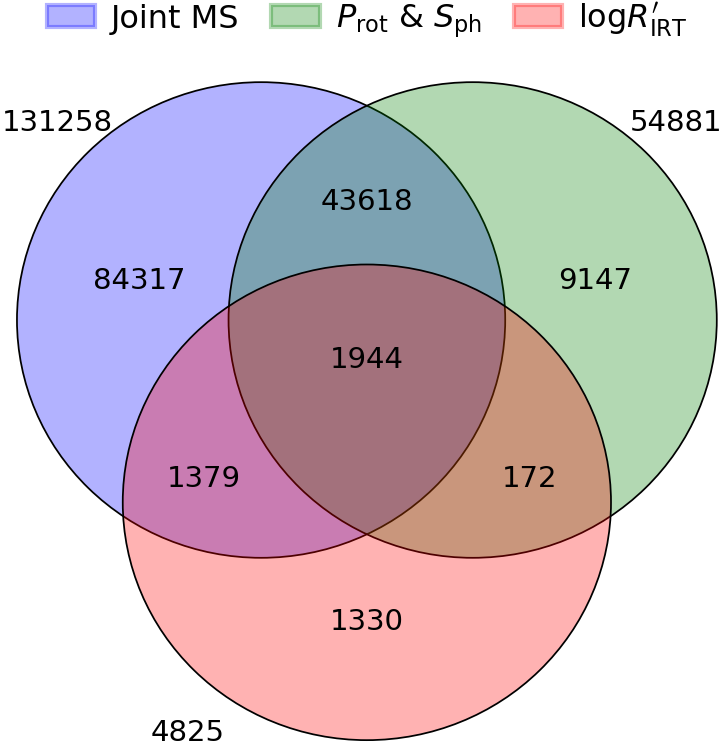} 
    \caption{Venn diagram for the subsets of stars classified as Joint MS (purple), with a measured rotation period and photometric activity index (green), and with a measured {\CaiiIRT} index (red).}
    \label{fig:Figure_characterization_overlap}
\end{figure}

All the above produced a set of {\logRIRT} and $\sigma_{\log{R^{\prime}_{\text{IRT}}}}$ values for 4,825 stars (2.5\% of the full sample), which are reported in Table~\ref{tab:table_catalog}. The top panel of Fig.~\ref{fig:Figure_data_calcium} shows the distribution of the {\CaiiIRT} index, which spans the range from {\logRIRT} $\approx -6.6$ to $\approx -4.5$, and peaks at {\logRIRT} $\approx -5.5$. Given that star formation occurs predominantly close to the Galactic plane, and {\kepler} observed stars away from it, we did not expect a large population of pre-MS stars or accreting T~Tauri stars, which would otherwise occupy the highest activity regime \citep{lanzafame23}.

For further reliability, we tested the significance of the above coefficient-interpolation assumption, as well as the use of different temperature and metallicity values (e.g., compared with the parameters by \citealt{berger20}), on the resulting {\CaiiIRT} index. We found the impact of all these choices to be small, with a median change of $|\Delta \log{R^{\prime}_{\mathrm{IRT}}}| \approx 0.02$. This value is well below the median index uncertainty of $\sigma_{\log{R^{\prime}_{\text{IRT}}}} \approx 0.04$. Along the same line, tests showed that the uncertainties $\sigma_{\log{R^{\prime}_{\text{IRT}}}}$ are heavily dominated by the $\sigma_{\alpha}$ term, and are only weakly dependent on the temperature and metallicity inputs.

The CMD projection of the {\CaiiIRT} sample is shown in the middle panel of Fig.~\ref{fig:Figure_data_calcium}. Despite their rapid rotation, earlier-type MS stars exhibit lower chromospheric activity due to their shallow convective envelopes. For MS stars cooler than the Kraft break, we observe a systematic increase in the {\logRIRT} index towards later spectral types, consistent with lower-mass stars being more chromospherically active, owing to their deeper convective envelopes. Interestingly, high-activity stars seem to be prominent on the photometric binary sequence, namely the region parallel to the MS at higher luminosities due to unresolved companions. This possibly hints at the role that multiplicity plays in enhancing magnetic activity (see Sect.~\ref{sec:discussion}).

The bottom panel of Fig.~\ref{fig:Figure_data_calcium} shows the {\prot} vs. color diagram for the Joint-MS stars, and the subset of them with {\CaiiIRT} values are color-coded by {\logRIRT}. For stars redder (cooler) than the Kraft break (vertical dotted line), at a fixed {\gaia} color, the slower rotators display lower activity values, in line with expectations of a typical unsaturated-regime rotation-activity relation.
\section{Characterization}
\label{sec:characterization}
\subsection{Sample overlap}
\label{subsec:characterization_overlap}

To understand the degree of overlap among the different data sets, Fig.~\ref{fig:Figure_characterization_overlap} shows the Venn diagram\footnote{Created with \texttt{venny4py} (\url{https://github.com/timyerg/venny4py/tree/main}).} of the subsets of stars classified as Joint MS, with {\prot} and {\sphaverage} values, and with {\CaiiIRT} activity index values. Naturally, the Joint-MS sample (purple; $N=131,258$) is the largest set, as the {\kepler} mission focused on observing Sun-like stars, and most of them have reliable {\gaia} data \citep{godoyrivera25}. Regarding the rotating sample (green; $N=54,881$), while a fraction of the targets are classified as Joint Evolved, the majority of the \citet{santos19,santos21} stars correspond to Joint-MS targets, as rotation is more easily detected for stars with shorter periods and higher photometric activity levels. Regarding the sample with {\CaiiIRT} measurements (red; $N=4,825$), while the majority of the stars correspond to Joint-MS targets (69\%), the {\gaia} RVS magnitude limit (apparent $G<13$ mag) resulted in many {\logRIRT} measurements belonging to the Joint-Evolved regime (31\%; see the middle panel of Fig.~\ref{fig:Figure_data_calcium}), i.e., targets that are intrinsically more luminous.

For the remainder of this paper, we focus on the intersection of the above samples, namely the subset of Joint-MS stars with measured rotation periods and photometric activity indices, as well as {\CaiiIRT} activity indices. This subset amounts to $N=1,944$ targets. Given the additional factors that dictate stellar evolution in evolved stars, we leave a thorough study of the Joint-Evolved targets and their rotation-activity relation (e.g., \citealt{dixon20,dixon25,lehtinen20,godoyrivera21b}) as future work.
\subsection{Photospheric vs. chromospheric magnetic activity}
\label{subsec:characterization_sphlogRIRT}

\begin{figure}[ht]
    \centering
    \includegraphics[width=\hsize]{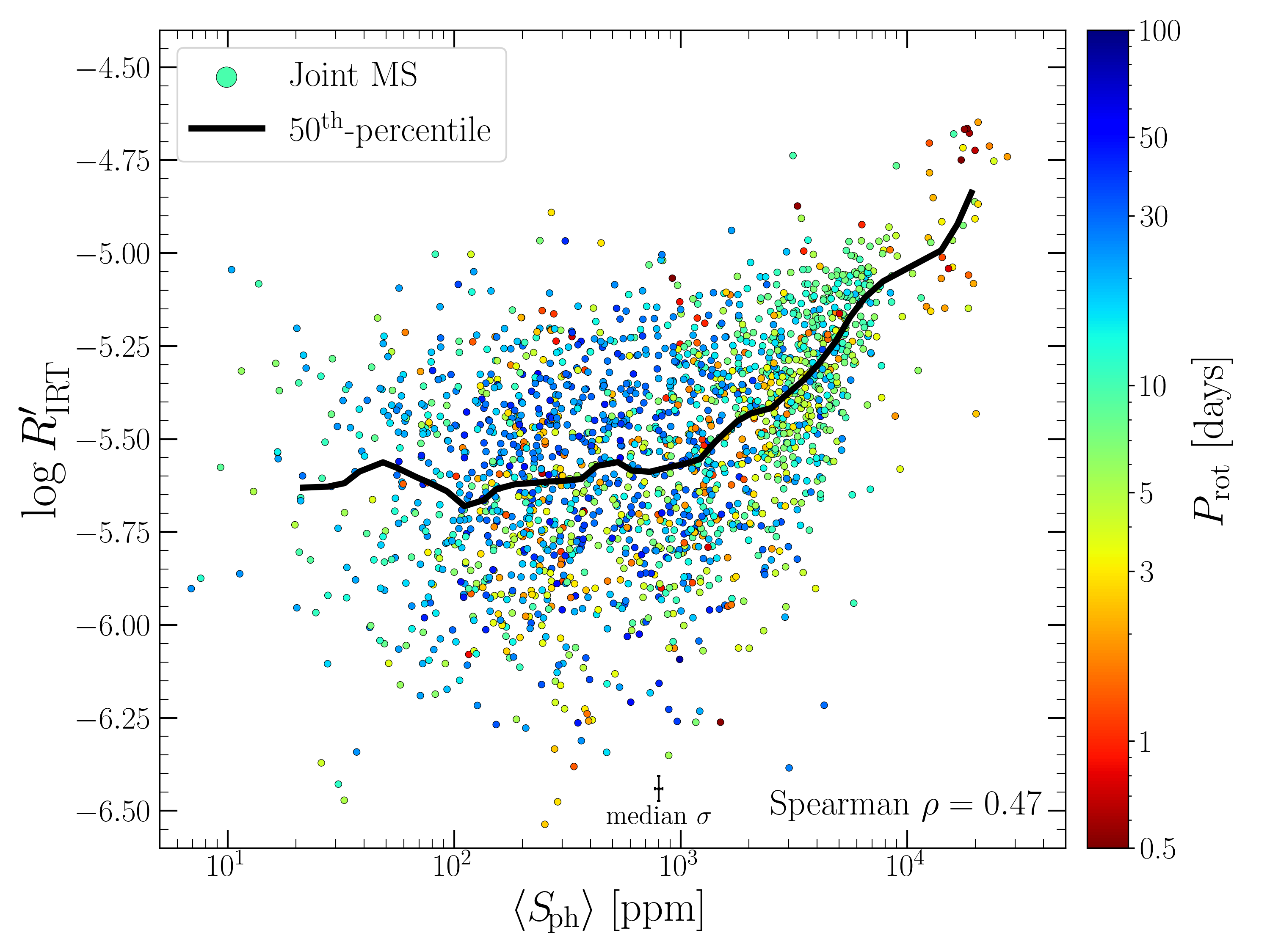}\\
    \includegraphics[width=\hsize]{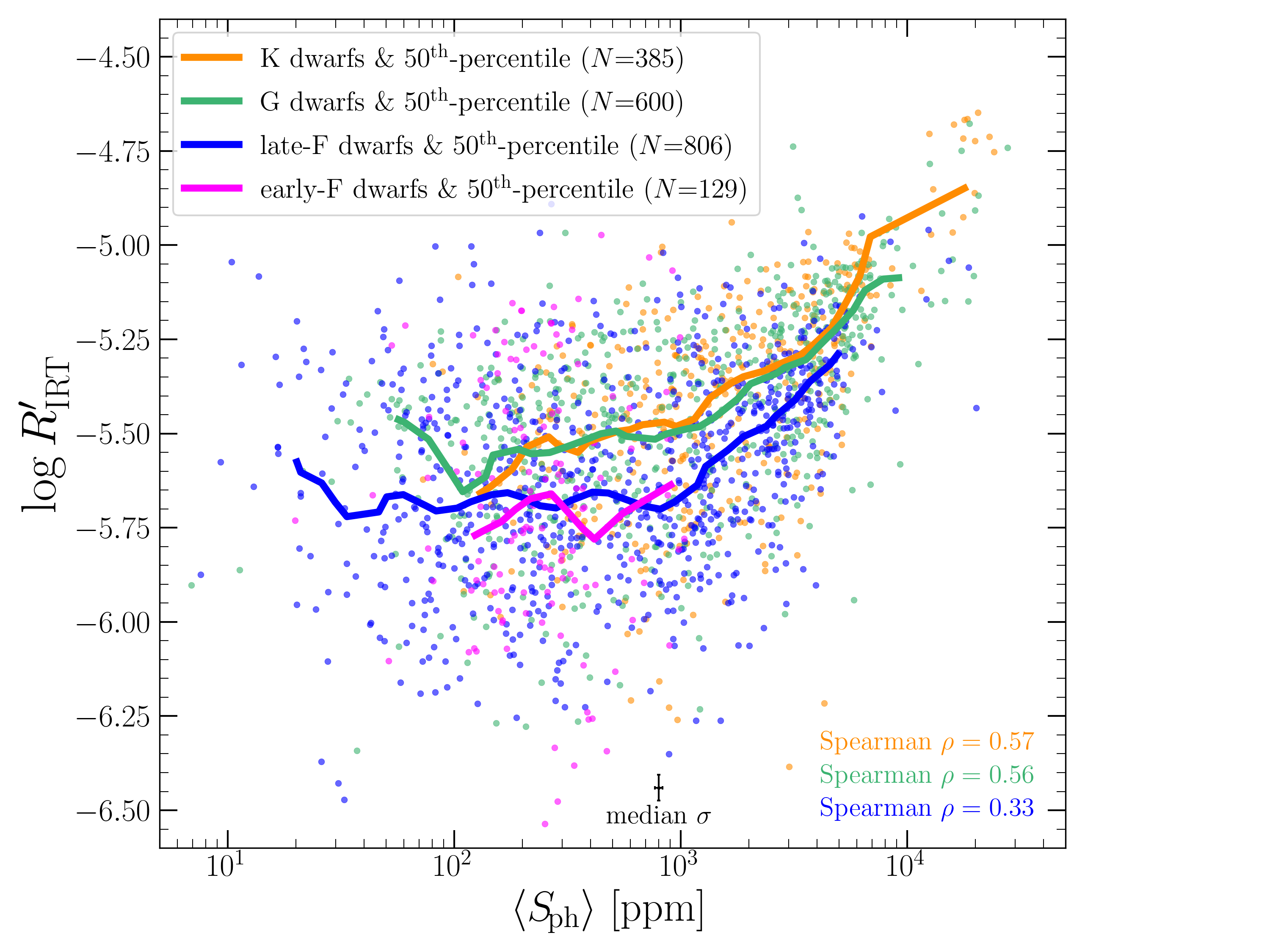}
    \caption{{\CaiiIRT} activity index vs. average photometric activity index for the Joint-MS sample. In the top panel the entire Joint-MS sample is color-coded by {\prot}, whereas in the bottom panel the color-coding reflects spectral type (K dwarfs in orange, G dwarfs in green, late-F dwarfs in blue, and early-F dwarfs in magenta). The solid lines show the respective 50$^{\mathrm{th}}$-percentiles, Spearman coefficients are listed in the bottom-right corners, and typical uncertainties are indicated by the median $\sigma$ symbols. A higher {\sphaverage} typically means a higher {\logRIRT}, with later spectral types exhibiting stronger correlations.}
    \label{fig:Figure_characterization_sphlogRIRT}
\end{figure}

Magnetic activity in solar-like stars gives rise to a variety of phenomena that originate at different depths within the star and its atmosphere. Accordingly, it is interesting to study the degree of correlation between different proxies. The top panel of Fig.~\ref{fig:Figure_characterization_sphlogRIRT} shows the comparison of the photospheric index {\sphaverage} (Sect.~\ref{subsec:data_targetrotation_ProtSph}) vs. the chromospheric index {\logRIRT} (Sect.~\ref{sec:data_calcium}) for the Joint-MS sample. For reference, the solid black line shows the 50$^{\mathrm{th}}$-percentile of the {\logRIRT} distribution in bins of {\sphaverage}. Further details on the binning procedure are described in Appendix~\ref{sec:app_binning}. There is a clear correlation between both proxies, especially in the regime of higher activity (i.e., {\logRIRT} $\gtrsim$ -5.6 and {\sphaverage} $\gtrsim 10^3$ ppm). This is supported by a Spearman correlation coefficient (SCC) of $\rho=0.47$ with a $p$-value of $8.79\times10^{-109}$, indicating a statistically significant, moderate relationship. 

Comparing our results with the literature, we note that they are in good agreement with the sample of {\gaia}/{\it K2} \citep{howell14} stars from \citet{breton25} in the high-activity regime (see their Fig.~11). Moreover, our global trend is qualitatively similar to the one seen in \citet{gehan24} (see their Fig.~3). While their sample is composed of RGB stars and ours is restricted to MS stars, both {\CaiiIRT} index vs. {\sphaverage} diagrams show a relation that steepens for {\sphaverage} $\gtrsim 10^3$ ppm.

We note that, at the very low activity levels in the top panel of Fig.~\ref{fig:Figure_characterization_sphlogRIRT}, stars may be reaching their basal chromospheric flux (e.g., \citealt{schrijver87,gondoin20}; Masseron et al. in preparation), thus impeding the correlation to extend further towards the bottom-left corner of the diagram. In principle, lower values of the {\CaiiIRT} index could have been retrieved from the raw {\gaia} data, specifically in the range of $-8 \lesssim$ {\logRIRT} $\lesssim -6.5$. However, these were excluded when applying the 3$\sigma$ quality cut on Sect.~\ref{sec:data_calcium}, which we adopted to ensure reliable measurements.

To expand our analysis, we test possible dependencies of the {\logRIRT} vs. {\sphaverage} correlation on rotation period. For this, the points in the top panel of Fig.~\ref{fig:Figure_characterization_sphlogRIRT} are color-coded by their {\prot} values. While the steepest part of the correlation ({\sphaverage} $\sim 3 \times 10^3$ ppm) is mostly populated by moderately rapid rotators ({\prot} $\sim$ 10 days), the overall trend holds for a range of rotation periods.

Despite this overall agreement, several factors can naturally introduce deviations between the two proxies. Although both indicators trace magnetic activity, they reflect phenomena occurring in different features and at different layers in the stellar atmosphere. The photometric {\sphaverage} is mostly sensitive to brightness variations caused by dark spots \citep{shapiro16,li24}, while $\log{R^{\prime}_\text{IRT}}$ measures chromospheric emission related to magnetic heating within plage regions (e.g., \citealt{shine72,linsky79,soderblom93}; but see also \citealt{andretta05,zills24}). 

Furthermore, the observing windows of both activity proxies were not contemporaneous, as {\kepler} observed from 2009 to 2013, while {\gaia} DR3 observed from 2014 to 2017. Since stars also exhibit variability in their magnetic activity levels (e.g., \citealt{wilson78,baliunas95,morris25}), the non-overlapping observations are an additional source of scatter in the top panel of Fig.~\ref{fig:Figure_characterization_sphlogRIRT}. Indeed, given the temporal offset between the {\kepler} and {\gaia} observations, since more rapidly rotating stars tend to have shorter magnetic activity cycles (e.g., \citealt{bohmvitense07}), they are more likely to exhibit the same cycle phase in {\logRIRT} and {\sphaverage} compared to slowly rotating stars. As shown by \citet{mathur25} based on solar data, this results in additional scatter being added for slowly rotating stars relative to rapidly rotating stars (see their Appendix~A). Another contributing factor to the scatter is the possible temporal offset between magnetic activity proxies (e.g., \citealt{salabert17}). In spite of the aforementioned caveats, taken together these results indicate that both proxies trace the same underlying magnetic activity, despite probing different atmospheric layers (e.g., \citealt{bjorgen18,zhang20}).

Because the depth of the convective envelope increases towards lower mass stars (e.g., \citealt{kippenhahn13}), and convection plays a central role in generating magnetic activity (e.g., \citealt{charbonneau14}), it is interesting to separate stars by their spectral type (SpT). As described in Appendix~\ref{sec:app_SpT_classification}, we split the Joint-MS sample into early-F, late-F, G, K, and M dwarfs. Their respective {\sphaverage} vs. {\logRIRT} diagrams are shown in the bottom panel of Fig.~\ref{fig:Figure_characterization_sphlogRIRT}. This revealed different levels of agreement between the photospheric and chromospheric indices. The correlation between {\sphaverage} and {\logRIRT} is strongest for K and G dwarfs (SCC of $\rho=0.57$ and $p$-value $= 1.09\times10^{-34}$, and SCC of $\rho=0.56$ and $p$-value $=1.58\times10^{-50}$, respectively) and weakest for late-F dwarfs (SCC of $\rho=0.33$; $p$-value $=2.36\times10^{-22}$). For early-F dwarfs, the correlation is not statistically significant ($p$-value of 0.90). M dwarfs were not analyzed, neither here nor in the upcoming diagrams, given their limited sample size ($< 25$ stars). 

One possible explanation is that magnetic cycle properties, including the cycle period, may depend on spectral type  or rotation period (e.g., \citealt{baliunas95,soon93,saar99,reinhold17,reinhold19,borosaikia18,kitchatinov18,olspert18,willamo20}). If cycle timescales vary across spectral types, magnetic activity levels may evolve on different timescales, potentially increasing the scatter when comparing non-contemporaneous photospheric and chromospheric measurements.

In addition, the relative contributions of spots and faculae/plage may also depend on spectral type and activity level. More active, typically late-type stars, are expected to exhibit a larger spot-to-facula/plage ratio than less active, typically early-type stars (e.g., \citealt{shapiro14}). Similarly, rapidly rotating stars are dominated by spots, while slowly rotating stars are dominated by faculae (e.g., \citealt{montet17,reinhold19}). These could also contribute to the varying {\logRIRT} vs. {\sphaverage} agreement due to the different sensitivity of these magnetic activity proxies, as described above.
\section{Discussion}
\label{sec:discussion}

\subsection{Impact of the intermediate-period gap on chromospheric activity across spectral types}
\label{subsec:discussion_rotact}

\begin{figure}[ht]
    \centering
    \includegraphics[width=\hsize]{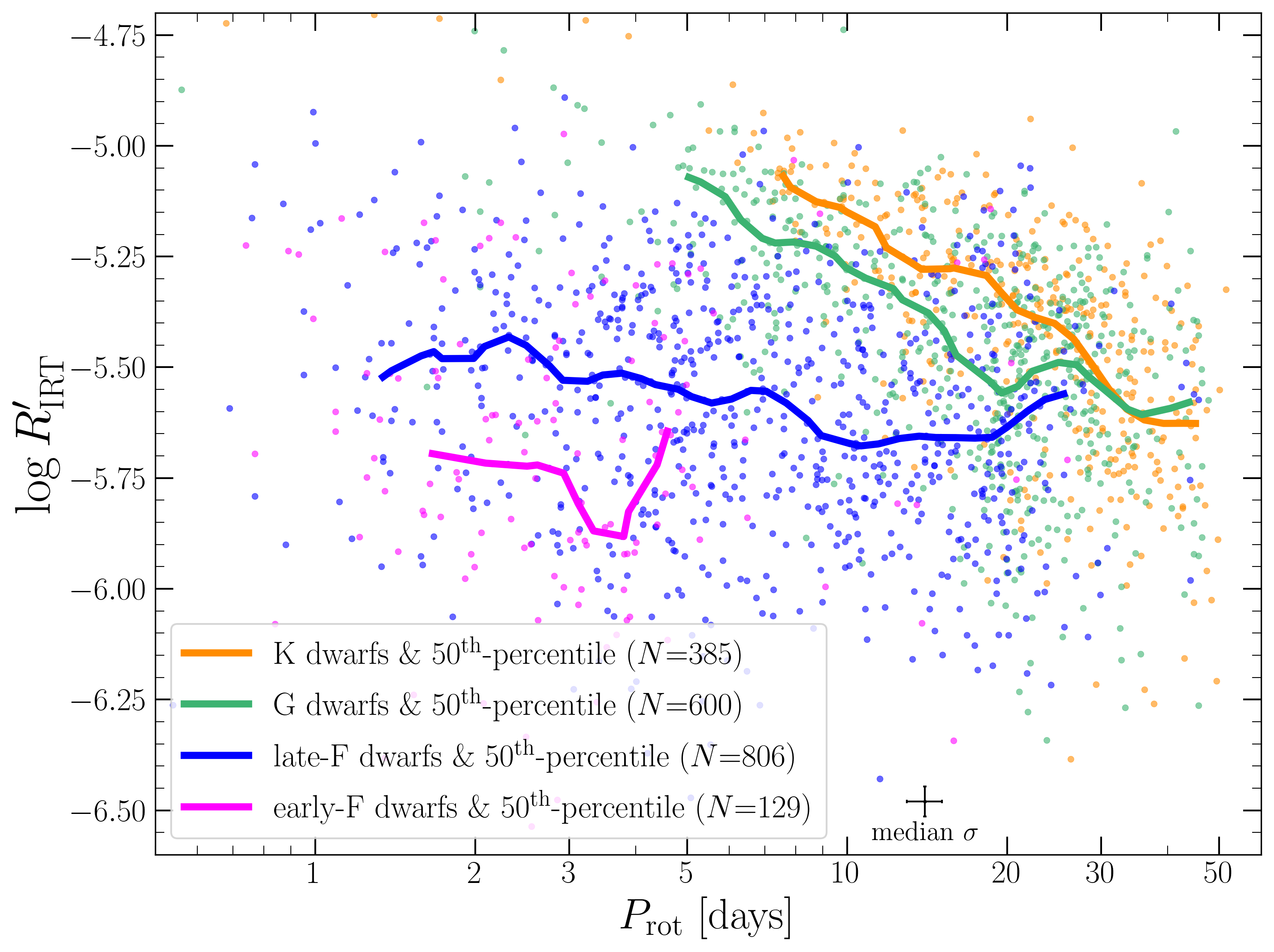}
    \caption{{\CaiiIRT} activity index vs. rotation period, as a function of spectral type (K dwarfs in orange, G dwarfs in green, late-F dwarfs in blue, and early-F dwarfs in magenta). Typical uncertainties are illustrated by the median $\sigma$ symbol. The lines show their respective 50$^{\mathrm{th}}$-percentiles, with their slopes getting steeper towards lower stellar masses. At a given rotation period, later spectral types typically have higher activity indices.}
    \label{fig:Figure_discussion_spectraltype}
\end{figure}

We now examine how the rotation-activity relation, as probed by the {\CaiiIRT} index, varies across spectral type on the MS (see Sect.~\ref{subsec:characterization_sphlogRIRT}), and how it is influenced by the intermediate-period gap. Figure~\ref{fig:Figure_discussion_spectraltype} shows the resulting rotation period vs. {\CaiiIRT} activity index diagram. The different spectral types are indicated by the colors, with the solid lines showing the respective 50$^{\mathrm{th}}$-percentile of the {\logRIRT} distribution in bins of {\prot}. The impact of spectral type on the range of spanned rotation periods is clear: the regime of rapid rotation (e.g., {\prot}$\lesssim 5$ days) is primarily populated by early-F and late-F dwarfs, while the regime of slow rotation (e.g., {\prot}$\gtrsim$ 20 days) is primarily populated by G and K dwarfs \citep{mcquillan14,santos21}. 

Across spectral types, we see an overall negative correlation, with magnetic activity decreasing with increasing {\prot} (i.e., towards slower rotation). The slopes of these correlations, as probed by the median lines, become more pronounced with decreasing stellar mass, likely due to the increasing depth of the convective envelope. They are relatively shallow for the early-F and late-F dwarfs, and are steeper for the G and K dwarfs. Moreover, at fixed rotation period in Fig.~\ref{fig:Figure_discussion_spectraltype}, later spectral types show higher levels of magnetic activity, in agreement with the literature (e.g., \citealt{mathur23,mathur25}). The variations across spectral types can be significant. For example, at a {\prot} $\approx 10$ days, the median activity goes from {\logRIRT} $\approx -5.65$ for late-F dwarfs, to {\logRIRT} $\approx -5.15$ for K dwarfs, half an order of magnitude of difference.

\begin{figure}[ht]
    \centering
    \includegraphics[width=0.90\hsize]{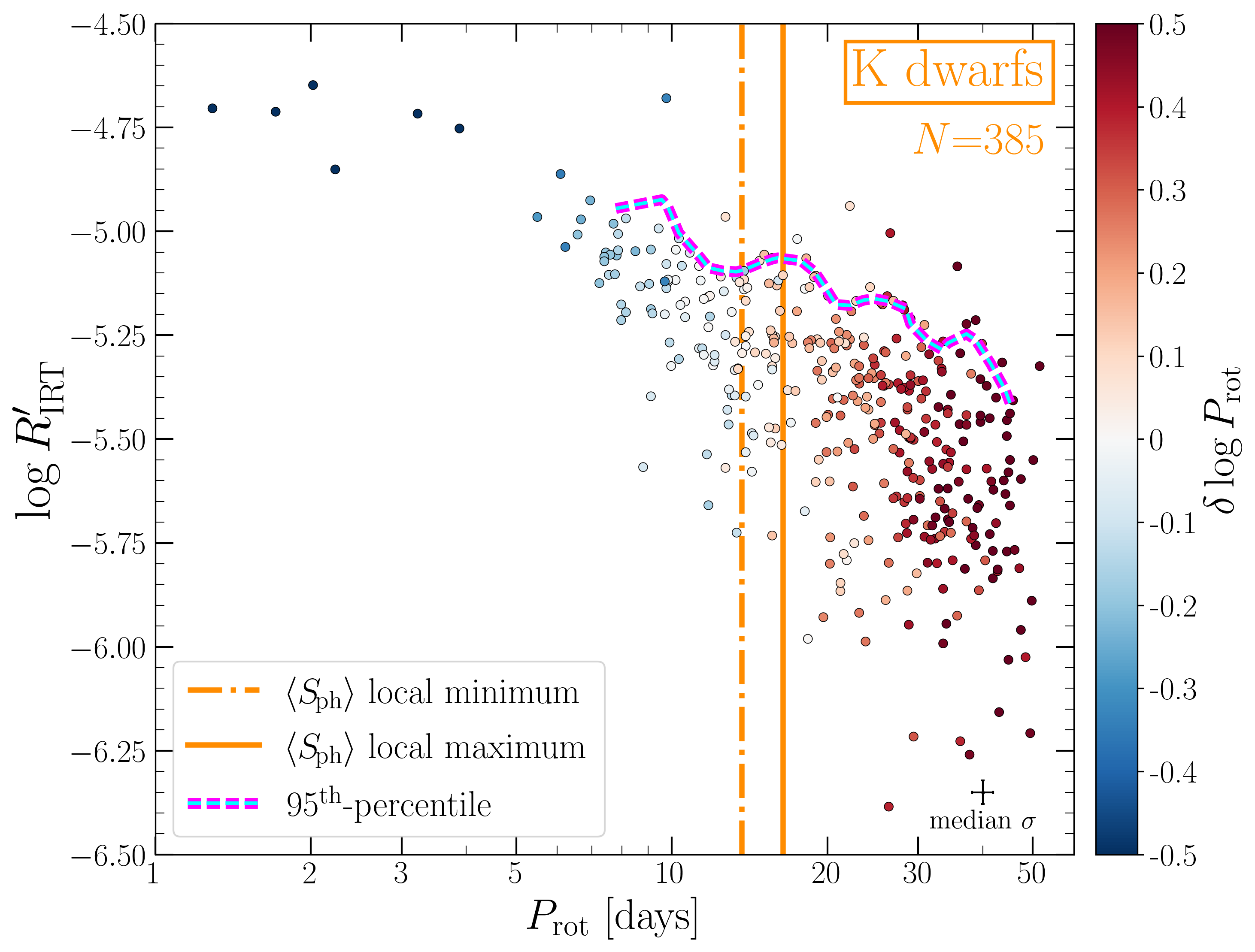}\\
    \includegraphics[width=0.90\hsize]{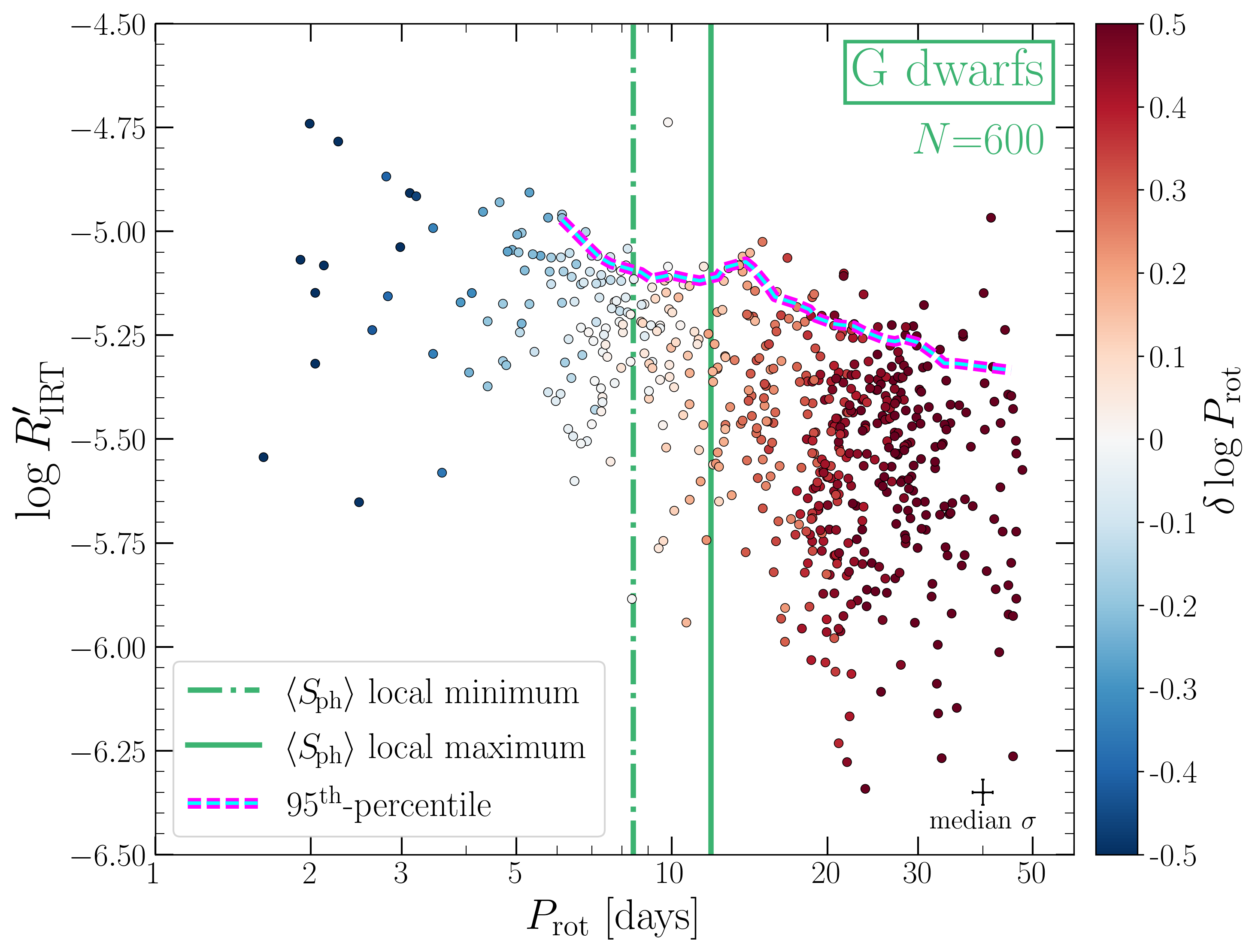}\\
    \includegraphics[width=0.90\hsize]{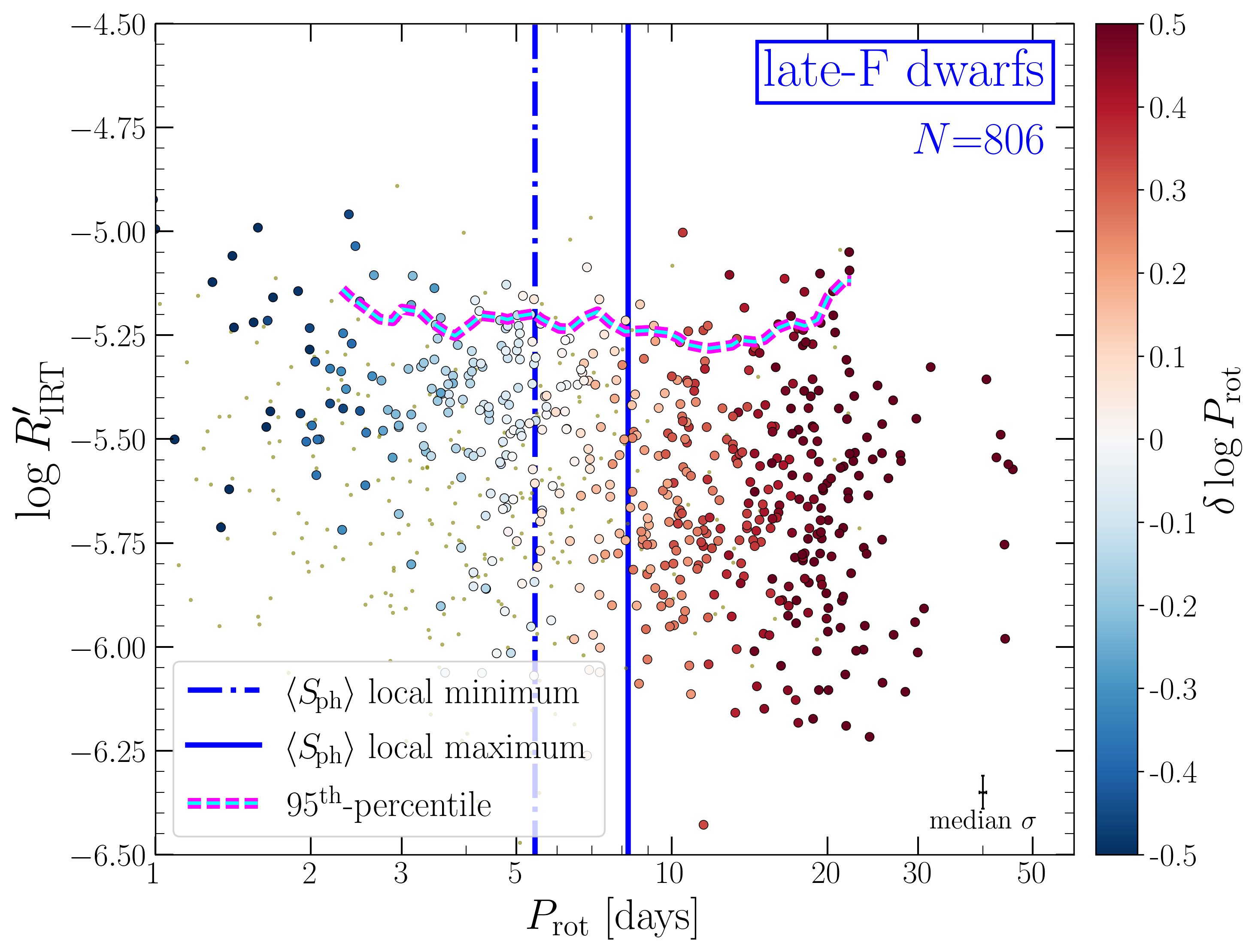}
    \caption{{\CaiiIRT} activity index vs. rotation period, separated by spectral type. From top to bottom these are: K dwarfs, G dwarfs, and late-F dwarfs. Stars are color-coded by their {\deltalogprot} values (if available). In each panel, the dashed line represents the 95$^\mathrm{th}$-percentile. Typical uncertainties are illustrated by the median $\sigma$ symbols. The vertical lines (see Table~\ref{tab:Sph_local_minima_maxima}) show the {\prot} values of the local minimum of {\sphaverage} at the intermediate-period gap (dash-dotted), and the local maximum after it (solid). Hints of enhanced {\CaiiIRT} activity after the gap are seen in the K and G dwarfs (and mildly in late-F dwarfs), and these occur at similar {\prot} values as the analogous {\sphaverage} features.}
    \label{fig:Figure_discussion_rotact}
\end{figure}

To examine the possible impact of the intermediate-period gap on the rotation-activity relation, in Fig.~\ref{fig:Figure_discussion_rotact} we repeat the {\logRIRT} vs. {\prot} diagram, showing one spectral type per panel. In this case, the stars are color-coded by their {\deltalogprot} values\footnote{The early-F dwarfs are not shown as the period gap itself is not defined in that color range (see Fig.~\ref{fig:Figure_data_targetrotation} and Appendix~\ref{sec:app_SpT_classification}). For the same reason, some late-F dwarfs lack {\deltalogprot} values (and therefore color-coding) in Fig.~\ref{fig:Figure_discussion_rotact}.} (see Sect.~\ref{sec:data_targetrotation}). Following \citet{santos23} and \citet{santos25} (see their Appendices B.3 and A, respectively), we quantified the upper envelope of the distribution by computing the 95$^\mathrm{th}$-percentile of {\logRIRT} in bins of {\prot}. These upper envelopes are shown as the dashed lines. Figure~\ref{fig:Figure_discussion_rotact} again shows that the rotation-activity diagram of late-F dwarfs is much flatter compared to those of the G and K dwarfs, as for the latter cases the upper envelope of {\CaiiIRT} activity falls rapidly with increasing rotation period.

For reference, the impact of the intermediate-period gap on the photometric magnetic activity proxy {\sphaverage} was recently studied by \citet{mathur25} and \citet{santos25}. Their results illustrated that, for field stars, the location of the gap coincides with a dip in magnetic activity (local minimum), followed by an enhancement of activity occurring after the gap (local maximum; see their Fig.~7 and 1, respectively). While our sample is significantly smaller than theirs (by a factor of $\sim 20$), due to the stringent brightness selection of the {\gaia} DR3 {\CaiiIRT} data (Sect.~\ref{sec:data_calcium}), it is interesting to compare them and assess whether chromospheric {\CaiiIRT} signatures occur at similar locations as the {\sphaverage} ones. Thus, for each spectral type, in Fig.~\ref{fig:Figure_discussion_rotact} the vertical lines indicate the location (in terms of {\prot}) of the local minimum (dash-dotted) and local maximum (solid) of photospheric activity as probed by the {\sphaverage} 95$^\mathrm{th}$-percentile around the intermediate-period gap. The values of these minima and maxima were determined following \citet{santos23} via bounded minimization (see their Appendix~B.3), and are reported in Table~\ref{tab:Sph_local_minima_maxima}.

\begin{table}
\caption{Local minimum and local maximum of the 95$^\mathrm{th}$-percentile of {\sphaverage} occurring at the intermediate-period gap (activity dip) and after the intermediate-period gap (activity enhancement), in terms of {\prot} for K, G, and late-F dwarfs (Fig.~\ref{fig:Figure_discussion_rotact}), and in terms of ($\mathrm{Ro}/\mathrm{Ro}_{\odot}$) for the Joint-MS sample (Fig.~\ref{fig:Figure_discussion_rossby}).}
\label{tab:Sph_local_minima_maxima}
\centering                         
\scriptsize 
    \begin{tabular}{cccc}
    	\hline
    	\hline
        Sample & Variable & {\sphaverage} local minimum & {\sphaverage} local maximum\\
        \hline
    K dwarfs & {\prot} & 13.6 & 16.4 \\
    G dwarfs & {\prot} & 8.4 & 11.9 \\
    Late-F dwarfs & {\prot} & 5.4 & 8.2\\
    Joint MS & ($\mathrm{Ro}/\mathrm{Ro}_{\odot}$) & 0.29 & 0.38 \\
       \hline
    \end{tabular}
\end{table}

For K dwarfs (top panel of Fig.~\ref{fig:Figure_discussion_rotact}), we observe a close correspondence of the {\sphaverage} vertical lines with a dip and a peak in the {\logRIRT} 95$^\mathrm{th}$-percentile. As illustrated by the {\deltalogprot} color-coding, these happen at the gap (white points) and after it (white-red points), respectively. While the period range between these features is narrow ($\approx$ 13 to 16 days), the fact that both activity proxies show their respective inflections at similar locations supports the idea that they trace a common underlying process. We therefore take this agreement as a hint of the impact of the intermediate-period gap on chromospheric activity.

For G dwarfs (middle panel of Fig.~\ref{fig:Figure_discussion_rotact}), the diagram shows again a decrease in the upper envelope of {\CaiiIRT} activity approximately at the intermediate-period gap ({\prot} $\approx$ 9 days; white points in {\deltalogprot}), followed by an increase after it ({\prot}$\approx 14$ days; white-red points in {\deltalogprot}). This behavior is qualitatively similar to that seen for {\sphaverage} in K dwarfs, although in this case the comparison with the respective local minimum and maximum shows a lesser degree of agreement, with the activity enhancement after the gap happening at a longer {\prot} in {\logRIRT} than in {\sphaverage} (see Table~\ref{tab:Sph_local_minima_maxima}).

For late-F dwarfs (bottom panel of Fig.~\ref{fig:Figure_discussion_rotact}), the decrease in {\CaiiIRT} activity with increasing rotation period is shallower, and consequently the 95$^\mathrm{th}$-percentile shows a much weaker period dependence. A modest dip in {\logRIRT} is seen at {\prot} $\approx 6$ days (white points in {\deltalogprot}), followed by a slight peak at {\prot} $\approx 7$ days (white-red points in {\deltalogprot}). The {\sphaverage} lines lie somewhat close to these features, however, given the flatter activity-rotation relation in this regime, defining these minima and maxima around the gap is more uncertain.

Throughout this analysis of Fig.~\ref{fig:Figure_discussion_rotact}, we have focused on the behavior of the upper envelope of {\logRIRT} vs. {\prot} as traced by the 95$^\mathrm{th}$-percentile. The motivation for this choice is that stars located near the upper edge of the activity vs. rotation diagram are observed in more favorable conditions. On the one hand, they are expected to be oriented at inclination angles close to $\sim 90^{\circ}$ (i.e., equator-on), such that they maximize the visibility of their active regions (e.g., \citealt{see21,sowmya21}), which for the Sun are typically formed within latitudes of $\sim \pm30^{\circ}$. On the other hand, the stars at the upper edge should be around the  maximum of their activity cycles. In this way, by adopting the 95$^\mathrm{th}$-percentile, we minimize the scatter introduced by stars observed at lower inclination angles and quieter phases of their cycles (e.g., see Sect.~2 and~4 of \citealt{mathur25}).
\subsection{Calcium IRT index vs. Rossby number}
\label{subsec:discussion_rossby}

A common approach to studying the evolution of magnetic activity is through the Rossby number ($\mathrm{Ro}$). This parameter is defined as the ratio of the rotation period to the convective overturn timescale ($\tau_\mathrm{c}$), i.e., $\mathrm{Ro} = P_\mathrm{rot}/\tau_\mathrm{c}$. The timescale $\tau_\mathrm{c}$ depends on the depth of the convective envelope, which in turn increases towards cooler MS stars (e.g., \citealt{corsaro21,bonanno25}). In this way, the Rossby number is a proxy for stellar rotation that accounts for the {\teff}-dependence, allowing comparisons among different spectral types.

While rotation periods can be measured observationally (in the case of our sample thanks to the {\kepler} photometry), estimates for convective overturn timescale are typically model-dependent. Thus, to study the rotation-activity relation in terms of the Rossby number, we went beyond the {\gaia} DR3 colors and adopted literature stellar properties. Specifically, given the large overlap with our sample, we adopted the  $\tau_\mathrm{c}$ values from \citet{mathur25}, who calculated these using the Yale Rotating Evolution Code (YREC; \citealt{pinsonneault89,bahcall01,demarque08}). By searching for the model that best fitted the observed parameters of the stars ({\teff}, $L$ or $\log g$, and [Fe/H]) with the \texttt{kiauhoku} optimization tool \citep{claytor20}, they obtained $\tau_\mathrm{c}$ and $\mathrm{Ro}$. For representation purposes, and to allow the comparison between different Rossby number prescriptions, these $\mathrm{Ro}$ values were normalized by the solar value in the YREC models ($\mathrm{Ro}_\odot = 2.16$). The modelling procedure is further detailed in \citet{mathur25}. 

We note that the temperature and metallicity values used by \citet{mathur25} to compute Ro differ from those we adopted in the {\logRIRT} calculation (Sect.~\ref{sec:data_calcium}). On the one hand, as currently no spectroscopic survey covers the entirety of the {\kepler} field \citep{godoyrivera26a}, \citet{mathur25} used a combination of spectroscopic and photometric inputs, namely: CFOP ({\kepler} Community Follow-up Observation Program; \citealt{furlan18}), APOGEE DR16 (Apache Point Observatory for Galactic Evolution Experiment; \citealt{ahumada20}), LAMOST DR7 (Large Sky Area Multi-ObjectFiber Spectroscopic Telescope; \citealt{zhao12,zong20}), and the two latest {\kepler} stellar properties catalogs \citep{mathur17,berger20}. On the other hand, to calculate the {\CaiiIRT} magnetic-activity index, we adopted the corresponding {\gaia} DR3 \texttt{gspspec} or \texttt{gspphot} parameters (see \citealt{lanzafame23}). As discussed in Sect.~\ref{sec:data_calcium}, a different set of stellar properties propagates to only a minor impact on the derived {\CaiiIRT} proxy (Equation~\ref{eqn:calcium_central}), which is typically within the uncertainties. Thus, although some heterogeneities are present in our {\CaiiIRT} vs. Ro analysis below, these do not impact any of the conclusions.

\begin{figure}[ht]
    \centering
    \includegraphics[width=\hsize]{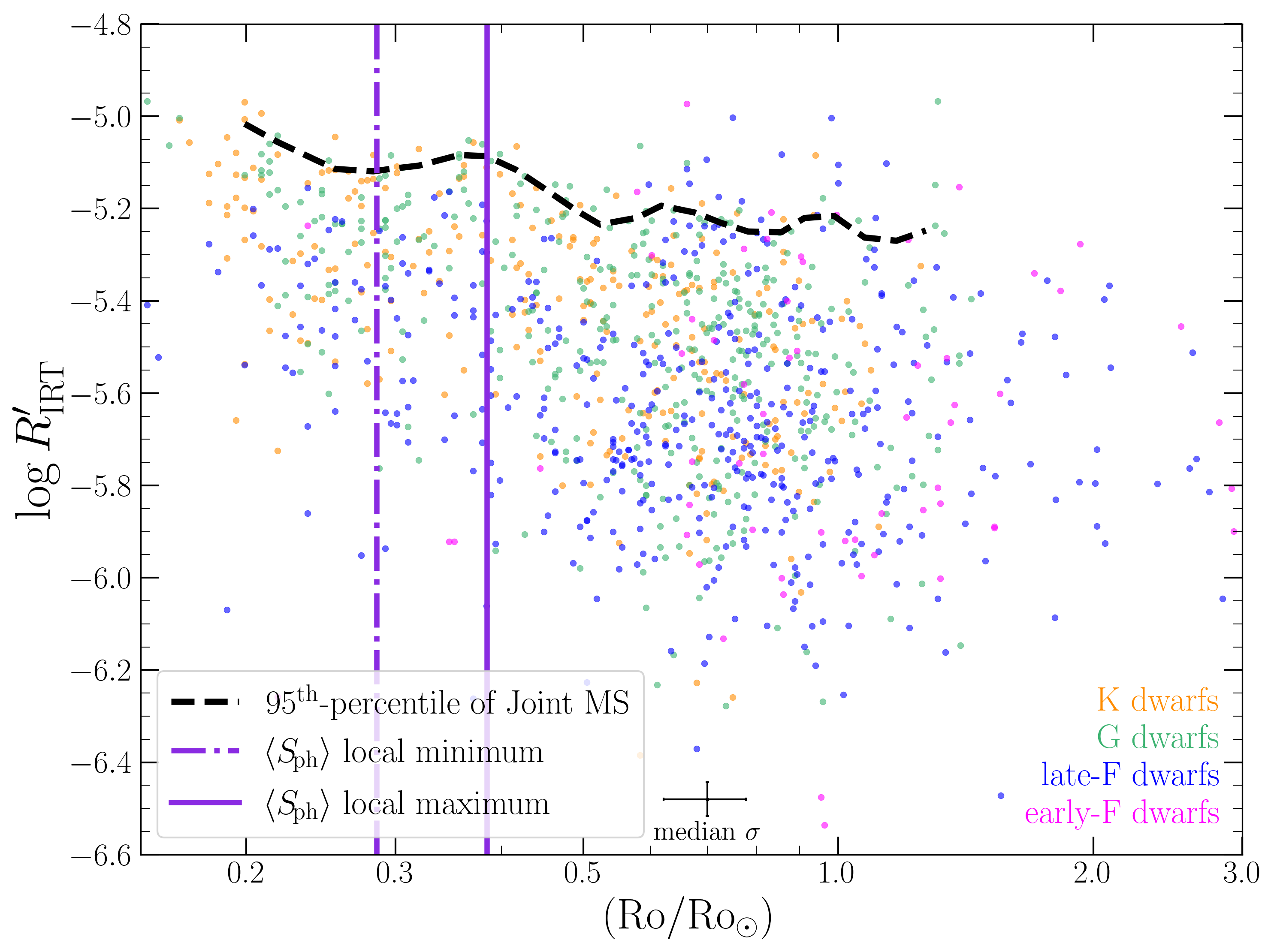}
    \caption{{\CaiiIRT} activity index vs. Rossby number ($\mathrm{Ro}=P_{\mathrm{rot}}/\tau$), normalized to the solar value ($\mathrm{Ro}/\mathrm{Ro}_{\odot}$). K dwarfs are shown in orange, G dwarfs in green, late-F dwarfs in blue, and early-F dwarfs in magenta. The dashed line shows the 95$^\mathrm{th}$-percentile for the combined Joint-MS sample. Typical uncertainties are illustrated by the median $\sigma$ symbol. A lower Rossby number typically means a higher {\logRIRT}. The vertical lines (see Table~\ref{tab:Sph_local_minima_maxima}) show the ($\mathrm{Ro}/\mathrm{Ro}_{\odot}$) values of the local minimum of {\sphaverage} at the intermediate-period gap (dash-dotted), and the local maximum after it (solid), which closely align with the analogous features observed in the {\CaiiIRT} index upper envelope.}
    \label{fig:Figure_discussion_rossby}
\end{figure}

Figure~\ref{fig:Figure_discussion_rossby} shows {\logRIRT} vs. the normalized Rossby number for the Joint-MS sample, with the different spectral types indicated by the colors. The 95$^{\mathrm{th}}$-percentile of the {\logRIRT} distribution in bins of $(\mathrm{Ro}/\mathrm{Ro}_{\odot})$ is shown as the black dashed line. Considering the sample as a whole, we observe an overall decrease in the chromospheric activity index with increasing Ro. Such a behavior is consistent with what has been found for other chromospheric and photospheric magnetic activity indices (e.g., \citealt{noyes84,zhang20,reiners22,brown22,cao22,mathur25,yang25}), as well as for the {\logRIRT} itself reported for cluster stars \citep{fritzewski21} and in the field \citep{freund25}. This trend is consistent with our targets being located in the aforementioned unsaturated regime, where the activity level is a strong function of {\prot} and thus $\mathrm{Ro}$, in agreement with other rotation-activity studies of the {\kepler} stars (e.g., \citealt{matt15}). We note that the saturated regime is only scarcely represented in the {\kepler} sample (e.g., \citealt{masuda22,santos24}), as expected for a predominantly spun-down field-star population.

The {\logRIRT} vs. $\mathrm{Ro}$ diagram of Fig.~\ref{fig:Figure_discussion_rossby} also probes the impact of the intermediate-period gap on chromospheric activity. In this context, we replicated the comparison between {\logRIRT} and {\sphaverage} in terms of the local minimum and maximum in activity tied to the gap (see Sect.~\ref{subsec:discussion_rotact}), but as a function of the Rossby number instead of rotation period. Using the \citet{mathur25} sample, we applied the analogous selection criteria (i.e., Joint-MS stars with measured {\sphaverage} and $\mathrm{Ro}$) and recomputed the positions of these local minimum and maximum features in terms of $(\mathrm{Ro}/\mathrm{Ro}_{\odot})$. These values are reported in Table~\ref{tab:Sph_local_minima_maxima} and shown as the vertical lines in Fig.~\ref{fig:Figure_discussion_rossby}.

Although again our {\CaiiIRT} sample is significantly smaller than the \citet{mathur25} one (by a factor of $\sim$ 30), interesting comparisons can be made. As traced by the 95$^{\mathrm{th}}$-percentile in {\logRIRT}, the local minimum at the intermediate-period gap appears to be slightly shifted towards a lower $\mathrm{Ro}$ compared to {\sphaverage} ($\mathrm{Ro}/\mathrm{Ro}_{\odot} \approx 0.25$). However, given our smaller sample size, the agreement is still reasonable. For the local maximum after the gap, the {\logRIRT} behavior shows a strong agreement with {\sphaverage}, coinciding at ($\mathrm{Ro}/\mathrm{Ro}_{\odot} \approx 0.38$). This comparison further demonstrates that, despite the differences discussed above (Sect.~\ref{sec:characterization} and ~\ref{subsec:discussion_rotact}), both {\logRIRT} and {\sphaverage} trace the same underlying phenomena, with their local minimum at the intermediate-period gap and local maximum after it appearing at similar values of the Rossby number.

The above analyses are in broad agreement with  \citet{ye25}, who studied the impact of the rotational stalling (see Sect.~\ref{sec:introduction}) in open clusters on chromospheric magnetic activity as probed by the \ion{Ca}{ii}~H\&K index. Specifically, our results of the {\CaiiIRT} index resemble those of \citet{ye25} for K-dwarf stars, with the intermediate-period gap and the stalling coinciding with a dip in chromospheric activity followed by an enhancement (see their Fig.~8c). While their comparison was focused on the location of the stalling in regards to K dwarfs, hints of the analogous behavior may be seen for their G dwarfs too. The comparison with their F-dwarf stars is not as straightforward, as while we decided to split between early-F and late-F dwarfs, \citet{ye25} combined them together into a single category.

All in all, the agreement across several magnetic activity proxies that span the photosphere and chromosphere paints a consistent picture: activity decreases at the intermediate-period gap, and increases after it. While \citet{reinhold19} and \citet{reinhold20} interpreted the photometric activity decrease at the gap as the cancellation between dark spots and bright faculae, the persistence of the activity dip and enhancement across photospheric and chromospheric diagnostics disfavor this hypothesis as an explanation to account for all the observables. Instead, with the existing data sets, the core-envelope coupling scenario appears as a more likely cause (e.g., \citealt{spada20,lu24}; see also Sect.~\ref{sec:introduction}).

Finally, at larger values of the Rossby number ($\mathrm{Ro}/\mathrm{Ro}_{\odot} \sim 0.7$ to $1.0$), Fig.~\ref{fig:Figure_discussion_rossby} shows a slight overdensity of stars. This could correspond to the pile-up found in the rotating {\kepler} sample, linked to the weakened magnetic braking phenomenon (e.g., \citealt{david22,metcalfe23}). Beyond the solar value ($\mathrm{Ro}/\mathrm{Ro}_{\odot} > 1.0$), the analysis of photometric and spectroscopic activity proxies from {\kepler} and {\it K2} data has suggested enhanced magnetic activity in G-type stars (\citealt{mathur25} and \citealt{brandenburg18}, respectively). This behavior may be related to anti-solar differential rotation according to magneto-hydrodynamical simulations \citep{karak20,brun22,noraz22}. While hints of a flattening in the {\CaiiIRT} index beyond $\mathrm{Ro}_{\odot}$ might be present in Fig.~\ref{fig:Figure_discussion_rossby}, unfortunately this regime of the diagram is too scarcely populated to draw any firm conclusions.
\subsection{The influence of multiplicity}
\label{subsec:discussion_multiplicity}

\begin{figure*}
    \centering
    \includegraphics[width=0.46\hsize]{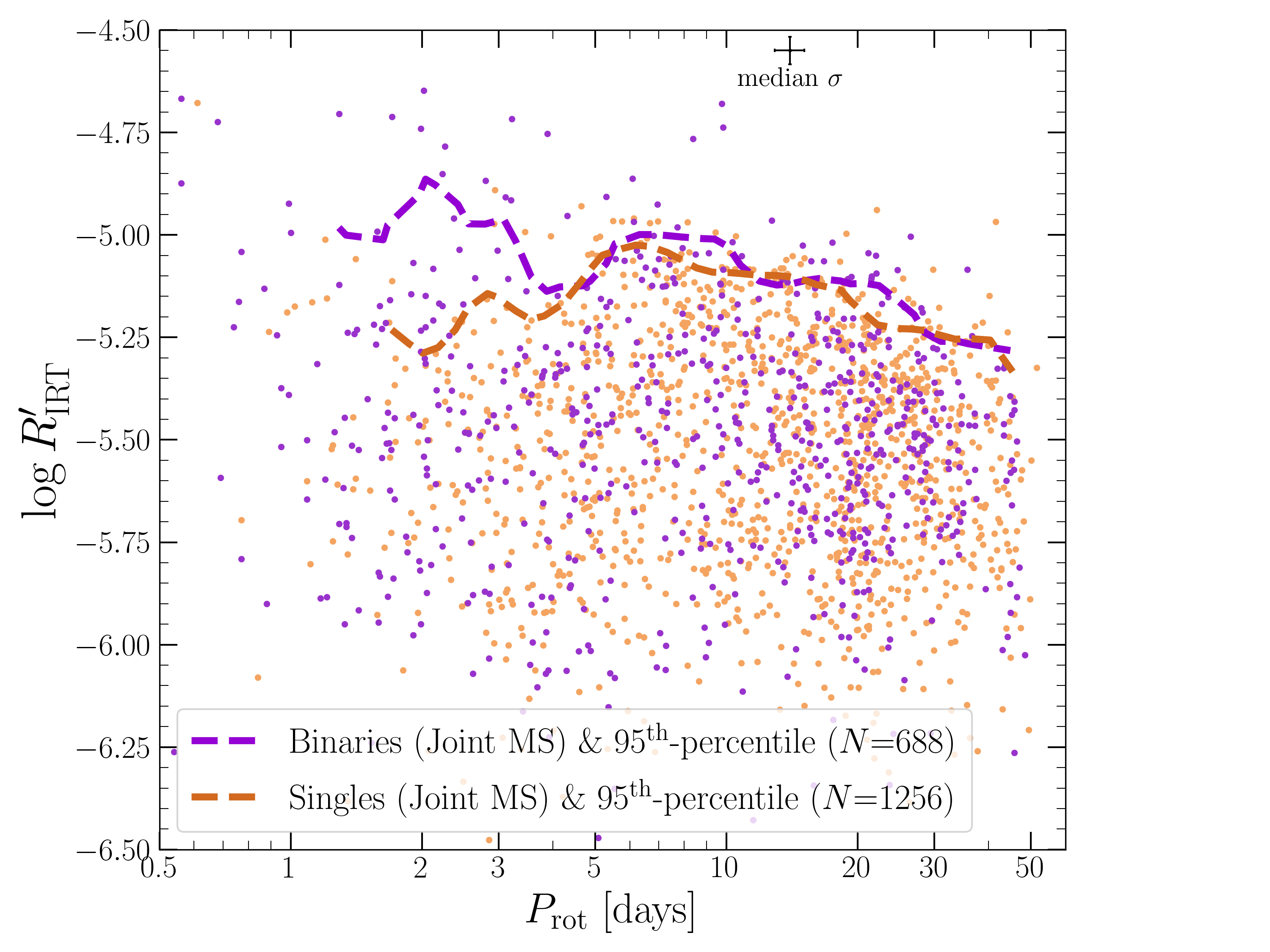}
    \includegraphics[width=0.46\hsize]{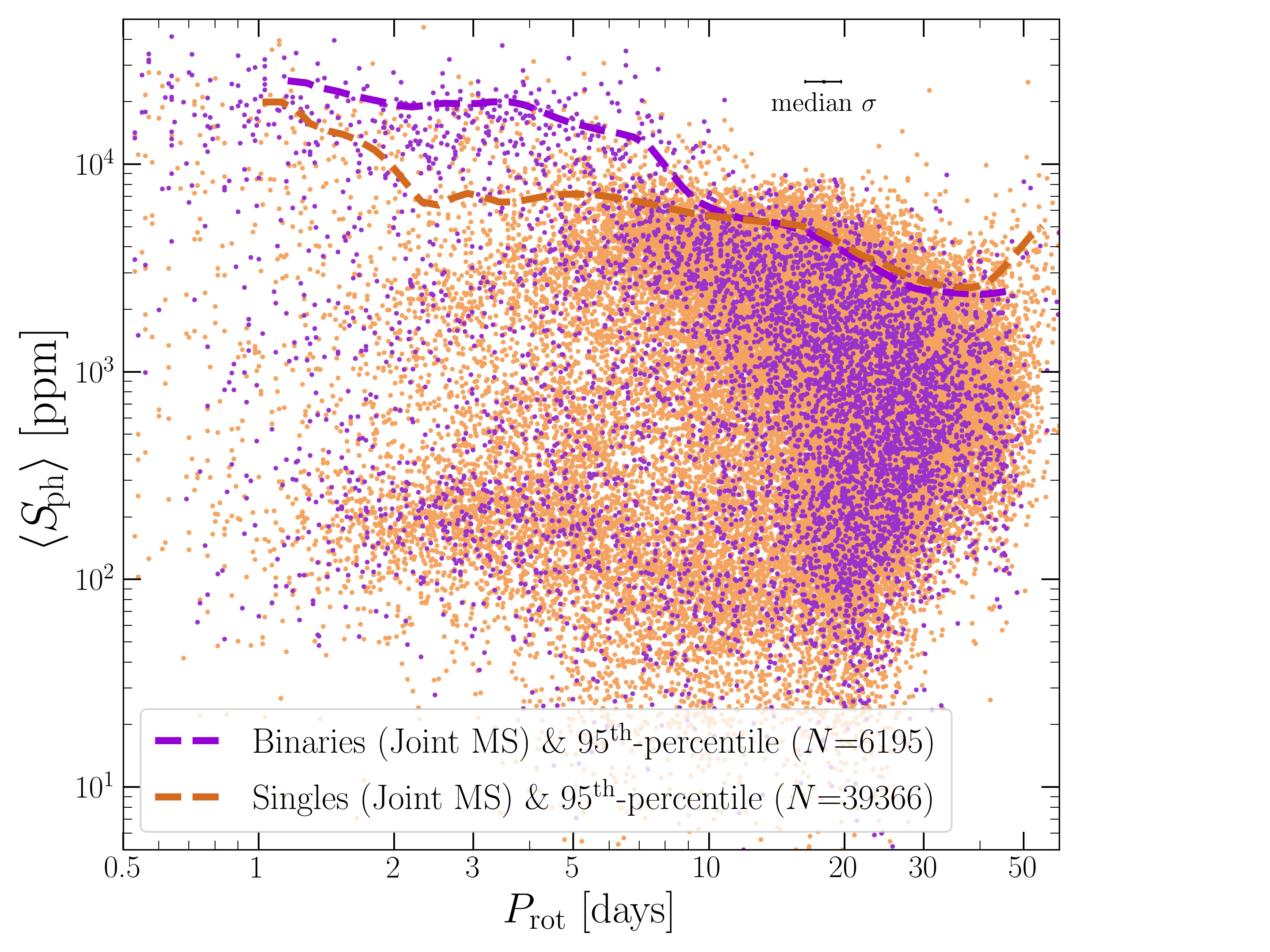}\\
    \includegraphics[width=0.46\hsize]{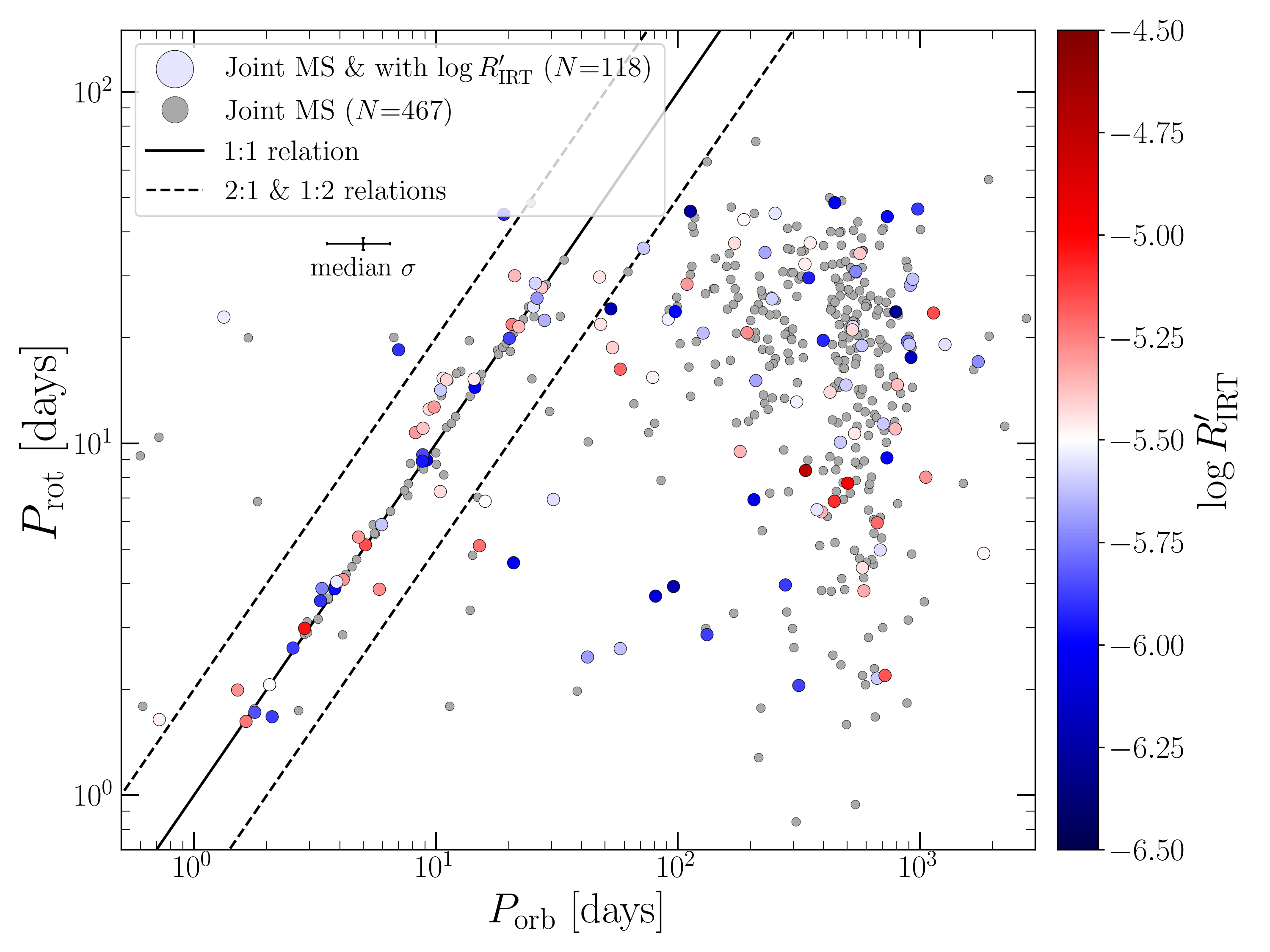}
    \includegraphics[width=0.46\hsize]{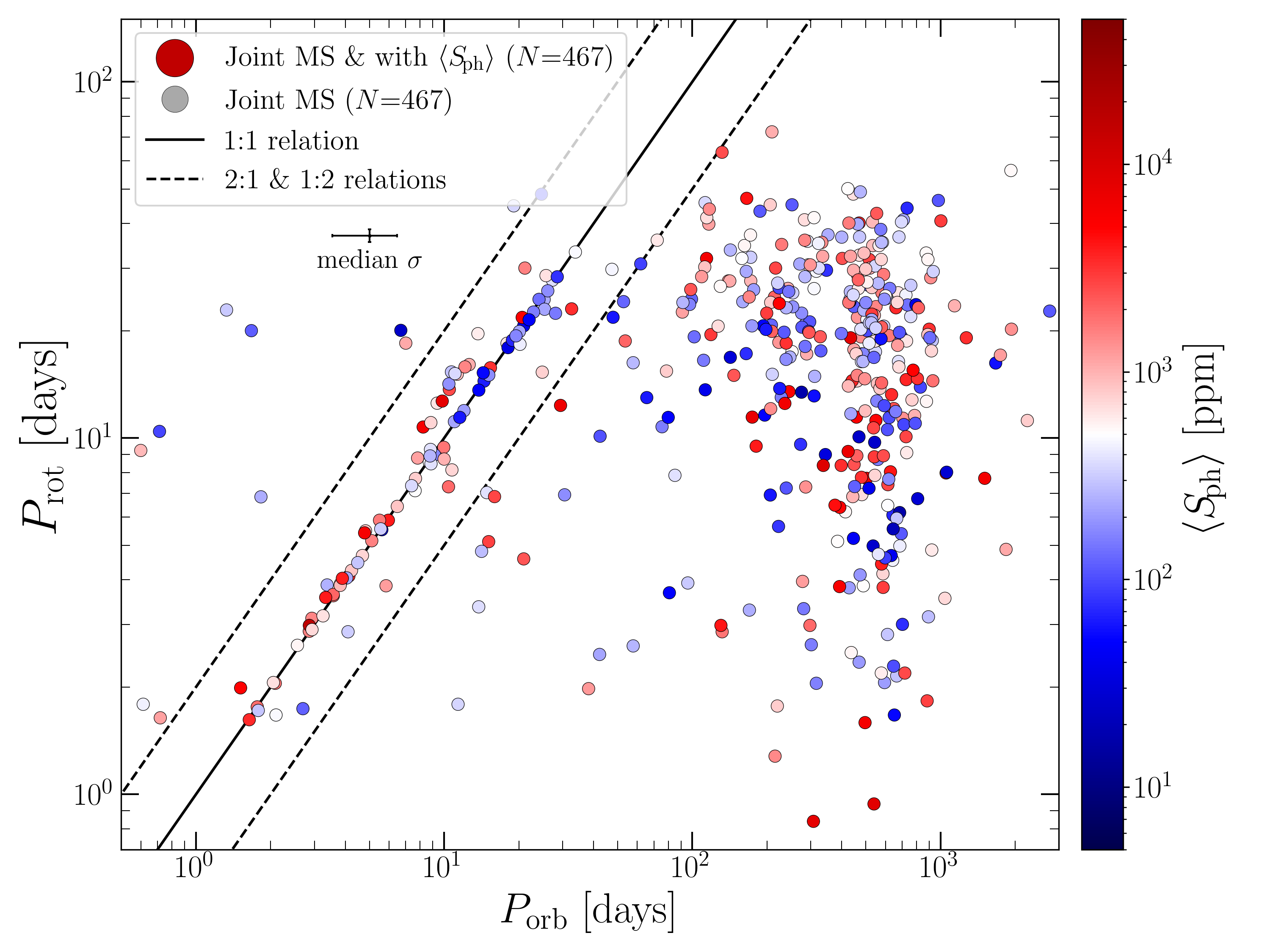}\\
    \includegraphics[width=0.46\hsize]{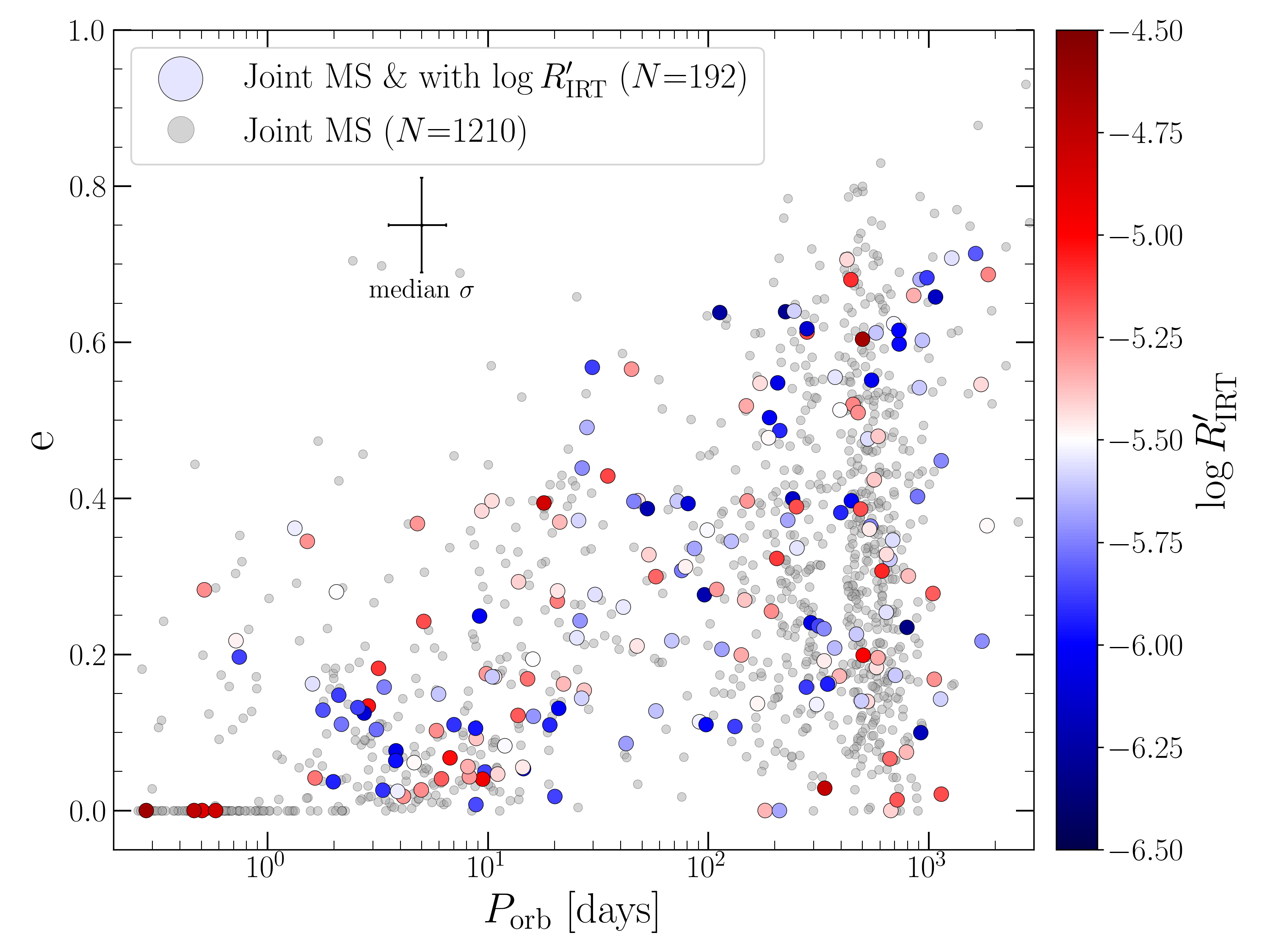}
    \includegraphics[width=0.46\hsize]{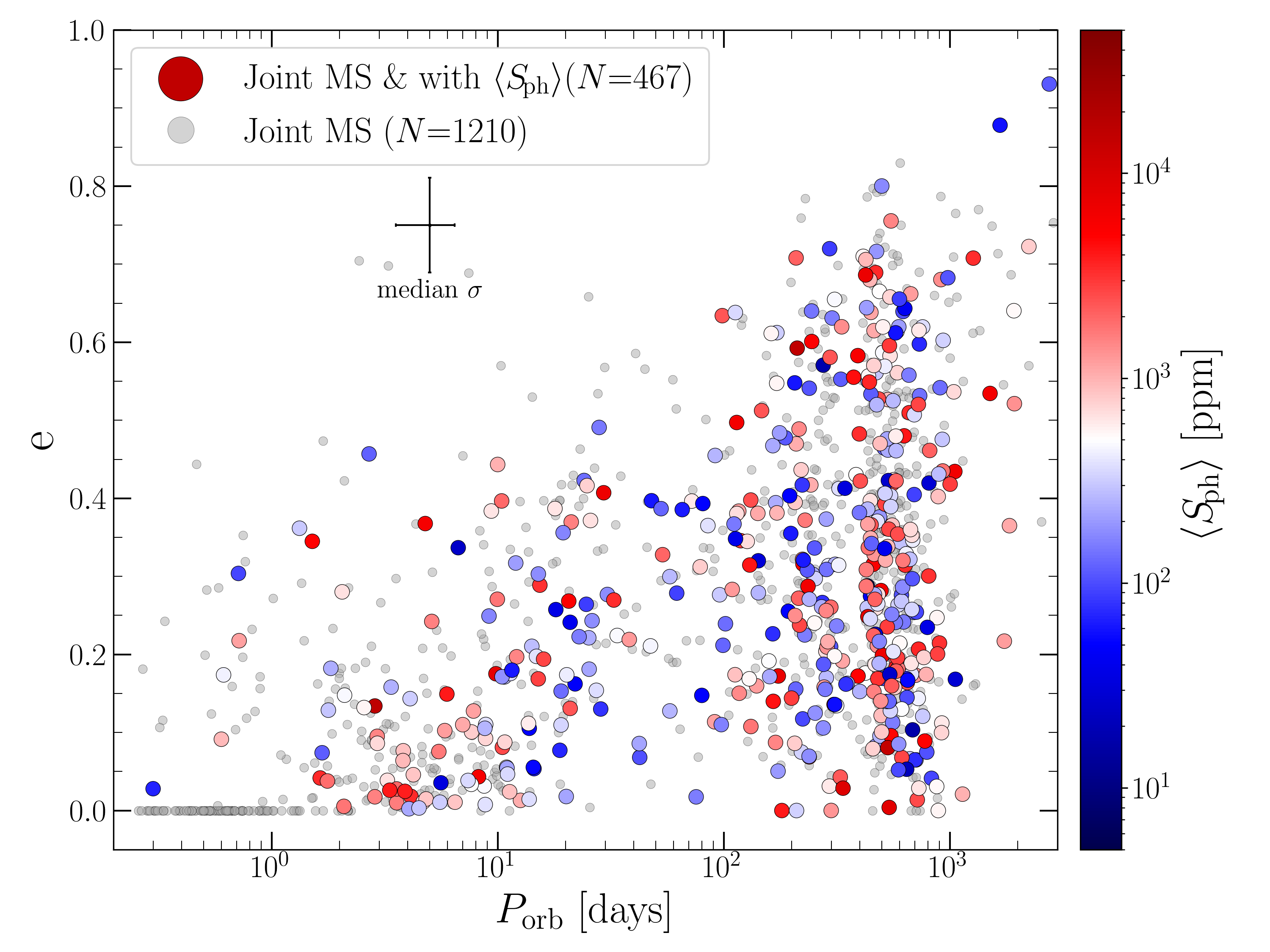}\\
    
    \caption{Impact of multiplicity on the magnetic activity of MS targets as probed by the chromospheric {\CaiiIRT} (left column) and the photospheric {\sphaverage} (right column). Top: activity index vs. rotation period, as a function of stellar multiplicity (singles in brown and binaries in violet). The dashed lines show the 95$^\mathrm{th}$-percentile of each distribution. In the regime of rapid rotators ($P_{\mathrm{rot}} \lesssim$ 5 {\textendash} 10 days), binary systems have a higher activity upper envelope than single stars at a given rotation period. Middle: rotation period vs. orbital period color-coded by activity index (if available). The solid lines illustrate the 1:1 relation, while the dashed lines illustrate the 2:1 and 1:2 relations. Along the {\prot}$=${\porb} line, especially for {\sphaverage}, enhanced activity can be seen for periods $\lesssim 10$ days. Bottom: orbital period vs. eccentricity color-coded by activity index (if available). Systems with {\porb} $\lesssim 10$ days and near circular orbits ($\mathrm{e} \lesssim 0.05$) often display high activity values. Typical uncertainties are illustrated by the median $\sigma$ symbols.}
    \label{fig:Figure_discussion_multiplicity}
\end{figure*}

As a final piece of the analysis, we examine the impact of multiplicity on magnetic activity. Binary and higher-order systems are fundamental astrophysical laboratories (e.g., \citealt{duchene13,beck26b}). They are common in solar-like stars \citep{duquennoy91,raghavan10}, and thus assessing their influence is key for our understanding of the rotation-activity relation \citep{schrijver91,poppenhaeger17,long25}. This is particularly relevant in the regime where close companions can modify the rotational evolution of stars (e.g., \citealt{kounkel22,kounkel23,godoyrivera18,silvabeyer23,gruner23a,gruner23b,beck26a}), yielding populations of rapid rotators in the field (e.g., \citealt{lurie17,simonian19,angus20,ding24,patton24}). A first hint of the impact of multiplicity on activity in our sample was shown on the CMD of Fig.~\ref{fig:Figure_data_calcium}, where several targets that populate the photometric binary region have high-activity levels ({\logRIRT} $\gtrsim -5.5$).

More broadly, a thorough search for candidate binary systems was carried out for the {\kepler} stars by \citet{godoyrivera25} using astrometric, photometric, and spectroscopic diagnostics on the {\gaia} data in combination with literature catalogs (see their Sect.~4 for details). For this work, we used their ``Flag Binary Union'' column as our binarity indicator, which we repeat in Table~\ref{tab:table_catalog} for simplicity. The targets not flagged by this column were taken to be single stars. We nonetheless note that, even with the high precision of current surveys, the observed binary fraction recovered from such data sets (e.g., \citealt{beck24}) remains lower than the expected value from population studies (e.g., \citealt{offner23,espinozarojas25}). Therefore, future data releases (e.g., {\gaia} DR4; \citealt{brown24}) will in all likelihood reveal additional multiple systems.

In the top panels of Fig.~\ref{fig:Figure_discussion_multiplicity}, we show the rotation-activity relation for our Joint-MS targets probed by {\logRIRT} (left column) and {\sphaverage} (right column), splitting the sample into singles (brown) and binaries (violet). For reference, the 95$^\mathrm{th}$-percentiles of their activity distributions in bins of {\prot} are shown as the dashed lines. 
For chromospheric activity, while at rotation periods longer than $\gtrsim 5$ days both populations show similar distributions, for {\prot} $\lesssim 4$ days binary systems stand out with a higher upper envelope of {\logRIRT} values than single stars at a given rotation period. A similar trend is seen for photospheric activity, where binaries display a higher upper envelope of {\sphaverage} values than single stars for {\prot} $\lesssim 10$ days. This is also reflected in the individual data points, with the majority of targets more active than {\logRIRT} $\gtrsim-5.0$ and {\sphaverage} $\gtrsim 10^4$ ppm being flagged as binaries. Still, MS binaries with short rotation periods span a range of activity values (e.g., Fig.~15 of \citealt{beck24}). 

Caveats of the above {\CaiiIRT} vs. {\sphaverage} comparison are that, on the one hand, the chromospheric activity index is available for a significantly smaller sample than the photospheric one (factor of $\sim$ 20). On the other hand, the sample with {\logRIRT} values is significantly brigther than its counterpart with {\sphaverage} values (median apparent $G \approx 11.9$ mag and $G \approx 14.7$, respectively, or a flux ratio of $\approx 13$), implying a higher completeness of the binary diagnostics for the former than the latter (e.g., see Fig.~7 of \citealt{godoyrivera25}). In spite of these, both proxies exhibit comparable qualitative behaviors.

These results show that the influence of multiplicity on enhancing magnetic activity relative to single stars is most prominent among fast rotators, in line with recent findings of chromospheric and photospheric activity proxies for MS and evolved targets (e.g., \citealt{gaulme20,gehan22,gehan24,beck24,shi25,godoyrivera26b}). Indeed, the range of periods ($\lesssim 10$ days) of the enhanced activity upper envelopes in the top panels of Fig.~\ref{fig:Figure_discussion_multiplicity} falls within the regime where tidal effects become significant (e.g., \citealt{lurie17}). Such strong tidal interactions drive binaries toward circularization and synchronization (e.g., \citealt{zahn13,offner23}), which can lead to strong magnetic activity as well as suppression of oscillation modes \citep{gaulme14,gaulme20,beck18,mathur19,benbakoura21,gehan24}.

To further explore this, we searched for orbital information in the {\gaia} DR3 non-single star two-body orbit catalog (NSS~TBO; \citealt{gaia23a,gaia23b}) and the eXtended Catalogue of Spectroscopic Binary Orbits (SBX; \citealt{merle26}). We found a total of 2,012 systems\footnote{Of these systems, 4 were found in both catalogs, showing good agreement between their solutions. For these, we adopted their SBX values.} with orbital periods ({\porb}), 58 coming from the SBX and 1,954 coming from the NSS~TBO\footnote{A subset of 30 targets had two solutions in the NSS~TBO. For these, we picked the \texttt{nss\_solution\_type} with the shortest orbital period, typically being an \texttt{SB1} over an \texttt{Orbital} solution.}. The majority of these (2,008 out of 2,012) also had values for the eccentricity ($\mathrm{e}$). These orbital parameters are reported in Table~\ref{tab:table_catalog}.

The middle panels of Fig.~\ref{fig:Figure_discussion_multiplicity} show the {\prot} vs. {\porb} diagram, with the subset of targets with measured activity indices color-coded by {\logRIRT} and {\sphaverage} (left and right columns, respectively). Below {\porb} $\lesssim 30$ days, binaries predominantly populate the {\prot} $=$ {\porb} line. Along this synchronization line, a trend of increased {\sphaverage} values can be seen for {\porb} $\lesssim 10$ days. For {\logRIRT} there is no such clear trend, although we note that the sample size is smaller (factor of $\sim$ 4). Comparing with the literature, activity enhancements have been seen for close binaries with respect to wide binaries on the MS (e.g., \citealt{yu25}), as well as for red giant binaries in spin-orbit resonances (e.g., \citealt{gaulme14,benbakoura21,gehan22,gehan24}).

The bottom panels of Fig.~\ref{fig:Figure_discussion_multiplicity} show the analogous eccentricity vs. orbital period diagram, color-coded by magnetic activity when available. Here, the {\logRIRT} index does show its highest values for the short-period circularized systems ({\porb} $< 1$ day and $\mathrm{e}\approx 0$), with the 1 day $<$ {\porb} $<$ 10 days regime showing both low- and high-activity systems. For {\sphaverage}, the regime of {\porb} $\lesssim 10$ days is again predominantly populated by high-activity targets in near-circular orbits. These trends are in broad agreement with the literature (e.g., \citealt{dixon25}).
\section{Conclusions}
\label{sec:conclusions}

Stellar magnetic activity indicators are key probes of the rotation-activity connection and of angular momentum evolution. In this paper, we studied chromospheric magnetic activity as probed by the {\CaiiIRT} index provided by {\gaia} DR3 for the {\kepler} field stars. The resulting catalog is reported in Table~\ref{tab:table_catalog}.

We combined our chromospheric {\logRIRT} indices with literature rotation periods and photospheric activity proxy {\sphaverage}, to obtain a coherent view of the magnetic properties of low-mass stars across the Hertzsprung-Russell diagram (Fig.~\ref{fig:Figure_data_calcium}). We compared both indices with each other, and found a statistically significant correlation (Fig.~\ref{fig:Figure_characterization_sphlogRIRT}). This reinforces the view that these proxies trace the same underlying dynamo-driven magnetic processes, despite probing different atmospheric layers and temporal windows, which mitigate their correspondence.

Our analysis shows that the intermediate-period gap leaves an imprint on the chromospheric activity level of MS stars that have recently experienced this phase. This is reflected as hints of an activity dip at the gap, followed by an enhancement after it in the {\logRIRT} index vs. {\prot} diagram (Fig.~\ref{fig:Figure_discussion_rotact}). The locations of these dip and enhancement closely track similar features previously identified in the literature using the photospheric {\sphaverage} index instead. This alignment is seen not only in terms of {\prot}, but also in the Rossby number (Fig.~\ref{fig:Figure_discussion_rossby}).

This agreement between photospheric and chromospheric diagnostics supports the interpretation of the intermediate-period gap as a genuine transition in stellar magnetic behavior, rather than an artifact limited to a particular activity proxy. Overall, the unsaturated regime probed by the {\CaiiIRT} proxy vs. Rossby number diagram confirms the presence of subtle structures associated with changes in braking efficiency. Moreover, although scarcely populated, our sample possibly hints at the possible weakened magnetic braking phenomenon around and beyond the solar Rossby number.

In terms of the impact of spectral type on the chromospheric {\logRIRT} index, later-type stars are found to display higher activity levels at a given rotation period (i.e., with K dwarfs being systematically more active than G dwarfs, which in turn are more active than F dwarfs; Fig.~\ref{fig:Figure_discussion_spectraltype}). The amplitude of these differences can reach half an order of magnitude in {\CaiiIRT} at fixed {\prot}, which has direct implications for the environments of exoplanets around K-type stars.

Multiplicity emerges as an additional ingredient in shaping chromospheric magnetic activity. Binary systems, especially in the regime of rapid rotators ({\prot} $<$ 5 {\textendash} 10 days), systematically show a higher upper envelope of {\CaiiIRT} and {\sphaverage} values than single stars at a given rotation period (Fig.~\ref{fig:Figure_discussion_multiplicity}), consistent with tidal interactions maintaining or enhancing rotation and magnetic activity. This highlights the important role of stellar architecture in shaping rotation–activity relations and age diagnostics.

All in all, this work sheds light on the impact of the intermediate-period gap on chromospheric magnetic activity. It represents a homogeneous reference data set for future studies of stellar magnetism, and opens the door to extending this type of analysis with complementary samples. These will benefit from the wide availability of the {\gaia} chromospheric index in current and upcoming data releases \citep{lanzafame23}, together with rotational measurements from ground- and space-based photometry such as the Transiting Exoplanet Survey Satellite (TESS; \citealt{ricker15}), the Zwicky Transient Facility (ZTF; \citealt{bellm19}), and the All Sky Automated Survey for SuperNovae (ASAS-SN; \citealt{shappee14}), as well as the upcoming Planetary Transits and Oscillations of Stars (PLATO; \citealt{rauer25}), the Earth~2.0 mission (ET; \citealt{ge24}), and the Rubin Observatory \citep{ivezic19}. All of the above will enable a better understanding of the age-rotation-activity relations, the transition between different braking regimes, and the magnetic environments of exoplanetary systems.
\section*{Data availability}

Table~\ref{tab:table_catalog} and Table~\ref{tab:gap_vs_color} are available in electronic form on Zenodo (\url{https://zenodo.org/records/21086333}) and at the CDS via anonymous ftp to cdsarc.u-strasbg.fr ([???]) or via http://cdsweb.u-strasbg.fr/cgi-bin/qcat?J/A+A/[???].


\begin{acknowledgements}
    We thank the referee for their constructive and insightful comments.
    We thank Thomas Masseron for valuable discussions.

    D.G.R. acknowledges support from the Spanish Ministry of Science and Innovation (MICINN) with the \emph{Juan de la Cierva} fellowship program under contract JDC2022-049054-I.
    
    D.G.R., A.R.G.S., S.M., and R.A.G. acknowledge support from the Spanish Ministry of Science and Innovation (MICINN) with the grant No. PID2023-149439NB-C41.

    D.G.R., R.A.G., and P.G.B. acknowledge support from the Spanish Ministry of Science and Innovation (MICINN) with the grant No. PID2023-146453NB-I00 (PLAtoSOnG, PI: Beck).

    T.R.Y. acknowledges support from the Instituto de Astrofísica de Canarias (IAC) Early Career Visitor Program.

    A.R.G.S acknowledges the support from Funda\c{c}\~ao para a Ci\^encia e a Tecnologia (FCT) through national funds by the grant UID/04434/2025 and work contract No. 2020.02480.CEECIND/CP1631/CT0001 (DOI: 10.54499/2020.02480.CEECIND/CP1631/CT0001). 

    
     R.A.G., A.R.G.S., and S.M. acknowledge financial support from the Centre national d’études spatiales (CNES), France (ROR: \url{https://ror.org/04h1h0y33}), within the framework of the GOLF/SoHO and PLATO space missions.
    
    D.H.G acknowledges the support of a fellowship from ”la Caixa” Foundation (ID 100010434). The fellowship code is LCF/BQ/DI23/11990068.


    P.G.B. acknowledges support by the Spanish Ministry of Science and Innovation with the \emph{Ramón y Cajal} fellowship (RYC-2021-033137-I, MRR4032204).

    This research was supported by the International Space Science Institute (ISSI) in Bern, through the ISSI International Team project 24-629 (``Multi-scale variability in solar and stellar magnetic cycles'').
    
    This paper includes data collected by the {\kepler} mission and obtained from the MAST data archive at the Space Telescope Science Institute (STScI). Funding for the {\kepler} mission is provided by the NASA Science Mission Directorate. STScI is operated by the Association of Universities for Research in Astronomy, Inc., under NASA contract NAS 5–26555.

    This work has made use of data from the European Space Agency (ESA) mission {\gaia} (\url{https://www.cosmos.esa.int/gaia}), processed by the {\gaia} Data Processing and Analysis Consortium (DPAC, \url{https://www.cosmos.esa.int/web/gaia/dpac/consortium}). Funding for the DPAC has been provided by national institutions, in particular the institutions participating in the {\gaia} Multilateral Agreement.

    This work made use of the \texttt{Gaia-Kepler.fun} crossmatch database created by Megan Bedell.

    This work made extensive use of TOPCAT \citep{taylor05}.
\end{acknowledgements}
\bibliographystyle{aa} 
\bibliography{kepler_activity_calcium_nuv} 

@ARTICLE{agueros18,
       author = {{Ag{\"u}eros}, M.~A. and {Bowsher}, E.~C. and {Bochanski}, J.~J. and {Cargile}, P.~A. and {Covey}, K.~R. and {Douglas}, S.~T. and {Kraus}, A. and {Kundert}, A. and {Law}, N.~M. and {Ahmadi}, A. and {Arce}, H.~G.},
        title = "{A New Look at an Old Cluster: The Membership, Rotation, and Magnetic Activity of Low-mass Stars in the 1.3 Gyr Old Open Cluster NGC 752}",
      journal = {\apj},
         year = 2018,
        month = jul,
       volume = {862},
       number = {1},
          eid = {33},
        pages = {33},
          doi = {10.3847/1538-4357/aac6ed},
archivePrefix = {arXiv},
       eprint = {1804.02016},
 primaryClass = {astro-ph.SR},
       adsurl = {https://ui.adsabs.harvard.edu/abs/2018ApJ...862...33A}
}

@ARTICLE{ahuir21,
       author = {{Ahuir}, J. and {Strugarek}, A. and {Brun}, A.-S. and {Mathis}, S.},
        title = "{Magnetic and tidal migration of close-in planets. Influence of secular evolution on their population}",
      journal = {\aap},
         year = 2021,
        month = jun,
       volume = {650},
          eid = {A126},
        pages = {A126},
          doi = {10.1051/0004-6361/202040173},
archivePrefix = {arXiv},
       eprint = {2104.01004},
 primaryClass = {astro-ph.EP},
       adsurl = {https://ui.adsabs.harvard.edu/abs/2021A&A...650A.126A}
}

@ARTICLE{ahumada20,
       author = {{Ahumada}, Romina and {Allende Prieto}, Carlos and {Almeida}, Andr{\'e}s and {Anders}, Friedrich and {Anderson}, Scott F. and {Andrews}, Brett H. and {Anguiano}, Borja and {Arcodia}, Riccardo and {Armengaud}, Eric and {Aubert}, Marie and {Avila}, Santiago and {Avila-Reese}, Vladimir and {Badenes}, Carles and {Balland}, Christophe and {Barger}, Kat and {Barrera-Ballesteros}, Jorge K. and {Basu}, Sarbani and {Bautista}, Julian and {Beaton}, Rachael L. and {Beers}, Timothy C. and {Benavides}, B. Izamar T. and {Bender}, Chad F. and {Bernardi}, Mariangela and {Bershady}, Matthew and {Beutler}, Florian and {Bidin}, Christian Moni and {Bird}, Jonathan and {Bizyaev}, Dmitry and {Blanc}, Guillermo A. and {Blanton}, Michael R. and {Boquien}, M{\'e}d{\'e}ric and {Borissova}, Jura and {Bovy}, Jo and {Brandt}, W.~N. and {Brinkmann}, Jonathan and {Brownstein}, Joel R. and {Bundy}, Kevin and {Bureau}, Martin and {Burgasser}, Adam and {Burtin}, Etienne and {Cano-D{\'\i}az}, Mariana and {Capasso}, Raffaella and {Cappellari}, Michele and {Carrera}, Ricardo and {Chabanier}, Sol{\`e}ne and {Chaplin}, William and {Chapman}, Michael and {Cherinka}, Brian and {Chiappini}, Cristina and {Doohyun Choi}, Peter and {Chojnowski}, S. Drew and {Chung}, Haeun and {Clerc}, Nicolas and {Coffey}, Damien and {Comerford}, Julia M. and {Comparat}, Johan and {da Costa}, Luiz and {Cousinou}, Marie-Claude and {Covey}, Kevin and {Crane}, Jeffrey D. and {Cunha}, Katia and {Ilha}, Gabriele da Silva and {Dai}, Yu Sophia and {Damsted}, Sanna B. and {Darling}, Jeremy and {Davidson}, Jr., James W. and {Davies}, Roger and {Dawson}, Kyle and {De}, Nikhil and {de la Macorra}, Axel and {De Lee}, Nathan and {Queiroz}, Anna B{\'a}rbara de Andrade and {Deconto Machado}, Alice and {de la Torre}, Sylvain and {Dell'Agli}, Flavia and {du Mas des Bourboux}, H{\'e}lion and {Diamond-Stanic}, Aleksandar M. and {Dillon}, Sean and {Donor}, John and {Drory}, Niv and {Duckworth}, Chris and {Dwelly}, Tom and {Ebelke}, Garrett and {Eftekharzadeh}, Sarah and {Davis Eigenbrot}, Arthur and {Elsworth}, Yvonne P. and {Eracleous}, Mike and {Erfanianfar}, Ghazaleh and {Escoffier}, Stephanie and {Fan}, Xiaohui and {Farr}, Emily and {Fern{\'a}ndez-Trincado}, Jos{\'e} G. and {Feuillet}, Diane and {Finoguenov}, Alexis and {Fofie}, Patricia and {Fraser-McKelvie}, Amelia and {Frinchaboy}, Peter M. and {Fromenteau}, Sebastien and {Fu}, Hai and {Galbany}, Llu{\'\i}s and {Garcia}, Rafael A. and {Garc{\'\i}a-Hern{\'a}ndez}, D.~A. and {Garma Oehmichen}, Luis Alberto and {Ge}, Junqiang and {Geimba Maia}, Marcio Antonio and {Geisler}, Doug and {Gelfand}, Joseph and {Goddy}, Julian and {Gonzalez-Perez}, Violeta and {Grabowski}, Kathleen and {Green}, Paul and {Grier}, Catherine J. and {Guo}, Hong and {Guy}, Julien and {Harding}, Paul and {Hasselquist}, Sten and {Hawken}, Adam James and {Hayes}, Christian R. and {Hearty}, Fred and {Hekker}, S. and {Hogg}, David W. and {Holtzman}, Jon A. and {Horta}, Danny and {Hou}, Jiamin and {Hsieh}, Bau-Ching and {Huber}, Daniel and {Hunt}, Jason A.~S. and {Ider Chitham}, J. and {Imig}, Julie and {Jaber}, Mariana and {Jimenez Angel}, Camilo Eduardo and {Johnson}, Jennifer A. and {Jones}, Amy M. and {J{\"o}nsson}, Henrik and {Jullo}, Eric and {Kim}, Yerim and {Kinemuchi}, Karen and {Kirkpatrick}, IV, Charles C. and {Kite}, George W. and {Klaene}, Mark and {Kneib}, Jean-Paul and {Kollmeier}, Juna A. and {Kong}, Hui and {Kounkel}, Marina and {Krishnarao}, Dhanesh and {Lacerna}, Ivan and {Lan}, Ting-Wen and {Lane}, Richard R. and {Law}, David R. and {Le Goff}, Jean-Marc and {Leung}, Henry W. and {Lewis}, Hannah and {Li}, Cheng and {Lian}, Jianhui and {Lin}, Lihwai and {Long}, Dan and {Longa-Pe{\~n}a}, Pen{\'e}lope and {Lundgren}, Britt and {Lyke}, Brad W. and {Mackereth}, J. Ted and {MacLeod}, Chelsea L. and {Majewski}, Steven R. and {Manchado}, Arturo and {Maraston}, Claudia and {Martini}, Paul and {Masseron}, Thomas and {Masters}, Karen L. and {Mathur}, Savita and {McDermid}, Richard M. and {Merloni}, Andrea and {Merrifield}, Michael and {M{\'e}sz{\'a}ros}, Szabolcs and {Miglio}, Andrea and {Minniti}, Dante and {Minsley}, Rebecca and {Miyaji}, Takamitsu and {Mohammad}, Faizan Gohar and {Mosser}, Benoit and {Mueller}, Eva-Maria and {Muna}, Demitri and {Mu{\~n}oz-Guti{\'e}rrez}, Andrea and {Myers}, Adam D. and {Nadathur}, Seshadri and {Nair}, Preethi and {Nandra}, Kirpal and {Correa do Nascimento}, Janaina and {Nevin}, Rebecca Jean and {Newman}, Jeffrey A. and {Nidever}, David L. and {Nitschelm}, Christian and {Noterdaeme}, Pasquier and {O'Connell}, Julia E. and {Olmstead}, Matthew D. and {Oravetz}, Daniel and {Oravetz}, Audrey and {Osorio}, Yeisson and {Pace}, Zachary J. and {Padilla}, Nelson and {Palanque-Delabrouille}, Nathalie and {Palicio}, Pedro A.},
        title = "{The 16th Data Release of the Sloan Digital Sky Surveys: First Release from the APOGEE-2 Southern Survey and Full Release of eBOSS Spectra}",
      journal = {\apjs},
         year = 2020,
        month = jul,
       volume = {249},
       number = {1},
          eid = {3},
        pages = {3},
          doi = {10.3847/1538-4365/ab929e},
archivePrefix = {arXiv},
       eprint = {1912.02905},
 primaryClass = {astro-ph.GA},
       adsurl = {https://ui.adsabs.harvard.edu/abs/2020ApJS..249....3A}
}

@ARTICLE{airapetian20,
       author = {{Airapetian}, V.~S. and {Barnes}, R. and {Cohen}, O. and {Collinson}, G.~A. and {Danchi}, W.~C. and {Dong}, C.~F. and {Del Genio}, A.~D. and {France}, K. and {Garcia-Sage}, K. and {Glocer}, A. and {Gopalswamy}, N. and {Grenfell}, J.~L. and {Gronoff}, G. and {G{\"u}del}, M. and {Herbst}, K. and {Henning}, W.~G. and {Jackman}, C.~H. and {Jin}, M. and {Johnstone}, C.~P. and {Kaltenegger}, L. and {Kay}, C.~D. and {Kobayashi}, K. and {Kuang}, W. and {Li}, G. and {Lynch}, B.~J. and {L{\"u}ftinger}, T. and {Luhmann}, J.~G. and {Maehara}, H. and {Mlynczak}, M.~G. and {Notsu}, Y. and {Osten}, R.~A. and {Ramirez}, R.~M. and {Rugheimer}, S. and {Scheucher}, M. and {Schlieder}, J.~E. and {Shibata}, K. and {Sousa-Silva}, C. and {Stamenkovi{\'c}}, V. and {Strangeway}, R.~J. and {Usmanov}, A.~V. and {Vergados}, P. and {Verkhoglyadova}, O.~P. and {Vidotto}, A.~A. and {Voytek}, M. and {Way}, M.~J. and {Zank}, G.~P. and {Yamashiki}, Y.},
        title = "{Impact of space weather on climate and habitability of terrestrial-type exoplanets}",
      journal = {International Journal of Astrobiology},
         year = 2020,
        month = apr,
       volume = {19},
       number = {2},
        pages = {136-194},
          doi = {10.1017/S1473550419000132},
archivePrefix = {arXiv},
       eprint = {1905.05093},
 primaryClass = {astro-ph.EP},
       adsurl = {https://ui.adsabs.harvard.edu/abs/2020IJAsB..19..136A}
}

@ARTICLE{allan26,
       author = {{Allan}, Andrew P. and {Vidotto}, Aline A. and {Sanz-Forcada}, Jorge and {Villarreal D'Angelo}, Carolina},
        title = "{The effects of stellar activity cycles on planetary atmospheric escape and the He I 1083 nm transit signature}",
      journal = {\mnras},
         year = 2026,
        month = jan,
       volume = {545},
       number = {2},
          eid = {staf1855},
        pages = {staf1855},
          doi = {10.1093/mnras/staf1855},
archivePrefix = {arXiv},
       eprint = {2510.23282},
 primaryClass = {astro-ph.EP},
       adsurl = {https://ui.adsabs.harvard.edu/abs/2026MNRAS.545f1855A}
}

@ARTICLE{andretta05,
       author = {{Andretta}, V. and {Bus{\`a}}, I. and {Gomez}, M.~T. and {Terranegra}, L.},
        title = "{The Ca II Infrared Triplet as a stellar activity diagnostic . I. Non-LTE photospheric profiles and definition of the R$_{IRT}$ indicator}",
      journal = {\aap},
         year = 2005,
        month = feb,
       volume = {430},
        pages = {669-677},
          doi = {10.1051/0004-6361:20041745},
       adsurl = {https://ui.adsabs.harvard.edu/abs/2005A&A...430..669A}
}

@ARTICLE{angus15,
       author = {{Angus}, Ruth and {Aigrain}, Suzanne and {Foreman-Mackey}, Daniel and {McQuillan}, Amy},
        title = "{Calibrating gyrochronology using Kepler asteroseismic targets}",
      journal = {\mnras},
         year = 2015,
        month = jun,
       volume = {450},
       number = {2},
        pages = {1787-1798},
          doi = {10.1093/mnras/stv423},
archivePrefix = {arXiv},
       eprint = {1502.06965},
 primaryClass = {astro-ph.EP},
       adsurl = {https://ui.adsabs.harvard.edu/abs/2015MNRAS.450.1787A}
}

@ARTICLE{angus19,
       author = {{Angus}, Ruth and {Morton}, Timothy D. and {Foreman-Mackey}, Daniel and {van Saders}, Jennifer and {Curtis}, Jason and {Kane}, Stephen R. and {Bedell}, Megan and {Kiman}, Rocio and {Hogg}, David W. and {Brewer}, John},
        title = "{Toward Precise Stellar Ages: Combining Isochrone Fitting with Empirical Gyrochronology}",
      journal = {\aj},
         year = 2019,
        month = nov,
       volume = {158},
       number = {5},
          eid = {173},
        pages = {173},
          doi = {10.3847/1538-3881/ab3c53},
archivePrefix = {arXiv},
       eprint = {1908.07528},
 primaryClass = {astro-ph.SR},
       adsurl = {https://ui.adsabs.harvard.edu/abs/2019AJ....158..173A}
}

@ARTICLE{angus20,
       author = {{Angus}, Ruth and {Beane}, Angus and {Price-Whelan}, Adrian M. and {Newton}, Elisabeth and {Curtis}, Jason L. and {Berger}, Travis and {van Saders}, Jennifer and {Kiman}, Rocio and {Foreman-Mackey}, Daniel and {Lu}, Yuxi Lucy and {Anderson}, Lauren and {Faherty}, Jacqueline K.},
        title = "{Exploring the Evolution of Stellar Rotation Using Galactic Kinematics}",
      journal = {\aj},
         year = 2020,
        month = aug,
       volume = {160},
       number = {2},
          eid = {90},
        pages = {90},
          doi = {10.3847/1538-3881/ab91b2},
archivePrefix = {arXiv},
       eprint = {2005.09387},
 primaryClass = {astro-ph.SR},
       adsurl = {https://ui.adsabs.harvard.edu/abs/2020AJ....160...90A}
}

@ARTICLE{bahcall01,
       author = {{Bahcall}, John N. and {Pinsonneault}, M.~H. and {Basu}, Sarbani},
        title = "{Solar Models: Current Epoch and Time Dependences, Neutrinos, and Helioseismological Properties}",
      journal = {\apj},
         year = 2001,
        month = jul,
       volume = {555},
       number = {2},
        pages = {990-1012},
          doi = {10.1086/321493},
archivePrefix = {arXiv},
       eprint = {astro-ph/0010346},
 primaryClass = {astro-ph},
       adsurl = {https://ui.adsabs.harvard.edu/abs/2001ApJ...555..990B}
}

@ARTICLE{baliunas95,
       author = {{Baliunas}, S.~L. and {Donahue}, R.~A. and {Soon}, W.~H. and {Horne}, J.~H. and {Frazer}, J. and {Woodard-Eklund}, L. and {Bradford}, M. and {Rao}, L.~M. and {Wilson}, O.~C. and {Zhang}, Q. and {Bennett}, W. and {Briggs}, J. and {Carroll}, S.~M. and {Duncan}, D.~K. and {Figueroa}, D. and {Lanning}, H.~H. and {Misch}, T. and {Mueller}, J. and {Noyes}, R.~W. and {Poppe}, D. and {Porter}, A.~C. and {Robinson}, C.~R. and {Russell}, J. and {Shelton}, J.~C. and {Soyumer}, T. and {Vaughan}, A.~H. and {Whitney}, J.~H.},
        title = "{Chromospheric Variations in Main-Sequence Stars. II.}",
      journal = {\apj},
         year = 1995,
        month = jan,
       volume = {438},
        pages = {269},
          doi = {10.1086/175072},
       adsurl = {https://ui.adsabs.harvard.edu/abs/1995ApJ...438..269B}
}

@ARTICLE{barnes03,
       author = {{Barnes}, Sydney A.},
        title = "{On the Rotational Evolution of Solar- and Late-Type Stars, Its Magnetic Origins, and the Possibility of Stellar Gyrochronology}",
      journal = {\apj},
         year = 2003,
        month = mar,
       volume = {586},
       number = {1},
        pages = {464-479},
          doi = {10.1086/367639},
archivePrefix = {arXiv},
       eprint = {astro-ph/0303631},
 primaryClass = {astro-ph},
       adsurl = {https://ui.adsabs.harvard.edu/abs/2003ApJ...586..464B}
}

@ARTICLE{barnes07,
       author = {{Barnes}, Sydney A.},
        title = "{Ages for Illustrative Field Stars Using Gyrochronology: Viability, Limitations, and Errors}",
      journal = {\apj},
         year = 2007,
        month = nov,
       volume = {669},
       number = {2},
        pages = {1167-1189},
          doi = {10.1086/519295},
archivePrefix = {arXiv},
       eprint = {0704.3068},
 primaryClass = {astro-ph},
       adsurl = {https://ui.adsabs.harvard.edu/abs/2007ApJ...669.1167B}
}

@ARTICLE{beck18,
       author = {{Beck}, P.~G. and {Mathis}, S. and {Gallet}, F. and {Charbonnel}, C. and {Benbakoura}, M. and {Garc{\'\i}a}, R.~A. and {do Nascimento}, J.-D.},
        title = "{Testing tidal theory for evolved stars by using red giant binaries observed by Kepler}",
      journal = {\mnras},
         year = 2018,
        month = sep,
       volume = {479},
       number = {1},
        pages = {L123-L128},
          doi = {10.1093/mnrasl/sly114},
archivePrefix = {arXiv},
       eprint = {1806.07208},
 primaryClass = {astro-ph.SR},
       adsurl = {https://ui.adsabs.harvard.edu/abs/2018MNRAS.479L.123B}
}

@ARTICLE{beck24,
       author = {{Beck}, P.~G. and {Grossmann}, D.~H. and {Steinwender}, L. and {Schimak}, L.~S. and {Muntean}, N. and {Vrard}, M. and {Patton}, R.~A. and {Merc}, J. and {Mathur}, S. and {Garcia}, R.~A. and {Pinsonneault}, M.~H. and {Rowan}, D.~M. and {Gaulme}, P. and {Allende Prieto}, C. and {Arellano-C{\'o}rdova}, K.~Z. and {Cao}, L. and {Corsaro}, E. and {Creevey}, O. and {Hambleton}, K.~M. and {Hanslmeier}, A. and {Holl}, B. and {Johnson}, J. and {Mathis}, S. and {Godoy-Rivera}, D. and {S{\'\i}mon-D{\'\i}az}, S. and {Zinn}, J.~C.},
        title = "{Constraining stellar and orbital co-evolution through ensemble seismology of solar-like oscillators in binary systems. A census of oscillating red giants and dwarf stars in Gaia DR3 binaries}",
      journal = {\aap},
         year = 2024,
        month = feb,
       volume = {682},
          eid = {A7},
        pages = {A7},
          doi = {10.1051/0004-6361/202346810},
archivePrefix = {arXiv},
       eprint = {2307.10812},
 primaryClass = {astro-ph.SR},
       adsurl = {https://ui.adsabs.harvard.edu/abs/2024A&A...682A...7B}
}

@ARTICLE{beck26a,
       author = {{Beck}, Paul G.},
        title = "{Tales of stellar and binary coevolution, told by stellar oscillations: Binary demographics and their impact on stellar mass, orbits, and age estimates in main-sequence and red-giant stars}",
      journal = {\aap},
         year = 2026,
        month = mar,
       volume = {707},
          eid = {A298},
        pages = {A298},
          doi = {10.1051/0004-6361/202555157},
archivePrefix = {arXiv},
       eprint = {2512.13581},
 primaryClass = {astro-ph.SR},
       adsurl = {https://ui.adsabs.harvard.edu/abs/2026A&A...707A.298B}
}

@ARTICLE{beck26b,
       author = {{Beck}, P.~G. and {Masseron}, T. and {Pavlovski}, K. and {Godoy-Rivera}, D. and {Mathur}, S. and {Grossmann}, D.~H. and {Hamy}, A. and {Palakkatharappil}, D.~B. and {Panetier}, E. and {Garc{\'\i}a}, R.~A. and {Merc}, J. and {Lu}, Y. and {Amestoy}, I. and {Deeg}, H.~J.},
        title = "{Dynamical mass of a solar-like oscillator at the main sequence turnoff from Gaia astrometry and ground-based spectroscopy}",
      journal = {\aap},
         year = 2026,
        month = feb,
       volume = {706},
          eid = {L19},
        pages = {L19},
          doi = {10.1051/0004-6361/202557452},
archivePrefix = {arXiv},
       eprint = {2601.14197},
 primaryClass = {astro-ph.SR},
       adsurl = {https://ui.adsabs.harvard.edu/abs/2026A&A...706L..19B}
}

@ARTICLE{bellm19,
       author = {{Bellm}, Eric C. and {Kulkarni}, Shrinivas R. and {Graham}, Matthew J. and {Dekany}, Richard and {Smith}, Roger M. and {Riddle}, Reed and {Masci}, Frank J. and {Helou}, George and {Prince}, Thomas A. and {Adams}, Scott M. and {Barbarino}, C. and {Barlow}, Tom and {Bauer}, James and {Beck}, Ron and {Belicki}, Justin and {Biswas}, Rahul and {Blagorodnova}, Nadejda and {Bodewits}, Dennis and {Bolin}, Bryce and {Brinnel}, Valery and {Brooke}, Tim and {Bue}, Brian and {Bulla}, Mattia and {Burruss}, Rick and {Cenko}, S. Bradley and {Chang}, Chan-Kao and {Connolly}, Andrew and {Coughlin}, Michael and {Cromer}, John and {Cunningham}, Virginia and {De}, Kishalay and {Delacroix}, Alex and {Desai}, Vandana and {Duev}, Dmitry A. and {Eadie}, Gwendolyn and {Farnham}, Tony L. and {Feeney}, Michael and {Feindt}, Ulrich and {Flynn}, David and {Franckowiak}, Anna and {Frederick}, S. and {Fremling}, C. and {Gal-Yam}, Avishay and {Gezari}, Suvi and {Giomi}, Matteo and {Goldstein}, Daniel A. and {Golkhou}, V. Zach and {Goobar}, Ariel and {Groom}, Steven and {Hacopians}, Eugean and {Hale}, David and {Henning}, John and {Ho}, Anna Y.~Q. and {Hover}, David and {Howell}, Justin and {Hung}, Tiara and {Huppenkothen}, Daniela and {Imel}, David and {Ip}, Wing-Huen and {Ivezi{\'c}}, {\v{Z}}eljko and {Jackson}, Edward and {Jones}, Lynne and {Juric}, Mario and {Kasliwal}, Mansi M. and {Kaspi}, S. and {Kaye}, Stephen and {Kelley}, Michael S.~P. and {Kowalski}, Marek and {Kramer}, Emily and {Kupfer}, Thomas and {Landry}, Walter and {Laher}, Russ R. and {Lee}, Chien-De and {Lin}, Hsing Wen and {Lin}, Zhong-Yi and {Lunnan}, Ragnhild and {Giomi}, Matteo and {Mahabal}, Ashish and {Mao}, Peter and {Miller}, Adam A. and {Monkewitz}, Serge and {Murphy}, Patrick and {Ngeow}, Chow-Choong and {Nordin}, Jakob and {Nugent}, Peter and {Ofek}, Eran and {Patterson}, Maria T. and {Penprase}, Bryan and {Porter}, Michael and {Rauch}, Ludwig and {Rebbapragada}, Umaa and {Reiley}, Dan and {Rigault}, Mickael and {Rodriguez}, Hector and {van Roestel}, Jan and {Rusholme}, Ben and {van Santen}, Jakob and {Schulze}, S. and {Shupe}, David L. and {Singer}, Leo P. and {Soumagnac}, Maayane T. and {Stein}, Robert and {Surace}, Jason and {Sollerman}, Jesper and {Szkody}, Paula and {Taddia}, F. and {Terek}, Scott and {Van Sistine}, Angela and {van Velzen}, Sjoert and {Vestrand}, W. Thomas and {Walters}, Richard and {Ward}, Charlotte and {Ye}, Quan-Zhi and {Yu}, Po-Chieh and {Yan}, Lin and {Zolkower}, Jeffry},
        title = "{The Zwicky Transient Facility: System Overview, Performance, and First Results}",
      journal = {\pasp},
         year = 2019,
        month = jan,
       volume = {131},
       number = {995},
        pages = {018002},
          doi = {10.1088/1538-3873/aaecbe},
archivePrefix = {arXiv},
       eprint = {1902.01932},
 primaryClass = {astro-ph.IM},
       adsurl = {https://ui.adsabs.harvard.edu/abs/2019PASP..131a8002B}
}

@ARTICLE{benbakoura21,
       author = {{Benbakoura}, M. and {Gaulme}, P. and {McKeever}, J. and {Sekaran}, S. and {Beck}, P.~G. and {Spada}, F. and {Jackiewicz}, J. and {Mathis}, S. and {Mathur}, S. and {Tkachenko}, A. and {Garc{\'\i}a}, R.~A.},
        title = "{Spectroscopic and seismic analysis of red giants in eclipsing binaries discovered by Kepler}",
      journal = {\aap},
         year = 2021,
        month = apr,
       volume = {648},
          eid = {A113},
        pages = {A113},
          doi = {10.1051/0004-6361/202037783},
archivePrefix = {arXiv},
       eprint = {2101.05351},
 primaryClass = {astro-ph.SR},
       adsurl = {https://ui.adsabs.harvard.edu/abs/2021A&A...648A.113B}
}

@ARTICLE{berger20,
       author = {{Berger}, Travis A. and {Huber}, Daniel and {van Saders}, Jennifer L. and {Gaidos}, Eric and {Tayar}, Jamie and {Kraus}, Adam L.},
        title = "{The Gaia-Kepler Stellar Properties Catalog. I. Homogeneous Fundamental Properties for 186,301 Kepler Stars}",
      journal = {\aj},
         year = 2020,
        month = jun,
       volume = {159},
       number = {6},
          eid = {280},
        pages = {280},
          doi = {10.3847/1538-3881/159/6/280},
archivePrefix = {arXiv},
       eprint = {2001.07737},
 primaryClass = {astro-ph.SR},
       adsurl = {https://ui.adsabs.harvard.edu/abs/2020AJ....159..280B}
}

@ARTICLE{beyer24,
       author = {{Beyer}, Alexa C. and {White}, Russel J.},
        title = "{The Kraft Break Sharply Divides Low-mass and Intermediate-mass Stars}",
      journal = {\apj},
         year = 2024,
        month = sep,
       volume = {973},
       number = {1},
          eid = {28},
        pages = {28},
          doi = {10.3847/1538-4357/ad6b0d},
archivePrefix = {arXiv},
       eprint = {2408.02638},
 primaryClass = {astro-ph.SR},
       adsurl = {https://ui.adsabs.harvard.edu/abs/2024ApJ...973...28B}
}

@ARTICLE{bjorgen18,
       author = {{Bj{\o}rgen}, Johan P. and {Sukhorukov}, Andrii V. and {Leenaarts}, Jorrit and {Carlsson}, Mats and {de la Cruz Rodr{\'\i}guez}, Jaime and {Scharmer}, G{\"o}ran B. and {Hansteen}, Viggo H.},
        title = "{Three-dimensional modeling of the Ca II H and K lines in the solar atmosphere}",
      journal = {\aap},
         year = 2018,
        month = mar,
       volume = {611},
          eid = {A62},
        pages = {A62},
          doi = {10.1051/0004-6361/201731926},
archivePrefix = {arXiv},
       eprint = {1712.01045},
 primaryClass = {astro-ph.SR},
       adsurl = {https://ui.adsabs.harvard.edu/abs/2018A&A...611A..62B}
}

@ARTICLE{bohmvitense07,
       author = {{B{\"o}hm-Vitense}, Erika},
        title = "{Chromospheric Activity in G and K Main-Sequence Stars, and What It Tells Us about Stellar Dynamos}",
      journal = {\apj},
         year = 2007,
        month = mar,
       volume = {657},
       number = {1},
        pages = {486-493},
          doi = {10.1086/510482},
       adsurl = {https://ui.adsabs.harvard.edu/abs/2007ApJ...657..486B}
}

@ARTICLE{bonanno25,
       author = {{Bonanno}, Alfio M. and {Corsaro}, Enrico and {Metcalfe}, Travis S. and {Breton}, Sylvain N. and {Creevey}, Orlagh L. and {Lindsay}, Christopher J.},
        title = "{Asteroseismic Calibration of the Rossby Number and Its Connection to the Stellar Dynamo and Fundamental Properties}",
      journal = {\apj},
         year = 2025,
        month = dec,
       volume = {995},
       number = {1},
          eid = {32},
        pages = {32},
          doi = {10.3847/1538-4357/ae12f2},
archivePrefix = {arXiv},
       eprint = {2510.12471},
 primaryClass = {astro-ph.SR},
       adsurl = {https://ui.adsabs.harvard.edu/abs/2025ApJ...995...32B}
}

@ARTICLE{borosaikia18,
       author = {{Boro Saikia}, S. and {Marvin}, C.~J. and {Jeffers}, S.~V. and {Reiners}, A. and {Cameron}, R. and {Marsden}, S.~C. and {Petit}, P. and {Warnecke}, J. and {Yadav}, A.~P.},
        title = "{Chromospheric activity catalogue of 4454 cool stars. Questioning the active branch of stellar activity cycles}",
      journal = {\aap},
         year = 2018,
        month = aug,
       volume = {616},
          eid = {A108},
        pages = {A108},
          doi = {10.1051/0004-6361/201629518},
archivePrefix = {arXiv},
       eprint = {1803.11123},
 primaryClass = {astro-ph.SR},
       adsurl = {https://ui.adsabs.harvard.edu/abs/2018A&A...616A.108B}
}

@ARTICLE{borucki10,
       author = {{Borucki}, William J. and {Koch}, David and {Basri}, Gibor and {Batalha}, Natalie and {Brown}, Timothy and {Caldwell}, Douglas and {Caldwell}, John and {Christensen-Dalsgaard}, J{\o}rgen and {Cochran}, William D. and {DeVore}, Edna and {Dunham}, Edward W. and {Dupree}, Andrea K. and {Gautier}, Thomas N. and {Geary}, John C. and {Gilliland}, Ronald and {Gould}, Alan and {Howell}, Steve B. and {Jenkins}, Jon M. and {Kondo}, Yoji and {Latham}, David W. and {Marcy}, Geoffrey W. and {Meibom}, S{\o}ren and {Kjeldsen}, Hans and {Lissauer}, Jack J. and {Monet}, David G. and {Morrison}, David and {Sasselov}, Dimitar and {Tarter}, Jill and {Boss}, Alan and {Brownlee}, Don and {Owen}, Toby and {Buzasi}, Derek and {Charbonneau}, David and {Doyle}, Laurance and {Fortney}, Jonathan and {Ford}, Eric B. and {Holman}, Matthew J. and {Seager}, Sara and {Steffen}, Jason H. and {Welsh}, William F. and {Rowe}, Jason and {Anderson}, Howard and {Buchhave}, Lars and {Ciardi}, David and {Walkowicz}, Lucianne and {Sherry}, William and {Horch}, Elliott and {Isaacson}, Howard and {Everett}, Mark E. and {Fischer}, Debra and {Torres}, Guillermo and {Johnson}, John Asher and {Endl}, Michael and {MacQueen}, Phillip and {Bryson}, Stephen T. and {Dotson}, Jessie and {Haas}, Michael and {Kolodziejczak}, Jeffrey and {Van Cleve}, Jeffrey and {Chandrasekaran}, Hema and {Twicken}, Joseph D. and {Quintana}, Elisa V. and {Clarke}, Bruce D. and {Allen}, Christopher and {Li}, Jie and {Wu}, Haley and {Tenenbaum}, Peter and {Verner}, Ekaterina and {Bruhweiler}, Frederick and {Barnes}, Jason and {Prsa}, Andrej},
        title = "{Kepler Planet-Detection Mission: Introduction and First Results}",
      journal = {Science},
         year = 2010,
        month = feb,
       volume = {327},
       number = {5968},
        pages = {977},
          doi = {10.1126/science.1185402},
       adsurl = {https://ui.adsabs.harvard.edu/abs/2010Sci...327..977B}
}

@ARTICLE{bouma23,
       author = {{Bouma}, Luke G. and {Palumbo}, Elsa K. and {Hillenbrand}, Lynne A.},
        title = "{The Empirical Limits of Gyrochronology}",
      journal = {\apjl},
         year = 2023,
        month = apr,
       volume = {947},
       number = {1},
          eid = {L3},
        pages = {L3},
          doi = {10.3847/2041-8213/acc589},
archivePrefix = {arXiv},
       eprint = {2303.08830},
 primaryClass = {astro-ph.SR},
       adsurl = {https://ui.adsabs.harvard.edu/abs/2023ApJ...947L...3B}
}

@ARTICLE{brandenburg18,
       author = {{Brandenburg}, Axel and {Giampapa}, Mark S.},
        title = "{Enhanced Stellar Activity for Slow Antisolar Differential Rotation?}",
      journal = {\apjl},
         year = 2018,
        month = mar,
       volume = {855},
       number = {2},
          eid = {L22},
        pages = {L22},
          doi = {10.3847/2041-8213/aab20a},
archivePrefix = {arXiv},
       eprint = {1802.08689},
 primaryClass = {astro-ph.SR},
       adsurl = {https://ui.adsabs.harvard.edu/abs/2018ApJ...855L..22B}
}

@ARTICLE{breton25,
       author = {{Breton}, S.~N. and {Distefano}, E. and {Lanzafame}, A.~C. and {Palakkatharappil}, D.~B.},
        title = "{Rotation of young solar-type stars as seen by Gaia and K2}",
      journal = {\aap},
         year = 2025,
        month = sep,
       volume = {701},
          eid = {A263},
        pages = {A263},
          doi = {10.1051/0004-6361/202553912},
archivePrefix = {arXiv},
       eprint = {2507.20909},
 primaryClass = {astro-ph.SR},
       adsurl = {https://ui.adsabs.harvard.edu/abs/2025A&A...701A.263B}
}

@ARTICLE{brown11,
       author = {{Brown}, Timothy M. and {Latham}, David W. and {Everett}, Mark E. and {Esquerdo}, Gilbert A.},
        title = "{Kepler Input Catalog: Photometric Calibration and Stellar Classification}",
      journal = {\aj},
         year = 2011,
        month = oct,
       volume = {142},
       number = {4},
          eid = {112},
        pages = {112},
          doi = {10.1088/0004-6256/142/4/112},
archivePrefix = {arXiv},
       eprint = {1102.0342},
 primaryClass = {astro-ph.SR},
       adsurl = {https://ui.adsabs.harvard.edu/abs/2011AJ....142..112B}
}

@ARTICLE{brown22,
       author = {{Brown}, E.~L. and {Jeffers}, S.~V. and {Marsden}, S.~C. and {Morin}, J. and {Boro Saikia}, S. and {Petit}, P. and {Jardine}, M.~M. and {See}, V. and {Vidotto}, A.~A. and {Mengel}, M.~W. and {Dahlkemper}, M.~N. and {the BCool Collaboration}},
        title = "{Linking chromospheric activity and magnetic field properties for late-type dwarf stars}",
      journal = {\mnras},
         year = 2022,
        month = aug,
       volume = {514},
       number = {3},
        pages = {4300-4319},
          doi = {10.1093/mnras/stac1291},
archivePrefix = {arXiv},
       eprint = {2205.03108},
 primaryClass = {astro-ph.SR},
       adsurl = {https://ui.adsabs.harvard.edu/abs/2022MNRAS.514.4300B}
}

@INPROCEEDINGS{brown24,
       author = {{Brown}, Anthony},
        title = "{Status of Gaia Data Release 4 processing}",
    booktitle = {EAS2024, European Astronomical Society Annual Meeting},
         year = 2024,
        month = jul,
          eid = {208},
        pages = {208},
       adsurl = {https://ui.adsabs.harvard.edu/abs/2024eas..conf..208B}
}

@ARTICLE{brun17,
       author = {{Brun}, Allan Sacha and {Browning}, Matthew K.},
        title = "{Magnetism, dynamo action and the solar-stellar connection}",
      journal = {Living Reviews in Solar Physics},
         year = 2017,
        month = dec,
       volume = {14},
       number = {1},
          eid = {4},
        pages = {4},
          doi = {10.1007/s41116-017-0007-8},
       adsurl = {https://ui.adsabs.harvard.edu/abs/2017LRSP...14....4B}
}

@ARTICLE{brun22,
       author = {{Brun}, Allan Sacha and {Strugarek}, Antoine and {Noraz}, Quentin and {Perri}, Barbara and {Varela}, Jacobo and {Augustson}, Kyle and {Charbonneau}, Paul and {Toomre}, Juri},
        title = "{Powering Stellar Magnetism: Energy Transfers in Cyclic Dynamos of Sun-like Stars}",
      journal = {\apj},
         year = 2022,
        month = feb,
       volume = {926},
       number = {1},
          eid = {21},
        pages = {21},
          doi = {10.3847/1538-4357/ac469b},
archivePrefix = {arXiv},
       eprint = {2201.13218},
 primaryClass = {astro-ph.SR},
       adsurl = {https://ui.adsabs.harvard.edu/abs/2022ApJ...926...21B}
}

@ARTICLE{busa07,
       author = {{Bus{\`a}}, I. and {Aznar Cuadrado}, R. and {Terranegra}, L. and {Andretta}, V. and {Gomez}, M.~T.},
        title = "{The Ca II infrared triplet as a stellar activity diagnostic. II. Test and calibration with high resolution observations}",
      journal = {\aap},
         year = 2007,
        month = may,
       volume = {466},
       number = {3},
        pages = {1089-1098},
          doi = {10.1051/0004-6361:20065588},
       adsurl = {https://ui.adsabs.harvard.edu/abs/2007A&A...466.1089B}
}

@ARTICLE{cao22,
       author = {{Cao}, Lyra and {Pinsonneault}, Marc H.},
        title = "{Star-spots and magnetism: testing the activity paradigm in the Pleiades and M67}",
      journal = {\mnras},
         year = 2022,
        month = dec,
       volume = {517},
       number = {2},
        pages = {2165-2189},
          doi = {10.1093/mnras/stac2706},
archivePrefix = {arXiv},
       eprint = {2209.10549},
 primaryClass = {astro-ph.SR},
       adsurl = {https://ui.adsabs.harvard.edu/abs/2022MNRAS.517.2165C}
}

@ARTICLE{cao23,
       author = {{Cao}, Lyra and {Pinsonneault}, Marc H. and {van Saders}, Jennifer L.},
        title = "{Core-envelope Decoupling Drives Radial Shear Dynamos in Cool Stars}",
      journal = {\apjl},
         year = 2023,
        month = jul,
       volume = {951},
       number = {2},
          eid = {L49},
        pages = {L49},
          doi = {10.3847/2041-8213/acd780},
archivePrefix = {arXiv},
       eprint = {2301.07716},
 primaryClass = {astro-ph.SR},
       adsurl = {https://ui.adsabs.harvard.edu/abs/2023ApJ...951L..49C}
}

@ARTICLE{carvalhosilva25,
       author = {{Carvalho-Silva}, Gabriela and {Mel{\'e}ndez}, Jorge and {Rathsam}, Anne and {Shejeelammal}, J. and {Martos}, Giulia and {Lorenzo-Oliveira}, Diego and {Spina}, Lorenzo and {Ribeiro Alves}, D{\'e}bora},
        title = "{A New Age{\textendash}Activity Relation For Solar Analogs that Accounts for Metallicity}",
      journal = {\apjl},
         year = 2025,
        month = apr,
       volume = {983},
       number = {2},
          eid = {L31},
        pages = {L31},
          doi = {10.3847/2041-8213/adc382},
archivePrefix = {arXiv},
       eprint = {2504.17482},
 primaryClass = {astro-ph.SR},
       adsurl = {https://ui.adsabs.harvard.edu/abs/2025ApJ...983L..31C}
}

@ARTICLE{chahal23,
       author = {{Chahal}, Deepak and {Kamath}, Devika and {de Grijs}, Richard and {Ventura}, Paolo and {Chen}, Xiaodian},
        title = "{Unravelling the period gap using LAMOST chromospheric activity indices}",
      journal = {\mnras},
         year = 2023,
        month = nov,
       volume = {525},
       number = {3},
        pages = {4026-4041},
          doi = {10.1093/mnras/stad2521},
archivePrefix = {arXiv},
       eprint = {2308.10539},
 primaryClass = {astro-ph.SR},
       adsurl = {https://ui.adsabs.harvard.edu/abs/2023MNRAS.525.4026C}
}

@ARTICLE{charbonneau14,
       author = {{Charbonneau}, Paul},
        title = "{Solar Dynamo Theory}",
      journal = {\araa},
         year = 2014,
        month = aug,
       volume = {52},
        pages = {251-290},
          doi = {10.1146/annurev-astro-081913-040012},
       adsurl = {https://ui.adsabs.harvard.edu/abs/2014ARA&A..52..251C}
}

@ARTICLE{charbonneau20,
       author = {{Charbonneau}, Paul},
        title = "{Dynamo models of the solar cycle}",
      journal = {Living Reviews in Solar Physics},
         year = 2020,
        month = dec,
       volume = {17},
       number = {1},
          eid = {4},
        pages = {4},
          doi = {10.1007/s41116-020-00025-6},
       adsurl = {https://ui.adsabs.harvard.edu/abs/2020LRSP...17....4C}
}

@ARTICLE{claytor20,
       author = {{Claytor}, Zachary R. and {van Saders}, Jennifer L. and {Santos}, {\^A}ngela R.~G. and {Garc{\'\i}a}, Rafael A. and {Mathur}, Savita and {Tayar}, Jamie and {Pinsonneault}, Marc H. and {Shetrone}, Matthew},
        title = "{Chemical Evolution in the Milky Way: Rotation-based Ages for APOGEE-Kepler Cool Dwarf Stars}",
      journal = {\apj},
         year = 2020,
        month = jan,
       volume = {888},
       number = {1},
          eid = {43},
        pages = {43},
          doi = {10.3847/1538-4357/ab5c24},
archivePrefix = {arXiv},
       eprint = {1911.04518},
 primaryClass = {astro-ph.SR},
       adsurl = {https://ui.adsabs.harvard.edu/abs/2020ApJ...888...43C}
}

@ARTICLE{claytor25,
       author = {{Claytor}, Zachary R. and {Tayar}, Jamie},
        title = "{New Rotation Periods from the Kepler Bonus Background Light Curves}",
      journal = {\apj},
         year = 2025,
        month = jul,
       volume = {987},
       number = {1},
          eid = {8},
        pages = {8},
          doi = {10.3847/1538-4357/add5f0},
archivePrefix = {arXiv},
       eprint = {2506.03248},
 primaryClass = {astro-ph.SR},
       adsurl = {https://ui.adsabs.harvard.edu/abs/2025ApJ...987....8C}
}

@ARTICLE{corsaro21,
       author = {{Corsaro}, E. and {Bonanno}, A. and {Mathur}, S. and {Garc{\'\i}a}, R.~A. and {Santos}, A.~R.~G. and {Breton}, S.~N. and {Khalatyan}, A.},
        title = "{A calibration of the Rossby number from asteroseismology}",
      journal = {\aap},
         year = 2021,
        month = aug,
       volume = {652},
          eid = {L2},
        pages = {L2},
          doi = {10.1051/0004-6361/202141395},
archivePrefix = {arXiv},
       eprint = {2107.08551},
 primaryClass = {astro-ph.SR},
       adsurl = {https://ui.adsabs.harvard.edu/abs/2021A&A...652L...2C}
}

@ARTICLE{curtis19,
       author = {{Curtis}, Jason Lee and {Ag{\"u}eros}, Marcel A. and {Douglas}, Stephanie T. and {Meibom}, S{\o}ren},
        title = "{A Temporary Epoch of Stalled Spin-down for Low-mass Stars: Insights from NGC 6811 with Gaia and Kepler}",
      journal = {\apj},
         year = 2019,
        month = jul,
       volume = {879},
       number = {1},
          eid = {49},
        pages = {49},
          doi = {10.3847/1538-4357/ab2393},
archivePrefix = {arXiv},
       eprint = {1905.06869},
 primaryClass = {astro-ph.SR},
       adsurl = {https://ui.adsabs.harvard.edu/abs/2019ApJ...879...49C}
}

@ARTICLE{curtis20,
       author = {{Curtis}, Jason Lee and {Ag{\"u}eros}, Marcel A. and {Matt}, Sean P. and {Covey}, Kevin R. and {Douglas}, Stephanie T. and {Angus}, Ruth and {Saar}, Steven H. and {Cody}, Ann Marie and {Vanderburg}, Andrew and {Law}, Nicholas M. and {Kraus}, Adam L. and {Latham}, David W. and {Baranec}, Christoph and {Riddle}, Reed and {Ziegler}, Carl and {Lund}, Mikkel N. and {Torres}, Guillermo and {Meibom}, S{\o}ren and {Aguirre}, Victor Silva and {Wright}, Jason T.},
        title = "{When Do Stalled Stars Resume Spinning Down? Advancing Gyrochronology with Ruprecht 147}",
      journal = {\apj},
         year = 2020,
        month = dec,
       volume = {904},
       number = {2},
          eid = {140},
        pages = {140},
          doi = {10.3847/1538-4357/abbf58},
archivePrefix = {arXiv},
       eprint = {2010.02272},
 primaryClass = {astro-ph.SR},
       adsurl = {https://ui.adsabs.harvard.edu/abs/2020ApJ...904..140C}
}

@ARTICLE{davenport17,
       author = {{Davenport}, James R.~A.},
        title = "{Rotating Stars from Kepler Observed with Gaia DR1}",
      journal = {\apj},
         year = 2017,
        month = jan,
       volume = {835},
       number = {1},
          eid = {16},
        pages = {16},
          doi = {10.3847/1538-4357/835/1/16},
archivePrefix = {arXiv},
       eprint = {1610.08563},
 primaryClass = {astro-ph.SR},
       adsurl = {https://ui.adsabs.harvard.edu/abs/2017ApJ...835...16D}
}

@ARTICLE{davenport18,
       author = {{Davenport}, James R.~A. and {Covey}, Kevin R.},
        title = "{Rotating Stars from Kepler Observed with Gaia DR2}",
      journal = {\apj},
         year = 2018,
        month = dec,
       volume = {868},
       number = {2},
          eid = {151},
        pages = {151},
          doi = {10.3847/1538-4357/aae842},
archivePrefix = {arXiv},
       eprint = {1807.09841},
 primaryClass = {astro-ph.SR},
       adsurl = {https://ui.adsabs.harvard.edu/abs/2018ApJ...868..151D}
}

@ARTICLE{david22,
       author = {{David}, Trevor J. and {Angus}, Ruth and {Curtis}, Jason L. and {van Saders}, Jennifer L. and {Colman}, Isabel L. and {Contardo}, Gabriella and {Lu}, Yuxi and {Zinn}, Joel C.},
        title = "{Further Evidence of Modified Spin-down in Sun-like Stars: Pileups in the Temperature-Period Distribution}",
      journal = {\apj},
         year = 2022,
        month = jul,
       volume = {933},
       number = {1},
          eid = {114},
        pages = {114},
          doi = {10.3847/1538-4357/ac6dd3},
archivePrefix = {arXiv},
       eprint = {2203.08920},
 primaryClass = {astro-ph.SR},
       adsurl = {https://ui.adsabs.harvard.edu/abs/2022ApJ...933..114D}
}

@ARTICLE{degott25,
       author = {{Degott}, L. and {Baudin}, F. and {Samadi}, R. and {Perri}, B. and {Pin{\c{c}}on}, C.},
        title = "{Activity of low-mass stars in the light of spot signature in the Fourier domain}",
      journal = {\aap},
         year = 2025,
        month = apr,
       volume = {696},
          eid = {A41},
        pages = {A41},
          doi = {10.1051/0004-6361/202451205},
archivePrefix = {arXiv},
       eprint = {2502.15329},
 primaryClass = {astro-ph.SR},
       adsurl = {https://ui.adsabs.harvard.edu/abs/2025A&A...696A..41D}
}

@ARTICLE{demarque08,
       author = {{Demarque}, P. and {Guenther}, D.~B. and {Li}, L.~H. and {Mazumdar}, A. and {Straka}, C.~W.},
        title = "{YREC: the Yale rotating stellar evolution code. Non-rotating version, seismology applications}",
      journal = {\apss},
         year = 2008,
        month = aug,
       volume = {316},
       number = {1-4},
        pages = {31-41},
          doi = {10.1007/s10509-007-9698-y},
archivePrefix = {arXiv},
       eprint = {0710.4003},
 primaryClass = {astro-ph},
       adsurl = {https://ui.adsabs.harvard.edu/abs/2008Ap&SS.316...31D}
}

@ARTICLE{dempsey93,
       author = {{Dempsey}, Robert C. and {Bopp}, Bernard W. and {Henry}, Gregory W. and {Hall}, Douglas S.},
        title = "{Observations of the CA II Infrared Triplet in Chromospherically Active Single and Binary Stars}",
      journal = {\apjs},
         year = 1993,
        month = may,
       volume = {86},
        pages = {293},
          doi = {10.1086/191779},
       adsurl = {https://ui.adsabs.harvard.edu/abs/1993ApJS...86..293D}
}

@ARTICLE{denissenkov10,
       author = {{Denissenkov}, Pavel A. and {Pinsonneault}, Marc and {Terndrup}, Donald M. and {Newsham}, Grant},
        title = "{Angular Momentum Transport in Solar-type Stars: Testing the Timescale for Core-Envelope Coupling}",
      journal = {\apj},
         year = 2010,
        month = jun,
       volume = {716},
       number = {2},
        pages = {1269-1287},
          doi = {10.1088/0004-637X/716/2/1269},
archivePrefix = {arXiv},
       eprint = {0911.1121},
 primaryClass = {astro-ph.SR},
       adsurl = {https://ui.adsabs.harvard.edu/abs/2010ApJ...716.1269D}
}

@ARTICLE{ding24,
       author = {{Ding}, Yuedan and {Zhang}, Shidi and {Han}, Henggeng and {Cui}, Wenyuan and {Wang}, Song and {Fang}, Min and {Gao}, Yawei},
        title = "{Double-edged Sword: The Influence of Tidal Interaction on Stellar Activity in Binaries}",
      journal = {\apj},
         year = 2024,
        month = dec,
       volume = {976},
       number = {2},
          eid = {243},
        pages = {243},
          doi = {10.3847/1538-4357/ad8eb9},
archivePrefix = {arXiv},
       eprint = {2410.15039},
 primaryClass = {astro-ph.SR},
       adsurl = {https://ui.adsabs.harvard.edu/abs/2024ApJ...976..243D}
}

@ARTICLE{dixon20,
       author = {{Dixon}, Don and {Tayar}, Jamie and {Stassun}, Keivan G.},
        title = "{Rotationally Driven Ultraviolet Emission of Red Giant Stars}",
      journal = {\aj},
         year = 2020,
        month = jul,
       volume = {160},
       number = {1},
          eid = {12},
        pages = {12},
          doi = {10.3847/1538-3881/ab9080},
archivePrefix = {arXiv},
       eprint = {2005.00577},
 primaryClass = {astro-ph.SR},
       adsurl = {https://ui.adsabs.harvard.edu/abs/2020AJ....160...12D}
}

@ARTICLE{dixon25,
       author = {{Dixon}, D. and {Stassun}, K.~G. and {Mathieu}, R.~D. and {Tayar}, J. and {Cao}, L.},
        title = "{Rotationally Driven Ultraviolet Emission of Red Giant Stars. II. Metallicity, Activity, Binarity, and Subsubgiants.}",
      journal = {\aj},
         year = 2025,
        month = jan,
       volume = {169},
          eid = {309},
        pages = {309},
          doi = {10.3847/1538-3881/adc92a},
archivePrefix = {arXiv},
       eprint = {2504.05561},
 primaryClass = {astro-ph.SR},
       adsurl = {https://ui.adsabs.harvard.edu/abs/2025AJ....169..309D}
}

@ARTICLE{duchene13,
       author = {{Duch{\^e}ne}, Gaspard and {Kraus}, Adam},
        title = "{Stellar Multiplicity}",
      journal = {\araa},
         year = 2013,
        month = aug,
       volume = {51},
       number = {1},
        pages = {269-310},
          doi = {10.1146/annurev-astro-081710-102602},
archivePrefix = {arXiv},
       eprint = {1303.3028},
 primaryClass = {astro-ph.SR},
       adsurl = {https://ui.adsabs.harvard.edu/abs/2013ARA&A..51..269D}
}

@ARTICLE{dungee22,
       author = {{Dungee}, Ryan and {van Saders}, Jennifer and {Gaidos}, Eric and {Chun}, Mark and {Garc{\'\i}a}, Rafael A. and {Magnier}, Eugene A. and {Mathur}, Savita and {Santos}, {\^A}ngela R.~G.},
        title = "{A 4 Gyr M-dwarf Gyrochrone from CFHT/MegaPrime Monitoring of the Open Cluster M67}",
      journal = {\apj},
         year = 2022,
        month = oct,
       volume = {938},
       number = {2},
          eid = {118},
        pages = {118},
          doi = {10.3847/1538-4357/ac90be},
archivePrefix = {arXiv},
       eprint = {2211.01377},
 primaryClass = {astro-ph.SR},
       adsurl = {https://ui.adsabs.harvard.edu/abs/2022ApJ...938..118D}
}

@ARTICLE{duquennoy91,
       author = {{Duquennoy}, A. and {Mayor}, M.},
        title = "{Multiplicity among Solar Type Stars in the Solar Neighbourhood - Part Two - Distribution of the Orbital Elements in an Unbiased Sample}",
      journal = {\aap},
         year = 1991,
        month = aug,
       volume = {248},
        pages = {485},
       adsurl = {https://ui.adsabs.harvard.edu/abs/1991A&A...248..485D}
}

@ARTICLE{epstein14,
       author = {{Epstein}, Courtney R. and {Pinsonneault}, Marc H.},
        title = "{How Good a Clock is Rotation? The Stellar Rotation-Mass-Age Relationship for Old Field Stars}",
      journal = {\apj},
         year = 2014,
        month = jan,
       volume = {780},
       number = {2},
          eid = {159},
        pages = {159},
          doi = {10.1088/0004-637X/780/2/159},
archivePrefix = {arXiv},
       eprint = {1203.1618},
 primaryClass = {astro-ph.SR},
       adsurl = {https://ui.adsabs.harvard.edu/abs/2014ApJ...780..159E}
}

@ARTICLE{espinozarojas25,
       author = {{Espinoza-Rojas}, Francisca and {Theme{\ss}l}, Nathalie and {Hekker}, Saskia},
        title = "{Red giant asteroseismic binaries in the Kepler field: Identifying gravitationally bound systems}",
      journal = {\aap},
         year = 2025,
        month = nov,
       volume = {703},
          eid = {A66},
        pages = {A66},
          doi = {10.1051/0004-6361/202555884},
archivePrefix = {arXiv},
       eprint = {2509.13412},
 primaryClass = {astro-ph.SR},
       adsurl = {https://ui.adsabs.harvard.edu/abs/2025A&A...703A..66E}
}

@ARTICLE{freund25,
       author = {{Freund}, S. and {Czesla}, S. and {Fuhrmeister}, B. and {Predehl}, P. and {Robrade}, J. and {Schneider}, P.~C. and {Schmitt}, J.~H.~M.~M.},
        title = "{The stellar corona-chromosphere connection: A comprehensive study of X-ray and Ca II IRT fluxes from eROSITA and Gaia}",
      journal = {\aap},
         year = 2025,
        month = may,
       volume = {697},
          eid = {A230},
        pages = {A230},
          doi = {10.1051/0004-6361/202451421},
archivePrefix = {arXiv},
       eprint = {2504.17593},
 primaryClass = {astro-ph.SR},
       adsurl = {https://ui.adsabs.harvard.edu/abs/2025A&A...697A.230F}
}

@ARTICLE{fritzewski21,
       author = {{Fritzewski}, D.~J. and {Barnes}, S.~A. and {James}, D.~J. and {J{\"a}rvinen}, S.~P. and {Strassmeier}, K.~G.},
        title = "{A detailed understanding of the rotation-activity relationship using the 300 Myr old open cluster NGC 3532}",
      journal = {\aap},
         year = 2021,
        month = dec,
       volume = {656},
          eid = {A103},
        pages = {A103},
          doi = {10.1051/0004-6361/202140896},
archivePrefix = {arXiv},
       eprint = {2112.03302},
 primaryClass = {astro-ph.SR},
       adsurl = {https://ui.adsabs.harvard.edu/abs/2021A&A...656A.103F}
}

@ARTICLE{furlan18,
       author = {{Furlan}, E. and {Ciardi}, D.~R. and {Cochran}, W.~D. and {Everett}, M.~E. and {Latham}, D.~W. and {Marcy}, G.~W. and {Buchhave}, L.~A. and {Endl}, M. and {Isaacson}, H. and {Petigura}, E.~A. and {Gautier}, III, T.~N. and {Huber}, D. and {Bieryla}, A. and {Borucki}, W.~J. and {Brugamyer}, E. and {Caldwell}, C. and {Cochran}, A. and {Howard}, A.~W. and {Howell}, S.~B. and {Johnson}, M.~C. and {MacQueen}, P.~J. and {Quinn}, S.~N. and {Robertson}, P. and {Mathur}, S. and {Batalha}, N.~M.},
        title = "{The Kepler Follow-up Observation Program. II. Stellar Parameters from Medium- and High-resolution Spectroscopy}",
      journal = {\apj},
         year = 2018,
        month = jul,
       volume = {861},
       number = {2},
          eid = {149},
        pages = {149},
          doi = {10.3847/1538-4357/aaca34},
archivePrefix = {arXiv},
       eprint = {1805.12089},
 primaryClass = {astro-ph.SR},
       adsurl = {https://ui.adsabs.harvard.edu/abs/2018ApJ...861..149F}
}

@ARTICLE{gaia16,
       author = {{Gaia Collaboration} and {Prusti}, T. and {de Bruijne}, J.~H.~J. and {Brown}, A.~G.~A. and {Vallenari}, A. and {Babusiaux}, C. and {Bailer-Jones}, C.~A.~L. and {Bastian}, U. and {Biermann}, M. and {Evans}, D.~W. and {Eyer}, L. and {Jansen}, F. and {Jordi}, C. and {Klioner}, S.~A. and {Lammers}, U. and {Lindegren}, L. and {Luri}, X. and {Mignard}, F. and {Milligan}, D.~J. and {Panem}, C. and {Poinsignon}, V. and {Pourbaix}, D. and {Randich}, S. and {Sarri}, G. and {Sartoretti}, P. and {Siddiqui}, H.~I. and {Soubiran}, C. and {Valette}, V. and {van Leeuwen}, F. and {Walton}, N.~A. and {Aerts}, C. and {Arenou}, F. and {Cropper}, M. and {Drimmel}, R. and {H{\o}g}, E. and {Katz}, D. and {Lattanzi}, M.~G. and {O'Mullane}, W. and {Grebel}, E.~K. and {Holland}, A.~D. and {Huc}, C. and {Passot}, X. and {Bramante}, L. and {Cacciari}, C. and {Casta{\~n}eda}, J. and {Chaoul}, L. and {Cheek}, N. and {De Angeli}, F. and {Fabricius}, C. and {Guerra}, R. and {Hern{\'a}ndez}, J. and {Jean-Antoine-Piccolo}, A. and {Masana}, E. and {Messineo}, R. and {Mowlavi}, N. and {Nienartowicz}, K. and {Ord{\'o}{\~n}ez-Blanco}, D. and {Panuzzo}, P. and {Portell}, J. and {Richards}, P.~J. and {Riello}, M. and {Seabroke}, G.~M. and {Tanga}, P. and {Th{\'e}venin}, F. and {Torra}, J. and {Els}, S.~G. and {Gracia-Abril}, G. and {Comoretto}, G. and {Garcia-Reinaldos}, M. and {Lock}, T. and {Mercier}, E. and {Altmann}, M. and {Andrae}, R. and {Astraatmadja}, T.~L. and {Bellas-Velidis}, I. and {Benson}, K. and {Berthier}, J. and {Blomme}, R. and {Busso}, G. and {Carry}, B. and {Cellino}, A. and {Clementini}, G. and {Cowell}, S. and {Creevey}, O. and {Cuypers}, J. and {Davidson}, M. and {De Ridder}, J. and {de Torres}, A. and {Delchambre}, L. and {Dell'Oro}, A. and {Ducourant}, C. and {Fr{\'e}mat}, Y. and {Garc{\'\i}a-Torres}, M. and {Gosset}, E. and {Halbwachs}, J.-L. and {Hambly}, N.~C. and {Harrison}, D.~L. and {Hauser}, M. and {Hestroffer}, D. and {Hodgkin}, S.~T. and {Huckle}, H.~E. and {Hutton}, A. and {Jasniewicz}, G. and {Jordan}, S. and {Kontizas}, M. and {Korn}, A.~J. and {Lanzafame}, A.~C. and {Manteiga}, M. and {Moitinho}, A. and {Muinonen}, K. and {Osinde}, J. and {Pancino}, E. and {Pauwels}, T. and {Petit}, J.-M. and {Recio-Blanco}, A. and {Robin}, A.~C. and {Sarro}, L.~M. and {Siopis}, C. and {Smith}, M. and {Smith}, K.~W. and {Sozzetti}, A. and {Thuillot}, W. and {van Reeven}, W. and {Viala}, Y. and {Abbas}, U. and {Abreu Aramburu}, A. and {Accart}, S. and {Aguado}, J.~J. and {Allan}, P.~M. and {Allasia}, W. and {Altavilla}, G. and {{\'A}lvarez}, M.~A. and {Alves}, J. and {Anderson}, R.~I. and {Andrei}, A.~H. and {Anglada Varela}, E. and {Antiche}, E. and {Antoja}, T. and {Ant{\'o}n}, S. and {Arcay}, B. and {Atzei}, A. and {Ayache}, L. and {Bach}, N. and {Baker}, S.~G. and {Balaguer-N{\'u}{\~n}ez}, L. and {Barache}, C. and {Barata}, C. and {Barbier}, A. and {Barblan}, F. and {Baroni}, M. and {Barrado y Navascu{\'e}s}, D. and {Barros}, M. and {Barstow}, M.~A. and {Becciani}, U. and {Bellazzini}, M. and {Bellei}, G. and {Bello Garc{\'\i}a}, A. and {Belokurov}, V. and {Bendjoya}, P. and {Berihuete}, A. and {Bianchi}, L. and {Bienaym{\'e}}, O. and {Billebaud}, F. and {Blagorodnova}, N. and {Blanco-Cuaresma}, S. and {Boch}, T. and {Bombrun}, A. and {Borrachero}, R. and {Bouquillon}, S. and {Bourda}, G. and {Bouy}, H. and {Bragaglia}, A. and {Breddels}, M.~A. and {Brouillet}, N. and {Br{\"u}semeister}, T. and {Bucciarelli}, B. and {Budnik}, F. and {Burgess}, P. and {Burgon}, R. and {Burlacu}, A. and {Busonero}, D. and {Buzzi}, R. and {Caffau}, E. and {Cambras}, J. and {Campbell}, H. and {Cancelliere}, R. and {Cantat-Gaudin}, T. and {Carlucci}, T. and {Carrasco}, J.~M. and {Castellani}, M. and {Charlot}, P. and {Charnas}, J. and {Charvet}, P. and {Chassat}, F. and {Chiavassa}, A. and {Clotet}, M. and {Cocozza}, G. and {Collins}, R.~S. and {Collins}, P. and {Costigan}, G.},
        title = "{The Gaia mission}",
      journal = {\aap},
         year = 2016,
        month = nov,
       volume = {595},
          eid = {A1},
        pages = {A1},
          doi = {10.1051/0004-6361/201629272},
archivePrefix = {arXiv},
       eprint = {1609.04153},
 primaryClass = {astro-ph.IM},
       adsurl = {https://ui.adsabs.harvard.edu/abs/2016A&A...595A...1G}
}

@ARTICLE{gaia21,
       author = {{Gaia Collaboration} and {Brown}, A.~G.~A. and {Vallenari}, A. and {Prusti}, T. and {de Bruijne}, J.~H.~J. and {Babusiaux}, C. and {Biermann}, M. and {Creevey}, O.~L. and {Evans}, D.~W. and {Eyer}, L. and {Hutton}, A. and {Jansen}, F. and {Jordi}, C. and {Klioner}, S.~A. and {Lammers}, U. and {Lindegren}, L. and {Luri}, X. and {Mignard}, F. and {Panem}, C. and {Pourbaix}, D. and {Randich}, S. and {Sartoretti}, P. and {Soubiran}, C. and {Walton}, N.~A. and {Arenou}, F. and {Bailer-Jones}, C.~A.~L. and {Bastian}, U. and {Cropper}, M. and {Drimmel}, R. and {Katz}, D. and {Lattanzi}, M.~G. and {van Leeuwen}, F. and {Bakker}, J. and {Cacciari}, C. and {Casta{\~n}eda}, J. and {De Angeli}, F. and {Ducourant}, C. and {Fabricius}, C. and {Fouesneau}, M. and {Fr{\'e}mat}, Y. and {Guerra}, R. and {Guerrier}, A. and {Guiraud}, J. and {Jean-Antoine Piccolo}, A. and {Masana}, E. and {Messineo}, R. and {Mowlavi}, N. and {Nicolas}, C. and {Nienartowicz}, K. and {Pailler}, F. and {Panuzzo}, P. and {Riclet}, F. and {Roux}, W. and {Seabroke}, G.~M. and {Sordo}, R. and {Tanga}, P. and {Th{\'e}venin}, F. and {Gracia-Abril}, G. and {Portell}, J. and {Teyssier}, D. and {Altmann}, M. and {Andrae}, R. and {Bellas-Velidis}, I. and {Benson}, K. and {Berthier}, J. and {Blomme}, R. and {Brugaletta}, E. and {Burgess}, P.~W. and {Busso}, G. and {Carry}, B. and {Cellino}, A. and {Cheek}, N. and {Clementini}, G. and {Damerdji}, Y. and {Davidson}, M. and {Delchambre}, L. and {Dell'Oro}, A. and {Fern{\'a}ndez-Hern{\'a}ndez}, J. and {Galluccio}, L. and {Garc{\'\i}a-Lario}, P. and {Garcia-Reinaldos}, M. and {Gonz{\'a}lez-N{\'u}{\~n}ez}, J. and {Gosset}, E. and {Haigron}, R. and {Halbwachs}, J.-L. and {Hambly}, N.~C. and {Harrison}, D.~L. and {Hatzidimitriou}, D. and {Heiter}, U. and {Hern{\'a}ndez}, J. and {Hestroffer}, D. and {Hodgkin}, S.~T. and {Holl}, B. and {Jan{\ss}en}, K. and {Jevardat de Fombelle}, G. and {Jordan}, S. and {Krone-Martins}, A. and {Lanzafame}, A.~C. and {L{\"o}ffler}, W. and {Lorca}, A. and {Manteiga}, M. and {Marchal}, O. and {Marrese}, P.~M. and {Moitinho}, A. and {Mora}, A. and {Muinonen}, K. and {Osborne}, P. and {Pancino}, E. and {Pauwels}, T. and {Petit}, J.-M. and {Recio-Blanco}, A. and {Richards}, P.~J. and {Riello}, M. and {Rimoldini}, L. and {Robin}, A.~C. and {Roegiers}, T. and {Rybizki}, J. and {Sarro}, L.~M. and {Siopis}, C. and {Smith}, M. and {Sozzetti}, A. and {Ulla}, A. and {Utrilla}, E. and {van Leeuwen}, M. and {van Reeven}, W. and {Abbas}, U. and {Abreu Aramburu}, A. and {Accart}, S. and {Aerts}, C. and {Aguado}, J.~J. and {Ajaj}, M. and {Altavilla}, G. and {{\'A}lvarez}, M.~A. and {{\'A}lvarez Cid-Fuentes}, J. and {Alves}, J. and {Anderson}, R.~I. and {Anglada Varela}, E. and {Antoja}, T. and {Audard}, M. and {Baines}, D. and {Baker}, S.~G. and {Balaguer-N{\'u}{\~n}ez}, L. and {Balbinot}, E. and {Balog}, Z. and {Barache}, C. and {Barbato}, D. and {Barros}, M. and {Barstow}, M.~A. and {Bartolom{\'e}}, S. and {Bassilana}, J.-L. and {Bauchet}, N. and {Baudesson-Stella}, A. and {Becciani}, U. and {Bellazzini}, M. and {Bernet}, M. and {Bertone}, S. and {Bianchi}, L. and {Blanco-Cuaresma}, S. and {Boch}, T. and {Bombrun}, A. and {Bossini}, D. and {Bouquillon}, S. and {Bragaglia}, A. and {Bramante}, L. and {Breedt}, E. and {Bressan}, A. and {Brouillet}, N. and {Bucciarelli}, B. and {Burlacu}, A. and {Busonero}, D. and {Butkevich}, A.~G. and {Buzzi}, R. and {Caffau}, E. and {Cancelliere}, R. and {C{\'a}novas}, H. and {Cantat-Gaudin}, T. and {Carballo}, R. and {Carlucci}, T. and {Carnerero}, M.~I. and {Carrasco}, J.~M. and {Casamiquela}, L. and {Castellani}, M. and {Castro-Ginard}, A. and {Castro Sampol}, P. and {Chaoul}, L. and {Charlot}, P. and {Chemin}, L. and {Chiavassa}, A. and {Cioni}, M.-R.~L. and {Comoretto}, G. and {Cooper}, W.~J. and {Cornez}, T. and {Cowell}, S. and {Crifo}, F. and {Crosta}, M. and {Crowley}, C. and {Dafonte}, C. and {Dapergolas}, A. and {David}, M. and {David}, P.},
        title = "{Gaia Early Data Release 3. Summary of the contents and survey properties}",
      journal = {\aap},
         year = 2021,
        month = may,
       volume = {649},
          eid = {A1},
        pages = {A1},
          doi = {10.1051/0004-6361/202039657},
archivePrefix = {arXiv},
       eprint = {2012.01533},
 primaryClass = {astro-ph.GA},
       adsurl = {https://ui.adsabs.harvard.edu/abs/2021A&A...649A...1G}
}

@ARTICLE{gaia23a,
       author = {{Gaia Collaboration} and {Vallenari}, A. and {Brown}, A.~G.~A. and {Prusti}, T. and {de Bruijne}, J.~H.~J. and {Arenou}, F. and {Babusiaux}, C. and {Biermann}, M. and {Creevey}, O.~L. and {Ducourant}, C. and {Evans}, D.~W. and {Eyer}, L. and {Guerra}, R. and {Hutton}, A. and {Jordi}, C. and {Klioner}, S.~A. and {Lammers}, U.~L. and {Lindegren}, L. and {Luri}, X. and {Mignard}, F. and {Panem}, C. and {Pourbaix}, D. and {Randich}, S. and {Sartoretti}, P. and {Soubiran}, C. and {Tanga}, P. and {Walton}, N.~A. and {Bailer-Jones}, C.~A.~L. and {Bastian}, U. and {Drimmel}, R. and {Jansen}, F. and {Katz}, D. and {Lattanzi}, M.~G. and {van Leeuwen}, F. and {Bakker}, J. and {Cacciari}, C. and {Casta{\~n}eda}, J. and {De Angeli}, F. and {Fabricius}, C. and {Fouesneau}, M. and {Fr{\'e}mat}, Y. and {Galluccio}, L. and {Guerrier}, A. and {Heiter}, U. and {Masana}, E. and {Messineo}, R. and {Mowlavi}, N. and {Nicolas}, C. and {Nienartowicz}, K. and {Pailler}, F. and {Panuzzo}, P. and {Riclet}, F. and {Roux}, W. and {Seabroke}, G.~M. and {Sordo}, R. and {Th{\'e}venin}, F. and {Gracia-Abril}, G. and {Portell}, J. and {Teyssier}, D. and {Altmann}, M. and {Andrae}, R. and {Audard}, M. and {Bellas-Velidis}, I. and {Benson}, K. and {Berthier}, J. and {Blomme}, R. and {Burgess}, P.~W. and {Busonero}, D. and {Busso}, G. and {C{\'a}novas}, H. and {Carry}, B. and {Cellino}, A. and {Cheek}, N. and {Clementini}, G. and {Damerdji}, Y. and {Davidson}, M. and {de Teodoro}, P. and {Nu{\~n}ez Campos}, M. and {Delchambre}, L. and {Dell'Oro}, A. and {Esquej}, P. and {Fern{\'a}ndez-Hern{\'a}ndez}, J. and {Fraile}, E. and {Garabato}, D. and {Garc{\'\i}a-Lario}, P. and {Gosset}, E. and {Haigron}, R. and {Halbwachs}, J.-L. and {Hambly}, N.~C. and {Harrison}, D.~L. and {Hern{\'a}ndez}, J. and {Hestroffer}, D. and {Hodgkin}, S.~T. and {Holl}, B. and {Jan{\ss}en}, K. and {Jevardat de Fombelle}, G. and {Jordan}, S. and {Krone-Martins}, A. and {Lanzafame}, A.~C. and {L{\"o}ffler}, W. and {Marchal}, O. and {Marrese}, P.~M. and {Moitinho}, A. and {Muinonen}, K. and {Osborne}, P. and {Pancino}, E. and {Pauwels}, T. and {Recio-Blanco}, A. and {Reyl{\'e}}, C. and {Riello}, M. and {Rimoldini}, L. and {Roegiers}, T. and {Rybizki}, J. and {Sarro}, L.~M. and {Siopis}, C. and {Smith}, M. and {Sozzetti}, A. and {Utrilla}, E. and {van Leeuwen}, M. and {Abbas}, U. and {{\'A}brah{\'a}m}, P. and {Abreu Aramburu}, A. and {Aerts}, C. and {Aguado}, J.~J. and {Ajaj}, M. and {Aldea-Montero}, F. and {Altavilla}, G. and {{\'A}lvarez}, M.~A. and {Alves}, J. and {Anders}, F. and {Anderson}, R.~I. and {Anglada Varela}, E. and {Antoja}, T. and {Baines}, D. and {Baker}, S.~G. and {Balaguer-N{\'u}{\~n}ez}, L. and {Balbinot}, E. and {Balog}, Z. and {Barache}, C. and {Barbato}, D. and {Barros}, M. and {Barstow}, M.~A. and {Bartolom{\'e}}, S. and {Bassilana}, J.-L. and {Bauchet}, N. and {Becciani}, U. and {Bellazzini}, M. and {Berihuete}, A. and {Bernet}, M. and {Bertone}, S. and {Bianchi}, L. and {Binnenfeld}, A. and {Blanco-Cuaresma}, S. and {Blazere}, A. and {Boch}, T. and {Bombrun}, A. and {Bossini}, D. and {Bouquillon}, S. and {Bragaglia}, A. and {Bramante}, L. and {Breedt}, E. and {Bressan}, A. and {Brouillet}, N. and {Brugaletta}, E. and {Bucciarelli}, B. and {Burlacu}, A. and {Butkevich}, A.~G. and {Buzzi}, R. and {Caffau}, E. and {Cancelliere}, R. and {Cantat-Gaudin}, T. and {Carballo}, R. and {Carlucci}, T. and {Carnerero}, M.~I. and {Carrasco}, J.~M. and {Casamiquela}, L. and {Castellani}, M. and {Castro-Ginard}, A. and {Chaoul}, L. and {Charlot}, P. and {Chemin}, L. and {Chiaramida}, V. and {Chiavassa}, A. and {Chornay}, N. and {Comoretto}, G. and {Contursi}, G. and {Cooper}, W.~J. and {Cornez}, T. and {Cowell}, S. and {Crifo}, F. and {Cropper}, M. and {Crosta}, M. and {Crowley}, C. and {Dafonte}, C. and {Dapergolas}, A. and {David}, M. and {David}, P. and {de Laverny}, P. and {De Luise}, F. and {De March}, R.},
        title = "{Gaia Data Release 3. Summary of the content and survey properties}",
      journal = {\aap},
         year = 2023,
        month = jun,
       volume = {674},
          eid = {A1},
        pages = {A1},
          doi = {10.1051/0004-6361/202243940},
archivePrefix = {arXiv},
       eprint = {2208.00211},
 primaryClass = {astro-ph.GA},
       adsurl = {https://ui.adsabs.harvard.edu/abs/2023A&A...674A...1G}
}

@ARTICLE{gaia23b,
       author = {{Gaia Collaboration} and {Arenou}, F. and {Babusiaux}, C. and {Barstow}, M.~A. and {Faigler}, S. and {Jorissen}, A. and {Kervella}, P. and {Mazeh}, T. and {Mowlavi}, N. and {Panuzzo}, P. and {Sahlmann}, J. and {Shahaf}, S. and {Sozzetti}, A. and {Bauchet}, N. and {Damerdji}, Y. and {Gavras}, P. and {Giacobbe}, P. and {Gosset}, E. and {Halbwachs}, J.-L. and {Holl}, B. and {Lattanzi}, M.~G. and {Leclerc}, N. and {Morel}, T. and {Pourbaix}, D. and {Re Fiorentin}, P. and {Sadowski}, G. and {S{\'e}gransan}, D. and {Siopis}, C. and {Teyssier}, D. and {Zwitter}, T. and {Planquart}, L. and {Brown}, A.~G.~A. and {Vallenari}, A. and {Prusti}, T. and {de Bruijne}, J.~H.~J. and {Biermann}, M. and {Creevey}, O.~L. and {Ducourant}, C. and {Evans}, D.~W. and {Eyer}, L. and {Guerra}, R. and {Hutton}, A. and {Jordi}, C. and {Klioner}, S.~A. and {Lammers}, U.~L. and {Lindegren}, L. and {Luri}, X. and {Mignard}, F. and {Panem}, C. and {Randich}, S. and {Sartoretti}, P. and {Soubiran}, C. and {Tanga}, P. and {Walton}, N.~A. and {Bailer-Jones}, C.~A.~L. and {Bastian}, U. and {Drimmel}, R. and {Jansen}, F. and {Katz}, D. and {van Leeuwen}, F. and {Bakker}, J. and {Cacciari}, C. and {Casta{\~n}eda}, J. and {De Angeli}, F. and {Fabricius}, C. and {Fouesneau}, M. and {Fr{\'e}mat}, Y. and {Galluccio}, L. and {Guerrier}, A. and {Heiter}, U. and {Masana}, E. and {Messineo}, R. and {Nicolas}, C. and {Nienartowicz}, K. and {Pailler}, F. and {Riclet}, F. and {Roux}, W. and {Seabroke}, G.~M. and {Sordo}, R. and {Th{\'e}venin}, F. and {Gracia-Abril}, G. and {Portell}, J. and {Altmann}, M. and {Andrae}, R. and {Audard}, M. and {Bellas-Velidis}, I. and {Benson}, K. and {Berthier}, J. and {Blomme}, R. and {Burgess}, P.~W. and {Busonero}, D. and {Busso}, G. and {C{\'a}novas}, H. and {Carry}, B. and {Cellino}, A. and {Cheek}, N. and {Clementini}, G. and {Davidson}, M. and {de Teodoro}, P. and {Nu{\~n}ez Campos}, M. and {Delchambre}, L. and {Dell'Oro}, A. and {Esquej}, P. and {Fern{\'a}ndez-Hern{\'a}ndez}, J. and {Fraile}, E. and {Garabato}, D. and {Garc{\'\i}a-Lario}, P. and {Haigron}, R. and {Hambly}, N.~C. and {Harrison}, D.~L. and {Hern{\'a}ndez}, J. and {Hestroffer}, D. and {Hodgkin}, S.~T. and {Jan{\ss}en}, K. and {Jevardat de Fombelle}, G. and {Jordan}, S. and {Krone-Martins}, A. and {Lanzafame}, A.~C. and {L{\"o}ffler}, W. and {Marchal}, O. and {Marrese}, P.~M. and {Moitinho}, A. and {Muinonen}, K. and {Osborne}, P. and {Pancino}, E. and {Pauwels}, T. and {Recio-Blanco}, A. and {Reyl{\'e}}, C. and {Riello}, M. and {Rimoldini}, L. and {Roegiers}, T. and {Rybizki}, J. and {Sarro}, L.~M. and {Smith}, M. and {Utrilla}, E. and {van Leeuwen}, M. and {Abbas}, U. and {{\'A}brah{\'a}m}, P. and {Abreu Aramburu}, A. and {Aerts}, C. and {Aguado}, J.~J. and {Ajaj}, M. and {Aldea-Montero}, F. and {Altavilla}, G. and {{\'A}lvarez}, M.~A. and {Alves}, J. and {Anders}, F. and {Anderson}, R.~I. and {Anglada Varela}, E. and {Antoja}, T. and {Baines}, D. and {Baker}, S.~G. and {Balaguer-N{\'u}{\~n}ez}, L. and {Balbinot}, E. and {Balog}, Z. and {Barache}, C. and {Barbato}, D. and {Barros}, M. and {Bartolom{\'e}}, S. and {Bassilana}, J.-L. and {Becciani}, U. and {Bellazzini}, M. and {Berihuete}, A. and {Bernet}, M. and {Bertone}, S. and {Bianchi}, L. and {Binnenfeld}, A. and {Blanco-Cuaresma}, S. and {Blazere}, A. and {Boch}, T. and {Bombrun}, A. and {Bossini}, D. and {Bouquillon}, S. and {Bragaglia}, A. and {Bramante}, L. and {Breedt}, E. and {Bressan}, A. and {Brouillet}, N. and {Brugaletta}, E. and {Bucciarelli}, B. and {Burlacu}, A. and {Butkevich}, A.~G. and {Buzzi}, R. and {Caffau}, E. and {Cancelliere}, R. and {Cantat-Gaudin}, T. and {Carballo}, R. and {Carlucci}, T. and {Carnerero}, M.~I. and {Carrasco}, J.~M. and {Casamiquela}, L. and {Castellani}, M. and {Castro-Ginard}, A. and {Chaoul}, L. and {Charlot}, P. and {Chemin}, L. and {Chiaramida}, V. and {Chiavassa}, A. and {Chornay}, N. and {Comoretto}, G.},
        title = "{Gaia Data Release 3. Stellar multiplicity, a teaser for the hidden treasure}",
      journal = {\aap},
         year = 2023,
        month = jun,
       volume = {674},
          eid = {A34},
        pages = {A34},
          doi = {10.1051/0004-6361/202243782},
archivePrefix = {arXiv},
       eprint = {2206.05595},
 primaryClass = {astro-ph.SR},
       adsurl = {https://ui.adsabs.harvard.edu/abs/2023A&A...674A..34G}
}

@ARTICLE{gallet13,
       author = {{Gallet}, F. and {Bouvier}, J.},
        title = "{Improved angular momentum evolution model for solar-like stars}",
      journal = {\aap},
         year = 2013,
        month = aug,
       volume = {556},
          eid = {A36},
        pages = {A36},
          doi = {10.1051/0004-6361/201321302},
archivePrefix = {arXiv},
       eprint = {1306.2130},
 primaryClass = {astro-ph.SR},
       adsurl = {https://ui.adsabs.harvard.edu/abs/2013A&A...556A..36G}
}

@ARTICLE{gallet15,
       author = {{Gallet}, F. and {Bouvier}, J.},
        title = "{Improved angular momentum evolution model for solar-like stars. II. Exploring the mass dependence}",
      journal = {\aap},
         year = 2015,
        month = may,
       volume = {577},
          eid = {A98},
        pages = {A98},
          doi = {10.1051/0004-6361/201525660},
archivePrefix = {arXiv},
       eprint = {1502.05801},
 primaryClass = {astro-ph.SR},
       adsurl = {https://ui.adsabs.harvard.edu/abs/2015A&A...577A..98G}
}

@ARTICLE{garcia11,
       author = {{Garc{\'\i}a}, R.~A. and {Hekker}, S. and {Stello}, D. and {Guti{\'e}rrez-Soto}, J. and {Handberg}, R. and {Huber}, D. and {Karoff}, C. and {Uytterhoeven}, K. and {Appourchaux}, T. and {Chaplin}, W.~J. and {Elsworth}, Y. and {Mathur}, S. and {Ballot}, J. and {Christensen-Dalsgaard}, J. and {Gilliland}, R.~L. and {Houdek}, G. and {Jenkins}, J.~M. and {Kjeldsen}, H. and {McCauliff}, S. and {Metcalfe}, T. and {Middour}, C.~K. and {Molenda-Zakowicz}, J. and {Monteiro}, M.~J.~P.~F.~G. and {Smith}, J.~C. and {Thompson}, M.~J.},
        title = "{Preparation of Kepler light curves for asteroseismic analyses}",
      journal = {\mnras},
         year = 2011,
        month = jun,
       volume = {414},
       number = {1},
        pages = {L6-L10},
          doi = {10.1111/j.1745-3933.2011.01042.x},
archivePrefix = {arXiv},
       eprint = {1103.0382},
 primaryClass = {astro-ph.SR},
       adsurl = {https://ui.adsabs.harvard.edu/abs/2011MNRAS.414L...6G}
}

@ARTICLE{garcia14a,
       author = {{Garc{\'\i}a}, R.~A. and {Ceillier}, T. and {Salabert}, D. and {Mathur}, S. and {van Saders}, J.~L. and {Pinsonneault}, M. and {Ballot}, J. and {Beck}, P.~G. and {Bloemen}, S. and {Campante}, T.~L. and {Davies}, G.~R. and {do Nascimento}, Jr., J.-D. and {Mathis}, S. and {Metcalfe}, T.~S. and {Nielsen}, M.~B. and {Su{\'a}rez}, J.~C. and {Chaplin}, W.~J. and {Jim{\'e}nez}, A. and {Karoff}, C.},
        title = "{Rotation and magnetism of Kepler pulsating solar-like stars. Towards asteroseismically calibrated age-rotation relations}",
      journal = {\aap},
         year = 2014,
        month = dec,
       volume = {572},
          eid = {A34},
        pages = {A34},
          doi = {10.1051/0004-6361/201423888},
archivePrefix = {arXiv},
       eprint = {1403.7155},
 primaryClass = {astro-ph.SR},
       adsurl = {https://ui.adsabs.harvard.edu/abs/2014A&A...572A..34G}
}

@ARTICLE{garcia14b,
       author = {{Garc{\'\i}a}, R.~A. and {Mathur}, S. and {Pires}, S. and {R{\'e}gulo}, C. and {Bellamy}, B. and {Pall{\'e}}, P.~L. and {Ballot}, J. and {Barcel{\'o} Forteza}, S. and {Beck}, P.~G. and {Bedding}, T.~R. and {Ceillier}, T. and {Roca Cort{\'e}s}, T. and {Salabert}, D. and {Stello}, D.},
        title = "{Impact on asteroseismic analyses of regular gaps in Kepler data}",
      journal = {\aap},
         year = 2014,
        month = aug,
       volume = {568},
          eid = {A10},
        pages = {A10},
          doi = {10.1051/0004-6361/201323326},
archivePrefix = {arXiv},
       eprint = {1405.5374},
 primaryClass = {astro-ph.SR},
       adsurl = {https://ui.adsabs.harvard.edu/abs/2014A&A...568A..10G}
}

@ARTICLE{garraffo18,
       author = {{Garraffo}, C. and {Drake}, J.~J. and {Dotter}, A. and {Choi}, J. and {Burke}, D.~J. and {Moschou}, S.~P. and {Alvarado-G{\'o}mez}, J.~D. and {Kashyap}, V.~L. and {Cohen}, O.},
        title = "{The Revolution Revolution: Magnetic Morphology Driven Spin-down}",
      journal = {\apj},
         year = 2018,
        month = jul,
       volume = {862},
       number = {1},
          eid = {90},
        pages = {90},
          doi = {10.3847/1538-4357/aace5d},
archivePrefix = {arXiv},
       eprint = {1804.01986},
 primaryClass = {astro-ph.SR},
       adsurl = {https://ui.adsabs.harvard.edu/abs/2018ApJ...862...90G}
}

@ARTICLE{gaulme14,
       author = {{Gaulme}, P. and {Jackiewicz}, J. and {Appourchaux}, T. and {Mosser}, B.},
        title = "{Surface Activity and Oscillation Amplitudes of Red Giants in Eclipsing Binaries}",
      journal = {\apj},
         year = 2014,
        month = apr,
       volume = {785},
       number = {1},
          eid = {5},
        pages = {5},
          doi = {10.1088/0004-637X/785/1/5},
archivePrefix = {arXiv},
       eprint = {1402.3027},
 primaryClass = {astro-ph.SR},
       adsurl = {https://ui.adsabs.harvard.edu/abs/2014ApJ...785....5G}
}

@ARTICLE{gaulme20,
       author = {{Gaulme}, Patrick and {Jackiewicz}, Jason and {Spada}, Federico and {Chojnowski}, Drew and {Mosser}, Beno{\^\i}t and {McKeever}, Jean and {Hedlund}, Anne and {Vrard}, Mathieu and {Benbakoura}, Mansour and {Damiani}, Cilia},
        title = "{Active red giants: Close binaries versus single rapid rotators}",
      journal = {\aap},
         year = 2020,
        month = jul,
       volume = {639},
          eid = {A63},
        pages = {A63},
          doi = {10.1051/0004-6361/202037781},
archivePrefix = {arXiv},
       eprint = {2004.13792},
 primaryClass = {astro-ph.SR},
       adsurl = {https://ui.adsabs.harvard.edu/abs/2020A&A...639A..63G}
}

@ARTICLE{ge24,
       author = {{Ge}, Jian and {Chen}, Wen and {Chen}, Yonghe and {Song}, Zongxi and {Wang}, Jian and {Zhang}, Hui and {Li}, Yan and {Zang}, Weicheng and {Zhou}, Dan and {Zhang}, Yongshuai and {Chen}, Kun and {Yang}, Yingquan and {Mao}, Shude and {Huang}, Chelsea and {Yao}, Xinyu and {Li}, Xinglong and {Jiang}, Haijiao and {Yu}, Yong and {Tang}, Zhenghong and {Dong}, Feng and {Gao}, Wei and {Zhang}, Hongfei and {Shen}, Chao and {Wang}, Fengtao and {Wei}, Chuanxin and {Yang}, Baoyu and {Li}, Yudong and {Wen}, Lin and {Zhang}, Pengjun and {Zhang}, Congcong and {Xie}, Jiwei and {Ma}, Bo and {Deng}, Hongping and {Liu}, Huigen and {Duan}, Xuliang and {Wang}, Haoyu and {Huang}, Jiangjiang and {Gao}, Yang and {Wang}, Yifei and {Wang}, Lei and {Qin}, Genjian and {Liu}, Xinyu and {Gao}, Jie},
        title = "{Search for a Second Earth - the Earth 2.0 (ET) Space Mission}",
      journal = {Chinese Journal of Space Science},
         year = 2024,
        month = may,
       volume = {44},
       number = {3},
        pages = {400-424},
          doi = {10.11728/cjss2024.03.yg05},
       adsurl = {https://ui.adsabs.harvard.edu/abs/2024ChJSS..44..400G}
}

@ARTICLE{gehan22,
       author = {{Gehan}, Charlotte and {Gaulme}, Patrick and {Yu}, Jie},
        title = "{Surface magnetism of rapidly rotating red giants: Single versus close binary stars}",
      journal = {\aap},
         year = 2022,
        month = dec,
       volume = {668},
          eid = {A116},
        pages = {A116},
          doi = {10.1051/0004-6361/202245083},
archivePrefix = {arXiv},
       eprint = {2211.01026},
 primaryClass = {astro-ph.SR},
       adsurl = {https://ui.adsabs.harvard.edu/abs/2022A&A...668A.116G}
}

@ARTICLE{gehan24,
       author = {{Gehan}, C. and {Godoy-Rivera}, D. and {Gaulme}, P.},
        title = "{Magnetic activity of red giants: Correlation between the amplitude of solar-like oscillations and chromospheric indicators}",
      journal = {\aap},
         year = 2024,
        month = jun,
       volume = {686},
          eid = {A93},
        pages = {A93},
          doi = {10.1051/0004-6361/202349008},
archivePrefix = {arXiv},
       eprint = {2401.13549},
 primaryClass = {astro-ph.SR},
       adsurl = {https://ui.adsabs.harvard.edu/abs/2024A&A...686A..93G}
}

@ARTICLE{godoyrivera18,
       author = {{Godoy-Rivera}, Diego and {Chanam{\'e}}, Julio},
        title = "{On the identification of wide binaries in the Kepler field}",
      journal = {\mnras},
         year = 2018,
        month = oct,
       volume = {479},
       number = {4},
        pages = {4440-4469},
          doi = {10.1093/mnras/sty1736},
archivePrefix = {arXiv},
       eprint = {1807.00009},
 primaryClass = {astro-ph.SR},
       adsurl = {https://ui.adsabs.harvard.edu/abs/2018MNRAS.479.4440G}
}

@ARTICLE{godoyrivera21a,
       author = {{Godoy-Rivera}, Diego and {Pinsonneault}, Marc H. and {Rebull}, Luisa M.},
        title = "{Stellar Rotation in the Gaia Era: Revised Open Clusters' Sequences}",
      journal = {\apjs},
         year = 2021,
        month = dec,
       volume = {257},
       number = {2},
          eid = {46},
        pages = {46},
          doi = {10.3847/1538-4365/ac2058},
archivePrefix = {arXiv},
       eprint = {2101.01183},
 primaryClass = {astro-ph.SR},
       adsurl = {https://ui.adsabs.harvard.edu/abs/2021ApJS..257...46G}
}

@ARTICLE{godoyrivera21b,
       author = {{Godoy-Rivera}, Diego and {Tayar}, Jamie and {Pinsonneault}, Marc H. and {Rodr{\'\i}guez Mart{\'\i}nez}, Romy and {Stassun}, Keivan G. and {van Saders}, Jennifer L. and {Beaton}, Rachael L. and {Garc{\'\i}a-Hern{\'a}ndez}, D.~A. and {Teske}, Johanna K.},
        title = "{Testing the Limits of Precise Subgiant Characterization with APOGEE and Gaia: Opening a Window to Unprecedented Astrophysical Studies}",
      journal = {\apj},
         year = 2021,
        month = jul,
       volume = {915},
       number = {1},
          eid = {19},
        pages = {19},
          doi = {10.3847/1538-4357/abf8ba},
archivePrefix = {arXiv},
       eprint = {2104.07679},
 primaryClass = {astro-ph.SR},
       adsurl = {https://ui.adsabs.harvard.edu/abs/2021ApJ...915...19G}
}

@ARTICLE{godoyrivera25,
       author = {{Godoy-Rivera}, D. and {Mathur}, S. and {Garc{\'\i}a}, R.~A. and {Pinsonneault}, M.~H. and {Santos}, {\^A}. R.~G. and {Beck}, P.~G. and {Grossmann}, D.~H. and {Schimak}, L. and {Bedell}, M. and {Merc}, J. and {Escorza}, A.},
        title = "{Kepler meets Gaia DR3: Homogeneous extinction-corrected color-magnitude diagram and binary classification}",
      journal = {\aap},
         year = 2025,
        month = apr,
       volume = {696},
          eid = {A243},
        pages = {A243},
          doi = {10.1051/0004-6361/202348735},
archivePrefix = {arXiv},
       eprint = {2501.18719},
 primaryClass = {astro-ph.SR},
       adsurl = {https://ui.adsabs.harvard.edu/abs/2025A&A...696A.243G}
}

@ARTICLE{godoyrivera26a,
       author = {{Godoy-Rivera}, Diego and {Grossmann}, Desmond H. and {Richey-Yowell}, Tyler and {Santos}, {\^A}ngela R.~G. and {Mathur}, Savita and {Garc{\'\i}a}, Rafael A.},
        title = "{Mining the Kepler Field: Atmospheric Parameters, Bolometric Corrections, and Luminosities}",
      journal = {Research Notes of the American Astronomical Society},
         year = 2026,
        month = mar,
       volume = {10},
       number = {3},
          eid = {53},
        pages = {53},
          doi = {10.3847/2515-5172/ae4fb6},
archivePrefix = {arXiv},
       eprint = {2603.18152},
 primaryClass = {astro-ph.SR},
       adsurl = {https://ui.adsabs.harvard.edu/abs/2026RNAAS..10...53G}
}

@article{godoyrivera26b,
  author       = {{Godoy-Rivera}, D. and {Richey-Yowell}, T. and {Santos}, A.~R.~G. and {Mathur}, S. and {Garc{\'\i}a}, R.~A. and {Grossmann}, D. H.},
  title        = {Completing the magnetic puzzle:
Photometric H-alpha as a stellar activity proxy},
  journal      = {Submitted to Frontiers in Astronomy and Space Sciences},
  year         = {2026}
}

@ARTICLE{gondoin18,
       author = {{Gondoin}, P.},
        title = "{Magnetic activity evolution on Sun-like stars}",
      journal = {\aap},
         year = 2018,
        month = sep,
       volume = {616},
          eid = {A154},
        pages = {A154},
          doi = {10.1051/0004-6361/201731541},
       adsurl = {https://ui.adsabs.harvard.edu/abs/2018A&A...616A.154G}
}

@ARTICLE{gondoin20,
       author = {{Gondoin}, P.},
        title = "{Chromospheric activity of nearby Sun-like stars. R$^{'}$$_{HK}$ index signature of a recent burst of star formation}",
      journal = {\aap},
         year = 2020,
        month = sep,
       volume = {641},
          eid = {A110},
        pages = {A110},
          doi = {10.1051/0004-6361/202038291},
       adsurl = {https://ui.adsabs.harvard.edu/abs/2020A&A...641A.110G}
}

@ARTICLE{gordon21,
       author = {{Gordon}, Tyler A. and {Davenport}, James R.~A. and {Angus}, Ruth and {Foreman-Mackey}, Daniel and {Agol}, Eric and {Covey}, Kevin R. and {Ag{\"u}eros}, Marcel A. and {Kipping}, David},
        title = "{Stellar Rotation in the K2 Sample: Evidence for Modified Spin-down}",
      journal = {\apj},
         year = 2021,
        month = may,
       volume = {913},
       number = {1},
          eid = {70},
        pages = {70},
          doi = {10.3847/1538-4357/abf63e},
archivePrefix = {arXiv},
       eprint = {2101.07886},
 primaryClass = {astro-ph.SR},
       adsurl = {https://ui.adsabs.harvard.edu/abs/2021ApJ...913...70G}
}

@ARTICLE{gruner20,
       author = {{Gruner}, D. and {Barnes}, S.~A.},
        title = "{Rotation periods for cool stars in the open cluster Ruprecht 147 (NGC 6774). Implications for gyrochronology}",
      journal = {\aap},
         year = 2020,
        month = dec,
       volume = {644},
          eid = {A16},
        pages = {A16},
          doi = {10.1051/0004-6361/202038984},
archivePrefix = {arXiv},
       eprint = {2010.02298},
 primaryClass = {astro-ph.SR},
       adsurl = {https://ui.adsabs.harvard.edu/abs/2020A&A...644A..16G}
}

@ARTICLE{gruner23a,
       author = {{Gruner}, D. and {Barnes}, S.~A. and {Janes}, K.~A.},
        title = "{Wide binaries demonstrate the consistency of rotational evolution between open cluster and field stars}",
      journal = {\aap},
         year = 2023,
        month = jul,
       volume = {675},
          eid = {A180},
        pages = {A180},
          doi = {10.1051/0004-6361/202346590},
archivePrefix = {arXiv},
       eprint = {2307.10836},
 primaryClass = {astro-ph.SR},
       adsurl = {https://ui.adsabs.harvard.edu/abs/2023A&A...675A.180G}
}

@ARTICLE{gruner23b,
       author = {{Gruner}, D. and {Barnes}, S.~A. and {Weingrill}, J.},
        title = "{New insights into the rotational evolution of near-solar age stars from the open cluster M 67}",
      journal = {\aap},
         year = 2023,
        month = apr,
       volume = {672},
          eid = {A159},
        pages = {A159},
          doi = {10.1051/0004-6361/202345942},
archivePrefix = {arXiv},
       eprint = {2305.16997},
 primaryClass = {astro-ph.SR},
       adsurl = {https://ui.adsabs.harvard.edu/abs/2023A&A...672A.159G}
}

@ARTICLE{gudel07,
       author = {{G{\"u}del}, Manuel},
        title = "{The Sun in Time: Activity and Environment}",
      journal = {Living Reviews in Solar Physics},
         year = 2007,
        month = dec,
       volume = {4},
       number = {1},
          eid = {3},
        pages = {3},
          doi = {10.12942/lrsp-2007-3},
archivePrefix = {arXiv},
       eprint = {0712.1763},
 primaryClass = {astro-ph},
       adsurl = {https://ui.adsabs.harvard.edu/abs/2007LRSP....4....3G}
}
\begin{appendix}
\section{Binning Procedure}
\label{sec:app_binning}

Part of the analysis presented throughout this paper is based on the trends obtained from percentiles of the distribution of {\logRIRT} in bins of {\sphaverage}, {\prot}, and $\mathrm{Ro}$. Here, we described the binning procedure used to obtain these percentiles. 

\begin{figure}[ht]
    \centering
    \includegraphics[width=0.92\hsize]{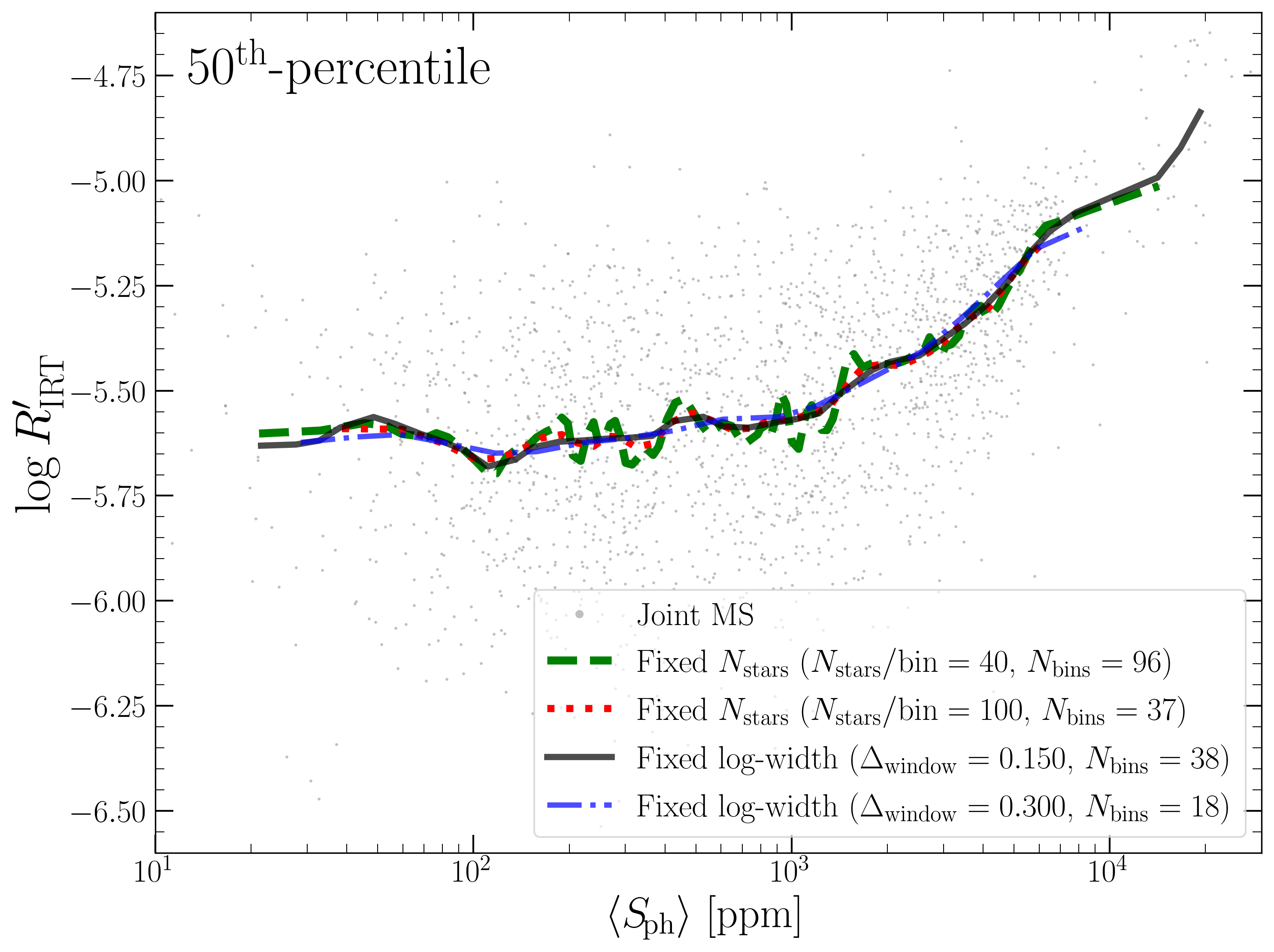}\\
    \includegraphics[width=0.92\hsize]{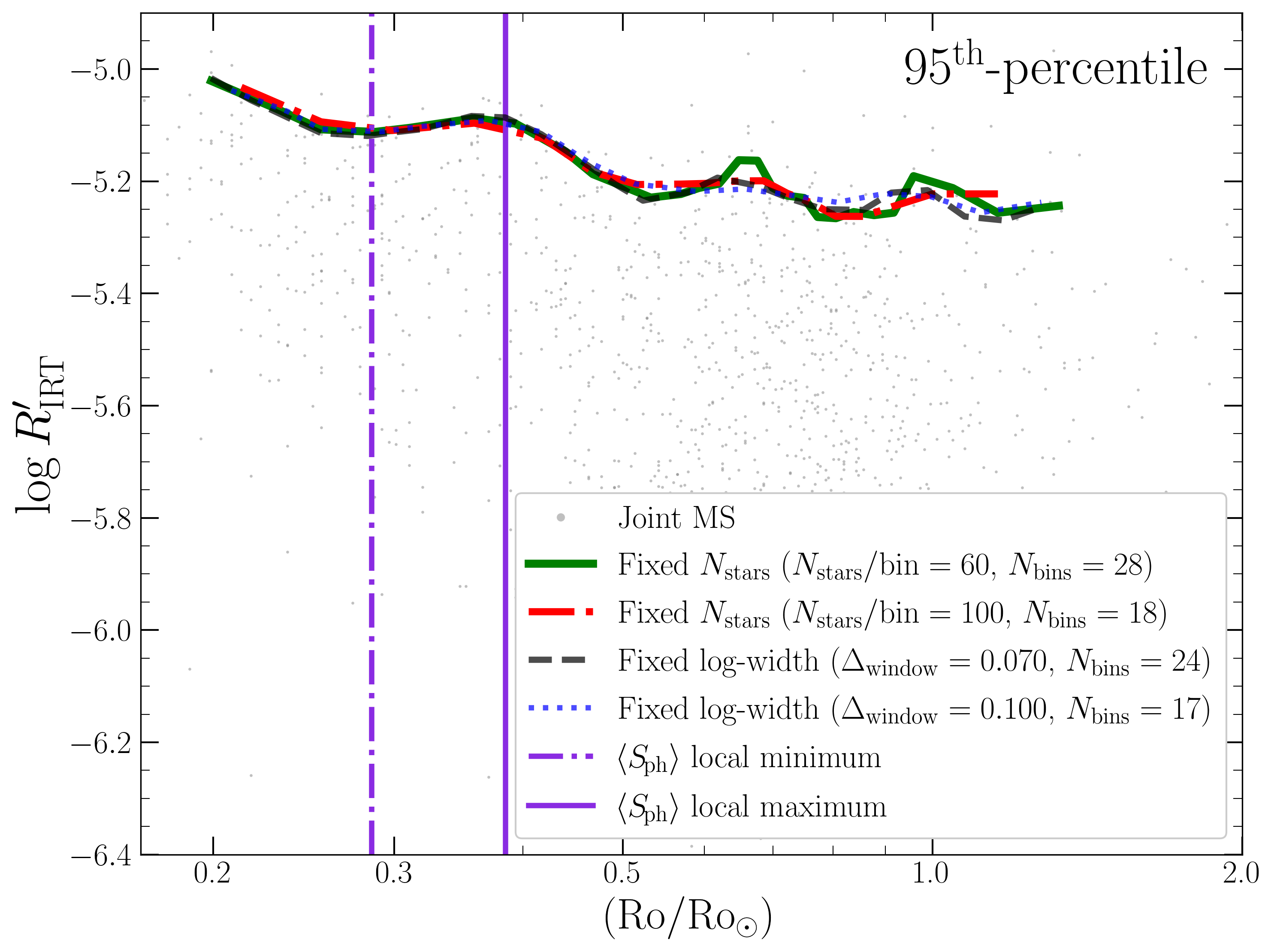}
    \caption{Impact of the binning procedure on the resulting {\logRIRT} trends and percentiles. Top: example with the 50$^{\mathrm{th}}$-percentile of {\logRIRT} vs. {\sphaverage} (from the top panel of Fig.~\ref{fig:Figure_characterization_sphlogRIRT}). Bottom: example with the 95$^{\mathrm{th}}$-percentile of {\logRIRT} vs. $\mathrm{Ro}$ (from Fig.~\ref{fig:Figure_discussion_rossby}). The two methods are fixed $N_\mathrm{stars}$ (green and red lines), and fixed logarithmic width (black and blue lines). Overall, the obtained trends are mostly insensitive to the specific binning method.}
    \label{fig:Figure_appendix_binning}
\end{figure}

The top panel of Fig.~\ref{fig:Figure_appendix_binning} illustrates this in the {\logRIRT} vs. {\sphaverage} diagram (see the top panel of Fig.~\ref{fig:Figure_characterization_sphlogRIRT}). From the scatter plot, we took two approaches to bin the data. In the first one, the stars were sorted by their {\sphaverage} values, and were binned by a fixed number of stars per bin ($N_{\mathrm{stars}}/\mathrm{bin}$). Within each bin, the 50$^{\mathrm{th}}$-percentile of the {\logRIRT} distribution was calculated, and the resulting trend was smoothed using a 1-D Gaussian convolution kernel. Two examples of this fixed $N_\mathrm{stars}$ approach are shown as the green and red lines. In the second approach, the data were binned by a fixed logarithmic {\sphaverage} width (of size $\Delta_\mathrm{window}$), and again the 50$^{\mathrm{th}}$-percentile of the {\logRIRT} distribution was calculated and the aforementioned smoothing was applied. Two examples of this fixed log-width approach are shown as the black and blue lines. The bottom panel of Fig.~\ref{fig:Figure_appendix_binning} shows the analogous exercise for the {\logRIRT} vs. $\mathrm{Ro}$ diagram (see Fig.~\ref{fig:Figure_discussion_rossby}), where the 95$^{\mathrm{th}}$-percentile of the {\logRIRT} distribution was calculated as a function of the different binning approaches.

Figure~\ref{fig:Figure_appendix_binning} shows that, regardless of the binning approach taken, or the specific parameters used, the same overall trends are seen in the diagrams. Throughout this paper, we chose the fixed logarithmic width approach (picking the parameters to account for the varying sample sizes and ranges covered), as it generally yielded smoother percentiles. All in all, the resulting trends are virtually insensitive to the choice of binning method.

\section{Classification of the Joint-MS sample into spectral types}
\label{sec:app_SpT_classification}

To separate the Joint-MS sample into spectral types, we used the de-reddened {\gaia} DR3 colors (Sect.~\ref{sec:data_targetrotation}) and the color-SpT table from \citet{pecaut13}, classifying the stars as: early-F dwarfs for $0.377\leq (BP-RP)_0 < 0.587$ mag, late-F dwarfs for $0.587\leq (BP-RP)_0 < 0.784$ mag, G dwarfs for $0.784\leq (BP-RP)_0 < 0.983$ mag, K dwarfs for $0.983\leq (BP-RP)_0 < 1.840$ mag, and  M dwarfs for $1.840\leq (BP-RP)_0 < 4.780$ mag. The distinction between early- and late-F dwarfs was taken as the SpT closest to the Kraft break location reported by \citet{beyer24}, corresponding to F5V (see also \citealt{wang26}). The assigned SpTs are reported as the ``Spectral Type'' column in Table~\ref{tab:table_catalog}.
\end{appendix}

\end{document}